\documentclass[11pt]{article}
\usepackage{amssymb}
\usepackage{amsmath}

\allowdisplaybreaks[3]

\makeatletter
\@addtoreset{equation}{section}
\makeatother

\def\dalemb#1#2{{\vbox{\hrule height .#2pt
        \hbox{\vrule width.#2pt height#1pt \kern#1pt
                \vrule width.#2pt}
        \hrule height.#2pt}}}

\def\cA{{\cal A}}

\def\cM{{\cal M}}

\def\0{{\sst{(0)}}}
\def\1{{\sst{(1)}}}
\def\2{{\sst{(2)}}}
\def\3{{\sst{(3)}}}
\def\4{{\sst{(4)}}}
\def\5{{\sst{(5)}}}
\def\6{{\sst{(6)}}}
\def\7{{\sst{(7)}}}
\def\8{{\sst{(8)}}}
\def\n{{\sst{(n)}}}

\def\ep{\epsilon}
\def\td{\tilde}

\def\half{{\textstyle{1\over2}}}
\def\hp{ \frac{1}{2}}
\def\qu{{\textstyle{1\over 4}}}

\let\a=\alpha \let\b=\beta \let\g=\gamma \let\d=\delta \let\e=\epsilon
\let\z=\zeta  \let\q=\theta  \let\k=\kappa
\let\l=\lambda \let\m=\mu \let\n=\nu  \let\r=\rho
\let\s=\sigma \let\t=\tau  \let\f=\phi  
\let\w=\omega    \let\L=\Lambda
    
 \let\W=\Omega   
\let\la=\label  
  
\def\nn{\nonumber} \def\bd{\begin{document}} \def\ed{\end{document}}
\def\ds{\documentstyle} \let\fr=\frac \let\bl=\bigl \let\br=\bigr
\let\Br=\Bigr \let\Bl=\Bigl
\let\bm=\bibitem
\let\na=\nabla
\let\pa=\partial \let\ov=\overline
\newcommand{\be}{\begin{equation}}
\newcommand{\ee}{\end{equation}}
\def\ba{\begin{array}}
\def\ea{\end{array}}
\def\ft#1#2{{\textstyle{{\scriptstyle #1}\over {\scriptstyle #2}}}}
\def\fft#1#2{{#1 \over #2}}
\def\del{\partial}
\def\sst#1{{\scriptscriptstyle #1}}
 \def\oneone{\rlap 1\mkern4mu{\rm l}}
\def\ie{{\it i.e.\ }}
\def\via{{\it via}}
\def\semi{{\ltimes}}
\def\str{{\rm str}}
\def\Dm{{{D_{\sst{max}}}}}
\def\vac{ \left | 0 \right \rangle }
\def\kvac{ \left | k \right \rangle }

\def\sp{\; \; \;}

\def\bol{ \left | B (p^+) \right \rangle}
\def\bo1{ \left | B^0 (p^+) \right \rangle}

\def\bolt{ \left | B (p^+) \right \rangle_{\t}}

\def\boxl{ \left | B (x^-) \right \rangle}

\def\<{ \langle }
\def\>{ \rangle }

\def\vf{\varphi}

\def\ls{{(l,0)}}
\def\lv{{(l,\pm1)}}
\def\lt{{(l,\pm2)}}

\def\lse#1{{(l_{#1},0)}}
\def\lve#1{{(l_{#1},\pm1)}}
\def\lte#1{{(l_{#1},\pm2)}}

\def\lsg#1{{5(l_{#1},0)}}
\def\lvg#1{{5(l_{#1},\pm1)}}
\def\ltg#1{{5(l_{#1},\pm2)}}

\def\lsi#1{{5{(#1,0)}}}
\def\lvi#1{{5{(#1,\pm1)}}}
\def\lti#1{{5{(#1,\pm2)}}}

\def\lsr#1{{1{(#1,0)}}}
\def\lvr#1{{1{(#1,\pm1)}}}
\def\ltr#1{{1{(#1,\pm2)}}}

\def\cD{{\cal D}}
\def\cE{{\cal E}}
\def\cF{{\cal F}}
\def\cG{{\cal G}}
\def\cH{{\cal H}}
\def\cK{{\cal K}}
\def\cO{{\cal O}}
\def\cP{{\cal P}}
\def\cQ{{\cal Q}}
\def\cR{{\cal R}}
\def\cS{{\cal S}}
\def\cT{{\cal T}}
\def\cU{{\cal U}}
\def\cV{{\cal V}}
\def\cW{{\cal W}}
\newcommand{\nono}{\nonumber}
\newcommand{\dtilde}[1]{\tilde{\tilde{#1}}}
\newcommand{\hatb}[1]{\hat{\ov{#1}}}
\newcommand{\hatt}[1]{\hat{\tilde{#1}}}
\newcommand{\emnr}{{e_\m}^{\n\r}}

\newcommand{\comment}[1]{}

\newcommand{\hsp}{\hspace{0.5cm}}

\newcommand{\ho}[1]{$\, ^{#1}$}
\newcommand{\hoch}[1]{$\, ^{#1}$}
\newcommand{\bea}{\begin{eqnarray}}
\newcommand{\eea}{\end{eqnarray}}
\newcommand{\ra}{\rightarrow}
\newcommand{\lra}{\longrightarrow}
\newcommand{\Lra}{\Leftrightarrow}
\newcommand{\ap}{\alpha^\prime}
\newcommand{\bp}{\tilde \beta^\prime}
\newcommand{\tr}{{\rm tr} }
\newcommand{\Tr}{{\rm Tr} }
\newcommand{\NP}{Nucl. Phys. }

\newcommand{\ams}{{\it Institute for Theoretical Physics,
University of Amsterdam, \\
Valckenierstraat 65, 1018XE Amsterdam, The Netherlands} \\
{\tt ingkanit, skenderi, taylor@science.uva.nl}}

\newcommand{\auth}{\large Ingmar Kanitscheider, Kostas Skenderis and Marika Taylor}

\begin{document}
\begin{flushright}
\hfill{ITFA-2007-09}
\end{flushright}

\vspace{15pt}

\begin{center}

{\Large \bf Fuzzballs with internal excitations}

\vspace{20pt}

\auth

\vspace{15pt}

\vspace{8pt}

{\ams}

\vspace{15pt}

\underline{ABSTRACT}
\end{center}

We construct general 2-charge D1-D5 horizon-free non-singular solutions of
IIB supergravity on $T^4$ and $K3$ describing fuzzballs with
excitations in the internal manifold; these excitations are
characterized by arbitrary curves. The solutions are obtained via dualities
from F1-P solutions of heterotic and type IIB on $T^4$ for the
$K3$ and $T^4$ cases, respectively. We compute the holographic
data encoded in these solutions, and show that the internal
excitations are captured by vevs of chiral primaries associated
with the middle cohomology of $T^4$ or $K3$. We argue that each
geometry is dual to a specific superposition of R ground states 
determined in terms of the Fourier coefficients of the curves defining
the supergravity solution. We compute vevs of 
chiral primaries associated with the middle
cohomology and show that they indeed acquire vevs 
in the superpositions corresponding to 
fuzzballs with internal excitations, in accordance with the
holographic results. We also address the
question of whether the fuzzball program can be implemented
consistently within supergravity.

\pagebreak

\tableofcontents
\pagebreak

\section{Introduction}

Over the last few years an interesting new proposal for the
gravitational nature of black hole microstates has emerged
\cite{Lunin:2001jy,Lunin:2002bj,Lunin:2002iz}; see also \cite{Lunin:2001fv,
Balasubramanian:2000rt, Maldacena:2000dr}, and \cite{Mathur:2005zp}. According
to this proposal there should exist non-singular horizon-free
geometries associated with the black hole microstates. These
so-called fuzzball geometries should asymptotically approach
the original black hole geometry and should generically differ from
each other around the horizon scale. In this scenario the black hole
provides only an average statistical description of the physics and
thus longstanding issues such as the information loss paradox
would be resolved. The underlying physics of the black hole would not be conceptually
different from that of a star, with the temperature and entropy being of
statistical origin. Given the importance of understanding black hole
physics and its implications for quantum gravity, this proposal
should be developed, explored and tested where possible.

Many issues need to be addressed to implement the fuzzball proposal at
a quantitative and precise level. The proposal requires the existence
of exponential numbers of horizon-free non-singular solutions for each
black hole. So the most basic of questions is whether one can find
such a number of solutions with the required properties and moreover
what precisely are the required properties for any given black hole.
Moreover one would like to show quantitatively how black hole
properties emerge upon coarse-graining; for this one needs to know the
precise relationship between geometries and microstates.

Much of the recent work on this proposal has been focused on
constructing fuzzball geometries for certain supersymmetric
black holes with macroscopic horizons; for a summary of progress in
this direction see \cite{Bena:2007kg}. The method of construction here
uses crucially supersymmetry and known classifications of
supersymmetric solutions: one looks for non-singular horizon-free
supersymmetric solutions with the correct charges to match those of
the black hole.

This method however has a number of limitations. One is that the
supersymmetric classifications are not sufficiently restrictive for
cases of interest and
thus one needs a specific ansatz to make progress. To date many of
the fuzzball geometries constructed are rather atypical; for example,
they have angular momenta much larger than those of a typical
black hole microstate. Whilst families of
typical geometries are presumably contained in the supersymmetric
classification,
finding an ansatz to construct families rather than isolated examples
is not easy.

Another key issue is that one
does not know precisely what is the relationship between a given
geometry and the black hole microstates. This in turn means that one
does not know whether one has constructed the correct geometries to
describe the black hole. Nor does one know whether one has enough geometries to
account for the black hole entropy upon geometric quantization, using the
methods of \cite{Rychkov:2005ji}. For example, in cases where the dual theory has
distinct Higgs and Coulomb branches, one needs to determine whether a
given fuzzball geometry describes Higgs or Coulomb branch physics.
More importantly, one would like to see explicitly how black hole
properties emerge upon coarse graining; to understand
how to do such a computation properly
the precise relation between the fuzzball geometries
and microstates is crucial.

To address this issue, we have advocated and developed
the use of AdS/CFT methods. That is, the supersymmetric
black holes of interest admit a dual CFT description and
the fuzzball geometries therefore
have a decoupling limit which is asymptotically AdS. One can
therefore use well-developed techniques of AdS/CFT, in particular
Kaluza-Klein holography \cite{Skenderis:2006uy}, to extract field
theory data from the geometry and diagnose precisely what the
geometry describes.

It is worth emphasizing at this point that the AdS/CFT correspondence
both motivates and supports the fuzzball picture. The gravity/gauge
theory dictionary relates a given asymptotically AdS geometry
to either a deformation of the CFT or the CFT in a non-trivial
vacuum characterized by the expectation values of gauge invariant operators.
Conversely, one expects that for any stable state of the CFT
(such as the BPS states) there exists
an asymptotically AdS solution, whose asymptotics encode the vevs
of gauge invariant operators in that state. If the field theory is in a pure
state, there is no entropy and one does not expect the corresponding
geometry to have a horizon, and hence entropy. AdS/CFT thus implies
that the field theory in a given pure stable (black hole) state should have a
geometric dual with no horizon; there is however no guarantee that the
geometry should be well-described by supergravity alone, i.e. weakly
curved everywhere\footnote{There are additional subtleties in low
dimensional quantum field theories due to the strong infrared fluctuations.
More properly  one should view a given
fuzzball geometry as dual to a wavefunction on the Higgs branch of the
field theory, but it seems in any case likely that such wavefunctions
would be localized around specific regions in the large $N$ limit and
thus that this issue does not play a key role at infinite $N$.}.

In our recent papers \cite{Skenderis:2006ah,Kanitscheider:2006zf},
we have discussed in some detail the case of the
D1-D5 system, for which (some) fuzzball geometries were constructed in
\cite{Lunin:2001jy}. Since this is a 2-charge system, there is no macroscopic
horizon: the naive geometry is singular, with the horizon believed to
form on taking into account $\a'$ corrections. Whilst this is not a
macroscopic black hole system, there are a number of reasons to
explore this case fully before moving on to supersymmetric macroscopic
black holes.

Firstly, one can obtain {\it{all}} fuzzball geometries in this system by
dualities from known solitonic solutions of F1-P systems. Thus one
should be able to account for all the entropy, and show how the
average black hole description emerges. Moreover, the dual description
of this black hole is the simplest and best understood: the black hole
entropy arises from the degeneracy of the Ramond ground states of the
dual $(4,4)$ CFT. This is an ideal system in which to address the
question of what is the precise correspondence between geometries and
microstates, and moreover how the properties of given microstates
determine and characterize the fuzzball geometries.

In the original work of \cite{Lunin:2001jy}, only a subset of the 2-charge
fuzzball geometries were constructed using dualities from F1-P
solutions. Recall that the D1-D5 system on $T^4$ is related by
dualities to the type II F1-P system, also on $T^4$, whilst the D1-D5 system
on $K3$ is related to the heterotic F1-P system on $T^4$; the exact
duality chains needed will be reviewed in sections \ref{t4} and
\ref{s-k3}. Now the solution for a fundamental string carrying momentum
in type II is characterized by 12
arbitrary curves, eight associated with transverse bosonic excitations and four
associated with the bosonization of eight fermionic excitations on the
string \cite{Taylor:2005db}. The corresponding heterotic string solution is characterized by 24
arbitrary curves, eight associated with transverse bosonic excitations
and 16 associated with charge waves on the string.

In the work of \cite{Lunin:2001jy}, the duality chain was carried out
for type II
F1-P solutions on $T^4$ for which only bosonic excitations in the
transverse $R^4$ are excited. That is, the solutions are characterized
by only four arbitrary curves; in the dual D1-D5 solutions these four
curves characterize the blow-up of the branes, which in the naive
solutions are sitting in the origin of the transverse
$R^4$, into a supertube. In this paper we carry out the dualities for generic
F1-P solutions in both the $T^4$ and $K3$ cases, to obtain generic
2-charge fuzzball solutions with internal excitations.
Note that
partial results for the $T^4$ case were previously given in the appendix of
\cite{Lunin:2002iz}; we will comment on the relation between our solutions
and theirs in section \ref{t4}.
The general solutions
are then characterized by arbitrary curves capturing excitations along
the compact manifold $M^4$, along with the four curves describing the
blow-up in $R^4$. They describe a bound state of D1 and D5-branes,
wrapped on the compact manifold $M^4$, blown up into a rotating supertube in
$R^4$ and with excitations along the part of the D5-branes wrapping
the $M^4$.

The duality chain that uses
string-string duality from heterotic on $T^4$ to type II on K3 provides
a route for obtaining fuzzball solutions that has
not been fully explored.  One of the results in this paper is to make
explicit all steps in this duality route. In particular, we work out the
reduction of type IIB on K3 and show how S-duality acts in six dimensions.
These results may be useful in obtaining fuzzball solution with more charges.
%Let us note one may also view our results as additional supporting evidence
%for string-string duality.

In our previous work \cite{Skenderis:2006ah,Kanitscheider:2006zf}, we made
a precise proposal for the
relationship between the 2-charge fuzzball geometries characterized by
four curves $F^i(v)$ and superpositions of R ground states: {\it a
  given geometry characterized by $F^i(v)$ is dual to a specific
  superposition of R vacua with the superposition determined by the
  Fourier coefficients of the curves $F^i(v)$}. In particular, note
that only geometries associated with circular curves are dual to a
single R ground state (in the usual basis, where the states are
eigenstates of the R-charge). This proposal has a straightforward extension
to generic 2-charge geometries, which we will spell out in section
\ref{s-field}, and the extended proposal passes all kinematical
and accessible dynamical tests, just as in
\cite{Skenderis:2006ah,Kanitscheider:2006zf}.

In particular, we extract one point functions for chiral primaries
from the asymptotically AdS region of the fuzzball solutions. We find
that chiral primaries associated with the middle cohomology of $M^4$
acquire vevs when there are both internal and transverse excitations;
these vevs hence characterize the internal excitations. Moreover,
there are selection rules for these vevs, in that the internal and
transverse curves must have common frequencies.

These properties of the holographic vevs follow directly from the
proposed dual superpositions of ground states. The vevs in these
ground states can be derived from three point functions between chiral
primaries at the conformal point. Selection rules for the latter,
namely charge conservation and conservation of the number of operators associated
with each middle cohomology cycle, lead to precisely the
features of the vevs found holographically.

To test the actual values of the kinematically allowed vevs would
require information about the three point functions of all chiral
primaries which is not currently known and is inaccessible in supergravity.
However, as in \cite{Kanitscheider:2006zf}, these vevs are
reproduced surprisingly well by simple approximations for the three
point functions, which follow from treating the operators as harmonic
oscillators. This suggests that the structure of the chiral ring may
simplify considerably in the large $N$ limit, and it would be
interesting to explore this question further.

An interesting feature of the solutions is that they collapse to
the naive geometry when there are internal but
no transverse excitations. One can understand this as follows.
Geometries with only internal excitations
are dual to superpositions of R ground states built from operators
associated with the middle cohomology of $M^4$. Such operators account
for a finite fraction of the entropy, but have zero R charges with
respect to the $SO(4)$ R symmetry group. This
means that they can only be characterized by the vevs of $SO(4)$
singlet operators but the only such operators visible in supergravity
are kinematically prevented from acquiring vevs. Thus it is consistent
that in supergravity one could not distinguish between such solutions: one
would need to go beyond supergravity to resolve them (by, for instance,
considering vevs of singlet operators dual to string states).

This brings us to a recurring question in the fuzzball program: can it
be implemented consistently within supergravity? As already mentioned,
rigorously testing the proposed correspondence between geometries and
superpositions of microstates requires information beyond
supergravity. Furthermore, the geometric duals of superpositions with
very small or zero R charges are not well-described in
supergravity. Even if one has geometries which are smooth supergravity
geometries, these may not be distinguishable from each other within
supergravity: for example, their vevs may differ only by terms of
order $1/N$, which cannot be reliably computed in supergravity.

The question of whether the fuzzball program can be implemented in
supergravity could first be phrased in the following way. Can one find
a complete basis of fuzzball geometries, each of which is well-described
everywhere by supergravity, which are distinguishable from each other within
supergravity and which together span the black hole microstates?
On general grounds one would expect this not to be possible
since many of the microstates carry small quantum numbers.
We quantify this discussion in the last section of this paper in the
context of both 2-charge and 3-charge systems.

To make progress within supergravity, however, it would suffice to
sample the black hole microstates in a controlled way. I.e. one could
try to find a basis of geometries which are well-described and distinguishable
in supergravity and which span the black hole microstates but for
which each basis element is assigned a measure. In this approach, one would deal
with the fact that many geometries are too similar to be distinguished
in supergravity by picking representative geometries with
appropriate measures. In constructing such a representative basis, the
detailed matching between geometries and black hole microstates would
be crucial, to correctly assign measures and to show that the basis
indeed spans all the black hole microstates.

\bigskip

The plan of this paper is as follows. In section \ref{t4} we determine
the fuzzball geometries for D1-D5 on $T^4$ from dualizing type II F1-P
solutions whilst in section \ref{s-k3} we obtain
fuzzball geometries for D1-D5 on $K3$ from dualizing heterotic F1-P
solutions. The resulting solutions are of the same form and are
summarized in section \ref{summary}; readers interested only in the
solutions may skip sections 2 and 3. In section \ref{s-vevs} we extract from
the asymptotically AdS regions the dual field theory data, one point
functions for chiral primaries. In section \ref{s-field} we discuss
the correspondence between geometries and R vacua, extending the
proposal of \cite{Skenderis:2006ah,Kanitscheider:2006zf} and using the
holographic vevs to test this proposal. In section \ref{final} we
discuss more generally the implications of our results for the
fuzzball proposal. Finally there are a number of appendices.
In appendix A we state our conventions for the field equations and duality
rules, in appendix B we discuss in detail the reduction of type IIB on
K3 and
appendix C summarizes relevant properties of spherical harmonics.
In appendix D we discuss fundamental string solutions with winding
along the torus, and the corresponding duals in the D1-D5
system. In appendix E we derive the density of ground states
with fixed R charges.

\section{Fuzzball solutions on $T^4$} \la{t4}

In this section we will obtain general 2-charge solutions for the
D1-D5 system on $T^4$ from type II F1-P solutions.

\subsection{Chiral null models}

Let us begin with a general chiral null model of ten-dimensional
supergravity, written in the form
\bea \la{cnm}
ds^2 &=& H^{-1} (x,v) dv (- du + K (x,v) dv + 2 A_{I} (x,v) dx^I) + dx^I
dx_I; \\
e^{-2 \Phi} &=& H(x,v); \hsp
B^{(2)}_{uv} = \half (H(x,v)^{-1} -1); \hsp
B^{(2)}_{vI} = H(x,v)^{-1} A_I (x,v). \nn
\eea
The conventions for the supergravity field equations are given in the
appendix \ref{fe}. The above is a solution of the equations of motion provided
that
the defining functions are harmonic in the transverse directions,
labeled by $x^I$:
\be
   \Box H (x,v) = \Box K (x,v) = \Box A_I (x,v) = (\pa_I A^I (x,v) - \pa_v
   H (x,v)) = 0.
\ee
Solutions of these equations appropriate for describing solitonic
fundamental strings carrying momentum were given in
\cite{Callan:1995hn,Dabholkar:1995nc}:
\be
   H = 1+\frac{Q}{|x-F(v)|^6},
\qquad A_I = -\frac{Q\dot{F}_I(v)}{|x-F(v)|^6}, \qquad
   K = \frac{Q^2 \dot{F}(v)^2 }{Q|x-F(v)|^6},
\ee
where $F^I(v)$ is an arbitrary null curve describing the transverse
location of the string, and $\dot{F}^I$ denotes $\pa_v F^I(v)$.
More general solutions appropriate for describing solitonic strings
with fermionic condensates were discussed in
\cite{Taylor:2005db}. Here we will dualise without using the explicit
forms of the functions, thus the resulting dual supergravity
solutions are applicable for all choices of harmonic functions.

The F1-P solutions described by such chiral null models can be
dualised to give corresponding solutions for the D1-D5 system as
follows. Compactify four of the transverse directions on a torus,
such that $x^i$ with $i=1,\cdots,4$ are coordinates
on $R^4$ and $x^{\r}$ with $\r = 5,\cdots,8$ are coordinates on $T^4$.
Then let $v = (t -y)$ and $u = (t+y)$ with the coordinate $y$ being
periodic with length $L_y \equiv 2 \pi R_y$,
and smear all harmonic functions over both this circle and
over the $T^4$, so that they satisfy
\be \la{equ_harmonics}
 \Box_{R^4} H (x) = \Box_{R^4} K (x) = \Box_{R^4} A_I (x) =0, \hsp
 \pa_{i} A^i = 0.
\ee
Thus the harmonic functions appropriate for describing strings with
only bosonic condensates are
\bea
H &=& 1 + \frac{Q}{L_y} \int_0^{L_y} \frac{dv}{|x-F(v)|^2};
\qquad A_i = -\frac{Q}{L_y}\int_0^{L_y} \frac{dv \dot{F}_i(v)}{|x-F(v)|^2}; \\
A_{\rho} &=& -\frac{Q}{L_y} \int_0^{L_y} \frac{dv \dot{F}_{\rho}(v)} {|x-F(v)|^2}; \qquad
   K = \frac{Q }{L_y} \int_0^{L_y} \frac{dv (\dot{F}_i(v)^2 +
\dot{ F}_{\rho} (v)^2)} {|x-F(v)|^2}. \nn
\eea
Here $|x-F(v)|^2$ denotes $\sum_{i} (x_i - F_i(v))^2$. Note that
neither $\dot{F}_{i}(v)$ nor $\dot{F}_{\rho} (v)$ have zero modes;
the asymptotic expansions of $A_I$ at large $|x|$ therefore begin at order
$1/|x|^3$. Closure of the curve in $R^4$ automatically implies
that $\dot{F}_i (v)$ has no zero modes.
The question of whether $\dot{F}_{\rho}(v)$ has zero modes is more
subtle: since the torus coordinate  $x^{\rho}$ is periodic, the curve
$F_{\rho}(v)$ could have winding modes. As we will discuss in appendix
\ref{wind}, however, such winding modes are possible only when the
worldsheet theory is deformed by constant $B$ fields. The
corresponding supergravity
solutions, and those obtained from them by dualities, should thus not be
included in describing BPS states in the original 2-charge systems.

The appropriate chain of dualities to the $D1-D5$ system is
\be \la{chain1}
\left ( \begin{array}{c} P_y \\ F1_y  \end{array} \right )
\stackrel{S}{\rightarrow}
\left ( \begin{array}{c} P_y \\ D1_y  \end{array} \right )
\stackrel{T5678}{\rightarrow}
\left ( \begin{array}{c} P_y \\ D5_{y5678}  \end{array} \right )
\stackrel{S}{\rightarrow}
\left ( \begin{array}{c} P_y \\ NS5_{y5678}  \end{array} \right )
\stackrel{Ty}{\rightarrow}
\left ( \begin{array}{c} F1_y \\ NS5_{y5678}  \end{array} \right ),
\ee
to map to the type IIA NS5-F1 system. The subsequent dualities
\be \la{chain2}
\left ( \begin{array}{c} F1_y \\ NS5_{y5678}  \end{array} \right )
\stackrel{T8}{\rightarrow}
\left ( \begin{array}{c} F1_y \\ NS5_{y5678}  \end{array} \right )
\stackrel{S}{\rightarrow}
\left ( \begin{array}{c} D1_y \\ D5_{y5678}  \end{array} \right )
\ee
result in a D1-D5 system. Here the subscripts of $Dp_{a_1\cdots a_p}$
denote the spatial directions wrapped by the brane. In carrying out
these dualities we use the rules reviewed in appendix \ref{duality}.
We will give details of the intermediate solution in the type IIA NS5-F1
system since it differs from that obtained in \cite{Lunin:2002iz}.

\subsection{The IIA F1-NS5 system}

By dualizing the chiral null model from the F1-P system in IIB to F1-NS5 in IIA
one obtains the solution
\bea
   ds^2 &=& \td{K}^{-1} [-(dt-A_i dx^i)^2+(dy-B_i dx^i)^2] + H dx_i dx^i + dx_{\rho}
dx^{\rho} \nono \\
   e^{2\Phi} &=& \td{K}^{-1} H , \qquad B^{(2)}_{ty} = \td{K}^{-1} -
   1, \\
%\qquad B^{(2)}_{ti} = -\td{K}^{-1} B_i, \nono \\
%   B^{(2)}_{yi} &=& \td{K}^{-1} A_i,
B^{(2)}_{\bar{\mu} i} &=& \td{K}^{-1} {\cal B}^{\bar{\mu}}_{i}, \qquad
\qquad B^{(2)}_{ij} = -c_{ij} + 2 \td{K}^{-1}
A_{[i}B_{j]} \nono \\
   C^{(1)}_{\rho} &=& H^{-1} A_{\rho}, \qquad  C^{(3)}_{ty\rho} = (H
   \td{K})^{-1} A_{\rho}, \qquad
C^{(3)}_{\bar{\mu} i \rho} =  (H \td{K})^{-1} {\cal B}^{\bar{\mu}} _i A_{\rho}, \nn
\\
%   C^{(3)}_{ti{\rho}} &=& -\frac{B_i \cA_{\rho}}{H \td{K}}, \qquad C^{(3)}_{yi{\rho}} = \frac{A_i
%A_{\rho}}{H \td{K}}, \nono \\
   C^{(3)}_{ij \rho} &=& (\l_{\rho})_{ij} + 2 (H \td{K})^{-1} A_{\rho} A_{[i}B_{j]},
   \qquad C^{(3)}_{\rho \sigma \tau} = \e_{\rho \sigma \tau \pi}H^{-1}A^{\pi}, \nono
\eea
where
\bea
   \td{K} &=& 1 + K - H^{-1} A_{\rho} A_{\rho}, \qquad dc = -\ast_4
   dH, \qquad dB = -\ast_4 dA, \\
d\l_{\rho} &=& \ast_4 dA_{\rho}, \qquad
{\cal B}^{\bar{\mu}}_i = (-B_i,A_i), \nn
\eea
with $\bar{\mu} = (t,y)$. Here
the transverse and torus directions are denoted by $(i,j)$ and
$(\rho,\sigma)$ respectively and $\ast_4$ denotes the Hodge dual in the flat
metric on $R^4$, with $\ep_{\rho \s \t \pi}$ denoting the Hodge dual in flat
$T^4$ metric. The defining functions satisfy the equations
given in (\ref{equ_harmonics}).

The RR field strengths corresponding to the above potentials are
\bea
   F^{(2)}_{i \rho} &=& \pa_i(H^{-1} A_{\rho}), \qquad F^{(4)}_{tyi {\rho}} =
\td{K}^{-1} \pa_i(H^{-1} A_{\rho}), \nn \\
 F^{(4)}_{\bar{\mu} ij {\rho}} &=& 2 \td{K}^{-1} {\cal B}^{\bar{\mu}}_{[i} \pa_{j]} (H^{-1}
   A_{\rho}) , \qquad
\qquad F^{(4)}_{i \r\s\t} = \e_{\r\s\t\pi} \pa_i
   (H^{-1} A^{\pi}), \\
%   F^{(4)}_{tij{\rho}} &=& -\frac{2 B_{[i} \pa_{j]} (H^{-1} \cA_{\rho})}{\td{K}},
%\qquad F^{(4)}_{yij{\rho}} = -\frac{2 A_{[i} \pa_{j]} (H^{-1}
%  \cA_{\rho})}{\td{K}}, \nn \\
   F^{(4)}_{ijk \rho} &=& \td{K}^{-1} \left ( 6 A_{[i}B_j \pa_{k]}(H^{-1} A_{\rho}) + H
     \e_{ijkl} \pa^l (H^{-1} A_{\rho}) \right ). \nono
\eea
Thus the solution describes NS5-branes wrapping the $y$ circle and the $T^4$,
bound to fundamental strings delocalized on the $T^4$ and
wrapping the $y$ circle, with additional
excitations on the $T^4$. These excitations break the $T^4$ symmetry
by singling out a direction within the torus, and source multipole
moments of the RR fluxes; the solution however has no net D-brane
charges.

Now let us briefly comment on the relation between this solution and
that presented in appendix B of \cite{Lunin:2002iz}\footnote{We thank Samir Mathur
  for discussions on this issue.}. The NS-NS sector
fields agree, but the RR fields are different; in \cite{Lunin:2002iz}
they are given as 1, 3 and 5-form potentials. The relation of these
potentials to field strengths (and the corresponding field equations) is
not given in \cite{Lunin:2002iz}. As
reviewed in appendix \ref{duality}, in the presence of both electric
and magnetic sources it is rather natural to use the so-called
democratic formalisms of supergravity \cite{Ber01}, in which one
includes $p$-form field strengths with $p > 5$ along with constraints
relating higher and lower form field strengths. Any solution written
in the democratic formalism can be rewritten in terms of the standard
formalism, appropriately eliminating the higher form field
strengths. If one interprets the RR forms of \cite{Lunin:2002iz} in
this way, one does not however obtain a supergravity solution in the
democratic formalism; the Hodge duality constraints between higher and
lower form field strengths are not satisfied. Furthermore, one would
not obtain from the RR fields of \cite{Lunin:2002iz} the solution
written here in the standard formalism, after eliminating the higher
forms.

\subsection{Dualizing further to the D1-D5 system}

The final steps in the duality chain are T-duality along a torus
direction, followed by S-duality. When T-dualizing further along a
torus direction to a F1-NS5 solution in
IIB, the excitations along the torus mean that
the dual solution depends explicitly on the chosen T-duality
cycle in the torus. We will discuss the physical interpretation of the
distinguished direction in section \ref{summary}.
In the following the T-duality is taken along the $x^8$
direction, resulting in the following D1-D5 system:
\bea
\label{equ_D1D5T4}
   ds^2 &=& \frac{{f}_1^{1/2} }{ f_5^{1/2}
     \td{f}_1} [-(dt-A_i dx^i)^2+(dy- B_i dx^i)^2] +
     f_1^{1/2} f_5^{1/2} dx_i dx^i + f_1^{1/2} f_5^{- 1/2}
dx_{\rho} dx^{\rho} \nono \\
   e^{2\Phi} &=& \frac{f_1^2}{f_5 \td{f}_1}, \qquad B^{(2)}_{ty} =
   \frac{\cA}{f_5 \td{f}_1},
\qquad B^{(2)}_{\bar{\mu}i} = \frac{\cA {\cal B}^{\bar{\mu}}_i}{f_5 \td{f}_1}, \\
B^{(2)}_{ij} &=&
\l_{ij} + \frac{2\cA A_{[i} B_{j]}} {f_5 \td{f}_1}, \qquad B^{(2)}_{\a\b} =
-\e_{\a\b\g} f_{5}^{-1} \cA^{\g},\qquad B^{(2)}_{\a8} = f_{5}^{-1} \cA_{\a}, \nono \\
   C^{(0)} &=& - f_{1}^{-1} \cA , \qquad
   C^{(2)}_{ty} = 1- \td{f}_1^{-1}, \qquad
C^{(2)}_{\bar{\mu}i} = - \td{f}_1^{-1} {\cal B}^{\bar{\mu}}_i, \nn \\
C^{(2)}_{ij} &=&  c_{ij} - 2 \td{f}_1^{-1}  A_{[i}B_{j]}, \qquad
   C^{(4)}_{tyij} = \l_{ij} + \frac{\cA}{ f_5 \td{f}_1}  (c_{ij} + 2
     A_{[i}B_{j]}), \nn \\
C^{(4)}_{\bar{\mu}ijk} &=& \frac{3\cA}{f_5 \td{f}_1} {\cal B}^{\bar{\mu}}_{[i}c_{jk]},
\qquad C^{(4)}_{ty\a\b} =
-\e_{\a\b\g} f_{5}^{-1} \cA^\g, \qquad C^{(4)}_{ty\a8} = f_{5}^{-1} \cA_\a, \nono
\\
   C^{(4)}_{\a\b\g8} &=& \e_{\a\b\g} f_{5}^{-1} \cA, \qquad C^{(4)}_{ij\a8} =
(\l_{\a})_{ij} + f_{5}^{-1} \cA_{\a} c_{ij}, \qquad C^{(4)}_{ij\a\b} =
-\e_{\a\b\g}({\l^{\g}}_{ij} + f_{5}^{-1} \cA^{\g} c_{ij}), \nono
\eea
where
\bea
&&   f_5 \equiv H, \qquad \td{f}_1 = 1 + K - H^{-1}(\cA_{\a} \cA_{\a}
+ (\cA)^2), \qquad f_1 = \td{f}_1 + H^{-1} (\cA)^2, \nn \\
&& dc = -\ast_4 dH, \qquad dB = -\ast_4 dA, \qquad
{\cal B}^{\bar{\mu}}_i = (-B_i,A_i), \la{redef} \\
&& d\l_{\a} = \ast_4
     d\cA_{\a}, \qquad d \l = \ast_4 d {\cal A}. \nn
\eea
Here $\bar{\mu} = (t,y)$ and we denote $A_8$ as ${\cal A}$ with
the remaining $A_{\rho}$ being denoted by $\cA_{\a}$ where
the index $\a$ runs over only  $5,6,7$. The Hodge dual over these
     coordinates is denoted by $\ep_{\a \b\g}$.
Explicit expressions for
these defining harmonic functions in terms of variables of the
D1-D5 system  will be given in section \ref{summary}.

The forms with components along the torus directions can be
written more compactly as follows. Introduce a basis of self-dual and
anti-self dual 2-forms on the torus such that
\be \la{tt1}
\omega^{\a_{\pm}} = \frac{1}{\sqrt{2}}
(dx^{4 + \a_{\pm}} \wedge dx^8 \pm \ast_{T^4} (dx^{4 + \a_{\pm}} \wedge dx^8) ),
\ee
with $\a_{\pm} = 1,2,3$. These forms are normalized such that
\be \la{tt2}
\int_{T^4} \omega^{\a_{\pm}} \wedge \omega^{\b_{\pm}} = \pm
(2 \pi)^4 V \d^{\a_{\pm} \b_{\pm}},
\ee
where $(2 \pi)^4 V$ is the volume of the torus.
Then the potentials wrapping the torus directions can be expressed as
\bea
B^{(2)}_{\rho \sigma} &=& C^{(4)}_{ty \rho \sigma} =
\sqrt{2} f_{5}^{-1} \cA^{\a_-} \w^{\a_-}_{\rho \sigma}, \\
C^{(4)}_{ij\rho \sigma}
&=& \sqrt{2} \left ((\l_{ij})^{\a_-} + f_{5}^{-1} \cA^{\a_-} c_{ij}
\right ) \w^{\a_-}_{\rho \sigma}, \nono \\
C^{(4)}_{\rho \sigma \tau \pi} &=& \e_{\rho \sigma \tau \pi}
f_{5}^{-1} \cA, \nn
\eea
with $\e_{\rho \sigma \tau \pi}$ being the Hodge dual in the flat metric on
$T^4$. Note that these fields are expanded only in the anti-self dual two-forms, with
neither the self dual two-forms nor the odd-dimensional forms on
the torus being switched on anywhere in the solution. As we will
discuss later, this means the corresponding six-dimensional solution can
be described in chiral $N = 4b$ six-dimensional supergravity. The
components of forms associated with the odd cohomology of $T^4$ reduce
to gauge fields in six dimensions which are contained in the full $N
=8$ six-dimensional supergravity, but not its truncation to $N = 4b$.

\section{Fuzzball solutions on $K3$} \la{s-k3}

In this section we will obtain general 2-charge solutions for the
D1-D5 system on $K3$ from F1-P solutions of the heterotic string.

\subsection{Heterotic chiral model in 10 dimensions}

The chiral model for the charged heterotic F1-P system in 10
dimensions is:
\bea \la{het1}
   ds^2 &=& H^{-1}(-dudv + (K-2\alpha^\prime
   H^{-1}N^{(c)}N^{(c)})dv^2+ 2A_I dx^I dv) + dx_I dx^I \nn \\
   \hat{B}^{(2)}_{uv} &=& \hp(H^{-1}-1), \qquad \hat{B}^{(2)}_{vI} = H^{-1} A_{I}, \\
\qquad \hat{\Phi} &=& -\hp \ln H,  \qquad
   \hat{V}_v^{(c)} = H^{-1} N^{(c)}, \nn
\eea
where $I=1,\cdots,8$ labels the transverse directions and $\hat{V}_m^{(c)}$ are
Abelian gauge fields, with
$((c)=1,\cdots,16)$ labeling the elements of the Cartan of the gauge group.
The fields are denoted with hats to distinguish them from the
six-dimensional fields used in the next subsection.
The equations of motion for the heterotic string are given in appendix
\ref{fe}; here again the defining functions satisfy
\be
   \Box H (x,v) = \Box K (x,v) = \Box A_I (x,v) = (\pa_I A^I (x,v) - \pa_v
   H (x,v)) = \Box N^{(c)} = 0.
\ee
For the solution to correspond to a solitonic charged heterotic
string, one takes the following solutions
\bea
\la{K3harmonics}
   H &=& 1+\frac{Q}{|x-F(v)|^6},
\qquad A_I = - \frac{Q\dot{F}_I(v)}{|x-F(v)|^6}, \qquad N^{(c)} =
\frac{q^{(c)}(v)}{|x-F(v)|^6}, \nono \\
   K &=& \frac{Q^2 \dot{F}(v)^2+2\alpha^\prime q^{(c)}q^{(c)}(v)}{Q|x-F(v)|^6},
\eea
where $F^I(v)$ is an arbitrary null curve in $R^8$; $q^{(c)}(v)$ is an
arbitrary charge wave and $\dot{F}_I(v)$ denotes $\pa_v F_I(v)$. Such
solutions were first discussed in
\cite{Callan:1995hn,Dabholkar:1995nc}, although the above has
a more generic charge wave, lying in $U(1)^{16}$ rather than $U(1)$.
In what follows it will be convenient to set $\a' = \qu$.

These solutions can be related by a duality chain to fuzzball
solutions in the D1-D5 system compactified on $K3$. The chain of
dualities is the following:
\be \la{chain3}
\left ( \begin{array}{c} P_y \\ F1_y  \end{array} \right )_{Het, T^4}
{\rightarrow}
\left ( \begin{array}{c} P_y \\ NS5_{ty,K3}  \end{array} \right
)_{IIA}
\stackrel{T_y}{\rightarrow}
\left ( \begin{array}{c} F1_y \\ NS5_{ty,K3}  \end{array} \right
)_{IIB}
\stackrel{S}{\rightarrow}
\left ( \begin{array}{c} D1_y \\ D5_{ty,K3}  \end{array} \right
)_{IIB}
\ee
The first step in the duality is string-string duality between the
heterotic theory on $T^4$ and type IIA on $K3$. Again the subscripts
of $Dp_{a_1 \cdots
  a_p}$ denote the spatial directions wrapped by the brane. To use
this chain of dualities on the charged solitonic strings given above,
the solutions must be smeared over the $T^4$ and over $v$, so
that the harmonic functions satisfy
\be
  \Box_{R^4} H = \Box_{R^4} K = \Box_{R^4} A_I  = \Box_{R^4} N^{(c)} =
  \pa_i A^i = 0
\ee
where $i = 1,\cdots,4$ labels the transverse $R^4$
directions. Note that although the chain of dualities is
shorter than in the previous case there are various subtleties
associated with it, related to the K3 compactification, which will be
discussed below.

\subsection{Compactification on $T^4$}
\la{compT4}

Compactification of the heterotic theory
on $T^4$ is straightforward, see \cite{Sen:1994fa, Maharana:1992my}
and the review \cite{You97}.
The 10-dimensional metric is reduced as
\begin{equation}
\hat{G}_{mn}=\left(\begin{array}{cc}
g_{MN}+
G_{{\rho}{\sigma}}V^{(1)\,\rho}_{{M}}V^{(1)\,\sigma}_{{N}} & V^{(1)\,\rho}_{{M}}
G_{\rho \sigma}  \cr  V^{(1)\,\sigma}_{N}G_{\rho \sigma} & G_{\rho \sigma} \end{array} \right),
\label{4dkk}
\end{equation}
where $V^{(1)\,\rho}_{M}$ with $\rho = 1,\cdots 4$, are KK gauge fields.
(Recall that the ten-dimensional quantities are denoted with
hats to distinguish them from six-dimensional quantities.)
The reduced theory contains the following bosonic fields: the graviton
$g_{MN}$, the six-dimensional dilaton $\Phi_6$, 24 Abelian gauge fields
$V^{(a)}_{M}\equiv (V^{(1)\,\rho}_{M},V^{(2)}_{M\,\rho},
V^{(3)\,(c)}_{M})$, a two form $B_{MN}$ and an $O(4,20)$ matrix of
scalars $M$. Note that
the index $(a),(b)$ for the $SO(4,20)$ vector runs from $(1,\cdots,24)$.
These six-dimensional fields are related to
the ten-dimensional fields as
\bea
\Phi_6 &=& \hat{\Phi}- \hp \ln \det G_{\rho \sigma}; \nn \\
 V^{(2)}_{M\,\rho} &=&  \hat{B}^{(2)}_{M \rho}
+\hat{B}^{(2)}_{\rho \sigma }V^{(1)\,\sigma}_{M}+ {1\over
  2}\hat{V}^{(c)}_{\rho} V^{(3)\,{(c)}}_{M}; \la{reduct} \\
V^{(3)\,{(c)}}_{M} &=& \hat{V}^{(c)}_{M} - \hat{V}^{(c)}_{\rho}
V^{(1)\,\rho}_{M}; \nn \\
H_{MNP} &=& 3 (\partial_{[M} \hat{B}^{(2)}_{NP]} -
{1\over 2} V^{(a)}_{[M}L_{(a)(b)}F(V)^{(b)}_{NP]}), \nn
\eea
with the metric $g_{MN}$ and $V^{(1)\,\rho}_{M}$ defined in
(\ref{4dkk}).  The matrix $L$ is given by
\be
\label{equ_Lmetric}
  L= \left(\begin{array}{cc}
        I_4 & 0 \\
        0 & -I_{20}
        \end{array}\right),
\ee
where $I_n$ denotes the $n \times n$ identity matrix.
The scalar moduli are defined via
\begin{equation}
M = \Omega_1^{T} \left ( \begin{array}{ccc} G^{-1} & -G^{-1}C & -G^{-1}V^T \cr
-C^T G^{-1} & G + C^T G^{-1}C +V^T V & C^T G^{-1}
V^T
+ V^T \cr -VG^{-1} & VG^{-1}C + V & I_{16} + VG^{-1}V^T
\end{array}\right ) \Omega_1,
\label{modulthree}
\end{equation}
where $G \equiv [\hat{G}_{\rho \sigma}]$, $C \equiv [{1\over 2}
\hat{V}^{(c)}_{{\rho}}\hat{V}^{(c)}_{\sigma}+\hat{B}^{(2)}_{\rho \sigma}]$ and
$V \equiv [\hat{V}^{(c)}_{{\rho}}]$ are defined in terms of the
components of the 10-dimensional fields along the torus. The constant
$O(4,20)$ matrix $\Omega_1$ is given by
\be
\Omega_1 = \frac{1}{\sqrt{2}} \left(\begin{array}{ccc}
        I_4 & I_4 & 0 \\
        - I_4 & I_4 & 0 \\
        0 & 0 & \sqrt{2} I_{16}
        \end{array}\right).
\ee
This matrix arises in \eqref{modulthree} as follows. In
\cite{Sen:1994fa,You97} the matrix $L$ was chosen to be off-diagonal,
but for our purposes it is useful for $L$ to be diagonal. An
off-diagonal choice is associated with an off-diagonal intersection
matrix for the self-dual and anti-self-dual forms of $K3$, but this is an
unnatural choice for our solutions, in which only anti-self-dual forms
are active. Thus
relative to the conventions of
\cite{Sen:1994fa,You97} we take $L \rightarrow \Omega_1^{T}
L \Omega_1$, which induces $M \rightarrow \Omega_1^{T} M \Omega_1$ and $F
\rightarrow \Omega_1^T F$.
The definitions of this and other constant matrices used throughout
the paper are summarized in appendix \ref{bc_matrices}.

These fields satisfy the equations of motion following from the action
\bea
\label{equ_acthet6d}
   S &=& \frac{1}{2 \k_6^2} \int d^6x \sqrt{-g} e^{-2\Phi_6}[R + 4 (\pa \Phi_6)^2
- \frac{1}{12} H_3^2 - \frac{1}{4}F(V)_{MN}^{(a)} (LML)_{(a)(b)} F(V)^{(b)MN}
\nono \\
   && \hsp \hsp \hsp + \frac{1}{8} \tr(\pa_M M L \pa^M M L)],
\eea
where $\alpha^\prime$ has been set to $1/4$ and $\k_6^2 =
\k_{10}^2/V_4$ with $V_4$ the volume of the torus.

The reduction of the heterotic solution to six dimensions is then
\bea
   ds^2 &=& H^{-1}\left[ -dudv + \left (K- H^{-1} (\half (N^{(c)})^2 +
       (A_{\rho})^2) \right )dv^2  + 2A_idx^idv \right] + dx_i dx^i, \nono \\
   B_{uv} &=& \half(H^{-1} -1), \qquad B_{vi} = H^{-1} A_i, \qquad \Phi_6 =
-\half \ln H \la{het-six} \\
   V_v^{(a)} &=& \left (0_4 , \sqrt{2} H^{-1} A_{\rho}, H^{-1} N^{(c)}
   \right ), \qquad M =
I_{24}, \nn
\eea
where $i=1,\cdots,4$ runs over the transverse $R^4$ directions
and $\rho=5,\cdots,8$ runs over the internal directions of the $T^4$. Thus
the six-dimensional solution has only one non-trivial scalar field, the dilaton,
with all other scalar fields being constant.

\subsection{String-string duality to P-NS5 (IIA) on $K3$}
\la{IIAK3}

Given the six-dimensional heterotic solution, the corresponding
IIA solution in six dimensions can be obtained as
follows. Compactification of type IIA on $K3$ leads to the following
six-dimensional theory \cite{Sen:1995cj}:
\bea \la{IIAact}
   S' & = & \frac{1}{2 \kappa_6^2} \int d^6x \sqrt{-g'} \left (
e^{-2\Phi'_6}[R' + 4 (\pa \Phi'_6)^2
- \frac{1}{12} {H'_3}^2 + \frac{1}{8} \tr(\pa_M M' L \pa^M M' L)]
\right . \\
 && \left . \hsp -\frac{1}{4}{F'(V)}_{MN}^{(a)} (LM'L)_{(a)(b)}
       {F'(V)}^{(b)MN} \right )
- 2 \int B_2' \wedge F_2'(V)^{(a)} \wedge F_2'(V)^{(b)} L_{(a)(b)}. \nn
\eea
%the explicit reduction on $K3$ will be discussed below in section \ref{uplift}.
The field content is the same as for the heterotic
theory in (\ref{equ_acthet6d}); note that in contrast to (\ref{reduct})
there is no Chern-Simons term in the
definition of the 3-form field strength, that is,
$H'_{MNP} = 3\pa_{[M} B'_{NP]}$.

The rules for string-string duality are \cite{Sen:1995cj}:
\bea
   \Phi_6^\prime &=& - \Phi_6, \qquad g^\prime_{MN} = e^{-2\Phi_6}
   g_{MN},
\qquad M^\prime = M, \qquad V_M^{\prime (a)} = V_M^{ (a)}, \nono \\
   H^\prime_3 &=& e^{-2\Phi_6} \ast_6H_3;
\eea
these transform the equations of motion derived from
\eqref{equ_acthet6d} into ones derived
from the action (\ref{IIAact}).

Acting with this string-string duality on the heterotic solutions
(\ref{het-six}) yields, dropping the primes on IIA fields:
\bea
\label{equ_PNS56d}
   ds^2 &=& -dudv + (K- H^{-1} ((N^{(c)})^2/2 + (A_{\rho})^2) )dv^2 + 2A_idx^idv + H
dx_idx^i, \nono \\
   H_{vij} &=& -\e_{ijkl} \pa^k A^l, \qquad H_{ijk} = \e_{ijkl} \pa^l
   H, \qquad \Phi_6 = \hp \ln H, \\
   V_v^{(a)} &=& \left (0_4, \sqrt{2} H^{-1} A_{\rho}, H^{-1} N^{(c)}
   \right ), \qquad M = I_{24}, \nn
\eea
with $\ep_{ijkl}$ denoting the dual in the flat $R^4$ metric.
This describes NS5-branes on type IIA, wrapped on K3 and on the
   circle direction $y$, carrying momentum along the circle
   direction.

\subsection{T-duality to F1-NS5 (IIB) on $K3$}

The next step in the duality chain is T-duality on the circle
direction $y$ to give an NS5-F1 solution of type IIB on $K3$.
It is most convenient to carry out this step
directly in six dimensions, using the results
of \cite{Behrndt:1995si} on T-duality of type II theories on $K3
\times S^1$.

Recall that type IIB compactified on $K3$ gives
$d=6$, $N=4b$ supergravity coupled to 21 tensor multiplets,
constructed by Romans
in \cite{Romans:1986er}. The bosonic field content of this theory is the
graviton $g_{MN}$, 5 self-dual and 21 anti-self dual tensor fields
and an O(5,21) matrix of scalars $\cM$ which can be written in terms
of a vielbein $\cM^{-1} = V^T V$.
Following the notation of \cite{Sez98}
the bosonic field equations may be written as
\bea \la{sugraIIBK3}
R_{MN} &=& 2 P^{nr}_M P^{nr}_N + H^n_{MPQ}
   {H^n_N}^{PQ}
+  H^r_{MPQ} {H^r_N}^{PQ}, \nono \\
   \na^M P_M^{nr} &=& Q^{M nm} P_M^{mr} + Q^{M rs} P_M^{ns}
   + \frac{\sqrt{2}}{3} H^{n MNP} H^r_{MNP},
\eea
along with Hodge duality conditions on the 3-forms
\be
\la{sugraIIBK3_hd}
   \ast_6 H^n_3 = H^n_3, \qquad \ast_6 H^r_3 = -H^r_3,
\ee
In these equations $(m,n)$ are $SO(5)$ vector indices running from
1 to 5 whilst $(r,s)$ are
$SO(21)$ vector indices running from 6 to 26.
The 3-form field strengths are given by
\be
\la{G3forms}
H^{n} = G^{A} V_{A}^n; \hsp
H^{r} = G^{A} V_{A}^r,
\ee
where $A \equiv \{n,r\} = 1, \cdots, 26$; $G^{A} = db^A$ are closed and the vielbein on the
coset space $SO(5,21)/(SO(5) \times SO(21))$ satisfies
\be
 V^T \eta V = \eta, \qquad V = \left(\begin{array}{c} {V^n}_A \\
   {V^r}_A \end{array}\right), \qquad \eta = \left(\begin{array}{cc}
   I_5 & 0 \\ 0 & -I_{21} \end{array}\right).
\ee
%with $\eta_{AB} = \rm{diag} (+++++--- \cdots -)$ and $V =
%\left(\begin{array}{c} {V^i}_K \\ {V^r}_K \end{array}\right)$.
The associated connection is
\be
\la{IIBK3conn}
d V V^{-1} =  \left ( \begin{array} {c c} Q^{mn} & \sqrt{2} P^{ms} \\
\sqrt{2} P^{rn} & Q^{rs} \end{array} \right ),
\ee
where $Q^{mn}$ and $Q^{rs}$ are antisymmetric and the off-diagonal
block matrices $P^{ms}$ and $P^{rn}$ are transposed to each other.
Note also that there is a freedom in choosing the vielbein;
$SO(5) \times SO(21)$ transformations acting on $H_3$ and $V$ as
\be
   V \rightarrow OV, \qquad H_3 \rightarrow OH_3,
\ee
leave $G_3$ and $\cM^{-1}$ unchanged. Note that
the field equations \eqref{sugraIIBK3} can also be derived from the
$SO(5,21)$ invariant Einstein frame pseudo-action \cite{Bergshoeff:1995sq}
\be
   S = \frac{1}{2 \k_6^2} \int d^6x \sqrt{-g}\left(R +
\frac{1}{8} \tr(\pa \cM^{-1} \pa \cM) - \frac{1}{3} G^A_{MNP}
\cM^{-1}_{AB} G^{B MNP}\right),
\ee
with the Hodge duality conditions \eqref{sugraIIBK3_hd} being imposed
independently.

Now let us consider the T-duality relating a six-dimensional
IIB solution to a six-dimensional IIA solution of (\ref{IIAact}); the
corresponding rules were derived in \cite{Behrndt:1995si}. Given that
the six-dimensional IIA supergravity has only an $SO(4,20)$ symmetry,
relating IIB to IIA requires explicitly breaking the $SO(5,21)$
symmetry of the IIB action down to $SO(4,20)$. That is, one defines a
conformal frame in which
only an $SO(4,20)$ subgroup is manifest and in which the action reads
\bea
\la{IIBK3string}
   S &=& \frac{1}{2 \k_6^2} \int d^6x \sqrt{-g} \left\{ e^{-2\Phi}
   \left(R + 4 (\pa \Phi)^2 + \frac{1}{8} \tr(\pa M^{-1} \pa M)\right)
   + \hp \pa l^{(a)} M^{-1}_{(a)(b)} \pa l^{(b)}\right. \nono \\
     && \left.- \frac{1}{3} G^A_{MNP} \cM^{-1}_{AB} G^{B MNP}\right\}.
\eea
The $SO(5,21)$ matrix $\cM^{-1}$ has now been split up into the dilaton
$\Phi$, an $SO(4,20)$ vector $l^{(a)}$ and an $SO(4,20)$ matrix
$M^{-1}_{(a)(b)}$, and we have chosen the parametrization
\be
\la{IIBK3Mscalars}
   \cM^{-1}_{AB} = \W_3^T\left( \begin{array}{ccc}
      e^{-2\Phi} + l^T M^{-1} l + \frac{1}{4} e^{2\Phi} l^4 & -
\hp e^{2\Phi} l^2 & (l^T M^{-1})_{(b)} + \hp e^{2\Phi} l^2 (l^T L)_{(b)} \\
      -\hp e^{2\Phi} l^2 & e^{2\Phi} & -e^{2\Phi} (l^T L)_{(b)} \\
      (M^{-1} l)_{(a)} + \hp e^{2\Phi}l^2 (L l)_{(a)} & -e^{2\Phi}
(L l)_{(a)} & M^{-1}_{(a)(b)} + e^{2\Phi} (Ll)_{(a)} (l^T L)_{(b)}  \\
                           \end{array} \right)\W_3,
\ee
where $l^2 = l^{(a)} l^{(b)} L_{(a)(b)}$, $L_{(a)(b)}$ was defined in
\eqref{equ_Lmetric} and $\W_3$ is a constant matrix
defined in appendix \ref{bc_matrices}.

The fields $\Phi$, $l^{(a)}$ and $M^{-1}$ and half of the 3-forms can
now be related to the IIA fields of section \ref{IIAK3} by the
following T-duality rules (given in terms of the 2-form potentials $b^A$)
\cite{Behrndt:1995si}:
\begin{align}
   &\tilde{g}_{yy} = g_{yy}^{-1}, & &
\tilde{b}^{1}_{y M} + \tilde{b}^{26}_{y M} =
\half g_{yy}^{-1} g_{y M}, \\
   &\tilde{g}_{y M} = g_{yy}^{-1} B_{y M}, & &\tilde{b}^{1}_{M
  N} + \tilde{b}^{26}_{MN} = \half g_{yy}^{-1} (B_{MN} + 2(g_{y[M}B_{N]y})), \nono \\
   &\tilde{g}_{MN} = g_{MN} - g_{yy}^{-1} (g_{y M} g_{y N} - B_{y M}
  B_{y N}), & & \tilde{l}^{(a)} = V_y^{(a)}, \nono \\
   &\tilde{\Phi} = \Phi - \hp \log |g_{yy}|, & &
 \tilde{M}^{-1}_{(a)(b)} = M^{-1}_{(a)(b)}, \nn \\
& \tilde{b}^{(a)+1}_{y
  M} = \frac{1}{\sqrt{8}}(V_M^{(a)} - g_{yy}^{-1} V_y^{(a)} g_{yM}), & & (1 \leq (a) \leq 24), \nono
%&
%&\tilde{b}^{(a)+1}_{y M} = \frac{1}{\sqrt{8}}(V_M^{(a)} -
%\frac{V_y^{(a)} g_{y M}}{g_{yy}}), & &(5 \leq (a) \leq 24), \nn
\end{align}
Here $y$ is the T-duality circle, the six-dimensional index $M$
excludes $y$ and IIB fields are denoted by
tildes to distinguish them from IIA fields. The other half of the
tensor fields, that is $\left ((\td{b}^{1}_{y M} - \td{b}^{26}_{y M}),
(\td{b}^{1}_{MN} - \td{b}^{26}_{MN}), \td{b}^{(a)+1}_{MN},
\td{b}^{(a)+1}_{MN}\right )$, can then be determined using the
Hodge duality constraints \eqref{sugraIIBK3_hd}.

We now have all the ingredients to obtain the T-dual of the IIA
solution \eqref{equ_PNS56d} along $y \equiv \half (u-v)$. The IIA
solution is expressed in terms of harmonic functions which also depend
on the null coordinate $v$, and thus one needs to smear the solutions
before dualizing. Note that it is the harmonic functions
$(H,K,A^I,N^{(c)})$ which must be smeared over $v$, rather than the
six-dimensional fields given in \eqref{equ_PNS56d}, since it is the
former that satisfy linear equations and can therefore be superimposed.

The Einstein frame metric and three forms are given by
\bea
   ds^2 &=& \frac{1}{\sqrt{H \td{K}}} [ -(dt-A_idx^i)^2 + (dy-B_i
     dx^i)^2)] + \sqrt{H \td{K} }
dx_idx^i, \nono \\
   G_{tyi}^A &=& \pa_i \left ( \frac{n^A}{H \td{K}} \right ) ,
\qquad
   G_{\bar{\mu}ij}^A = - 2 \pa_{[i} \left( \frac{n^A}{H \td{K} } {\cal B}^{\bar{\mu}}_{j]}\right), \\
%   G_{yij}^A &=& -2 \pa_{[i} \left(\frac{n^A}{H \td{K} }A_{j]}\right), \nono \\
   G_{ijk}^A &=& \e_{ijkl} \pa^l n^A + 6\pa_{[i}
     \left(\frac{n^A}{H \td{K} } A_j B_{k]}\right), \nn
\eea
where
\bea
   n^{m} &=& \frac{1}{4}\left(H+K+1, 0_4\right),
\qquad n^{r} = \frac{1}{4}\left(
- 2 A_{\rho},- \sqrt{2} N^{(c)}, H-K-1 \right), \\
  \qquad \td{K} &=& 1+ K - H^{-1} ( \half (N^{(c)})^2 + (A_{\rho})^2),
     \qquad dB= -\ast_4 dA, \qquad {\cal B}^{\mu}_i = (-B_i,A_i).  \nn
\eea
Recall that $n=1,\cdots,5$ and $r=6,\cdots, 26$ and $\ast_4$ denotes
     the dual on flat $R^4$; $\bar{\mu} = (t,y)$.  The $SO(4,20)$ scalars are given by
\be
\la{F1NS5scalars}
   \Phi = \hp \ln \frac{H}{ \td{K} }, \qquad l^{(a)} = \left (0_4,
\sqrt{2} H^{-1} A_{\rho}, H^{-1} N^{(c)} \right ), \qquad  M = I_{24}.
\ee
The $SO(5,21)$ scalar matrix $\cM^{-1} = V^T V$ in
\eqref{IIBK3Mscalars} can then conveniently be expressed in terms of the vielbein
\be
   V = \W_3^T \left( \begin{array}{ccc}
                       \sqrt{H^{-1} \td{K}} & 0 & 0 \\
                       - (\sqrt{H^3 \td{K}})^{-1} (A_{\rho}^2 + \hp
     (N^{(c)})^2) &
\sqrt{H \td{K}^{-1}}  & - \sqrt{H \td{K}^{-1}} l^{(b)} \\
                       l^{(a)} & 0 & I_{24}
                   \end{array} \right) \W_3.
\ee

\subsection{S-duality to D1-D5 on $K3$} \la{D1D5K3}

One further step in the duality chain is required to obtain the
D1-D5 solution in type IIB, namely S duality. However, in the previous
section the type II solutions have been given in six rather than ten
dimensions. To carry out S duality one needs to specify the
relationship between six and ten dimensional fields. Whilst
the ten-dimensional $SL(2,R)$ symmetry is part of the six-dimensional
symmetry group, its embedding into the full six-dimensional symmetry
group is only defined once one specifies the uplift to ten
dimensions. The details of the dimensional reduction are given in appendix
\ref{IIBK3}, with the six-dimensional S duality rules being given in
(\ref{6dSdual}); the S duality leaves the Einstein frame metric
invariant, and acts as a constant rotation and similarity transformation on
the three forms $G^A$ and the matrix of scalars ${\cal {M}}$ respectively.
The S-dual solution is thus
\bea
\la{D1D56d}
   ds^2 &=& \frac{1}{\sqrt{f_5 \td{f}_1}} [ -(dt-A_idx^i)^2 + (dy-B_i
     dx^i)^2)] + \sqrt{f_5 \td{f}_1}
dx_idx^i, \\
   G_{tyi}^A &=& \pa_i \left ( \frac{m^A}{f_5 \td{f}_1 } \right ) , \qquad
   G_{\bar{\mu}ij}^A = - 2 \pa_{[i} \left( \frac{m^A}{f_5 \td{f}_1 } 
{\cal B}^{\bar{\mu}}_{j]}\right), \nono \\
%   G_{yij}^A &=& -2 \pa_{[i} \left(\frac{m^A}{f_5 \td{f}_1 } A_{j]}\right), \nono \\
   G_{ijk}^A &=& \e_{ijkl} \pa^l m^A + 6\pa_{[i} \left(\frac{m^A}{f_5
       \td{f}_1 } A_j B_{k]}\right), \nn
\eea
with
\bea
   m^{n} &=& \left(0_4,\qu (f_5 + F_1) \right),\\
 m^{r} &=& \frac{1}{4}\left((f_5 -  F_1),
 - 2 A_{\a} ,
- \sqrt{2} N^{(c)}, 2 A_5 \right) \nono \\
&\equiv & \frac{1}{4}\left((f_5 - F_1),  - 2 {\cA}^{\a_-}, 2 \cA
\right). \nn
\eea
Here the index $\a=6,7,8$. Note that the specific reduction used here,
see appendix \ref{IIBK3},\comment{, in particular the embedding of
  the $SO(3,19)$ matrix $d_{AB}$ in the $SO(5,21)$ matrix $\cM^{-1}$,}
distinguished $A_5$ from the other $A_{\rho}$ and $N^{(c)}$. A different
embedding would single out a different harmonic function, and hence a different
vector, and it is thus convenient to introduce $(\cA, \cA^{\a_-})$ to
denote the choice of splitting more abstractly.
Also as in (\ref{redef}) it is convenient to introduce the following
combinations of harmonic functions:
\bea
f_{5} &=& H, \qquad
\td{f}_1 = 1 + K - H^{-1} (\cA^2 + \cA^{\a_-} \cA^{\a_-}), \\
F_1 &=& 1 + K, \qquad
f_1 = \td{f}_1 + H^{-1} \cA^2. \nn
\eea
The vielbein of scalars is given by
\be
\la{D1D5vielbein}
  V = \W_4^T \left(\begin{array}{ccccc}
            \sqrt {f_{1}^{-1} \td{f_1}} & 0 & 0 & 0 & 0
                       \\
                      G \cA^2 & \sqrt{\td{f}_1^{-1} f_1} & - G \cA F_1
       & (\sqrt{f_1 \td{f_1}})^{-1} \cA & - G \cA \, k^{\gamma} \\
%
%(\sqrt{f_1 \td{f}_1} f_5)^{-1}  \cA^2 &
%                      \sqrt{\td{f}_1^{-1} f_1} &
%- (\sqrt{f_1 \td{f}_1} f_5)^{-1}\cA F_1  & (\sqrt{f_1
%    \td{f_1}})^{-1} \cA & - (\sqrt{f_1
%       \td{f}_1} f_5)^{-1} \cA \, k^{\gamma} \\
                       - F \cA & 0 &
                       \sqrt{f_{5}^{-1} f_1}& 0 & 0 \\
                       F \cA & 0 & -\hp
f_5^{-1} F (k^{\gamma})^2 &
\sqrt{f_5 f_1^{-1}} & - F k^{\gamma} \\
                       0 & 0 & f_{5}^{-1} k^{\gamma}  & 0 & I_{22}
                     \end{array}\right) \W_4,
\ee
where to simplify notation quantities $(F,G)$ are defined as
\be
F = (f_1 f_5)^{-1/2}, \qquad
G = (f_1 \td{f_1} f^2_5)^{-1/2}.
\ee
We also define the 22-dimensional vector $k^{\gamma}$ as
\be
k^{\gamma} = (0_3, \sqrt{2} \cA^{\a_-}).
\ee
Here $\gamma = 1,\cdots,b^2$ where the second Betti number is $b^2 =
22$ for K3.
Using the reduction formulae \eqref{redIIBscalars} and
\eqref{redIIBfs6d}, the six-dimensional solution
\eqref{D1D56d}, \eqref{D1D5vielbein}
can be lifted to ten dimensions, resulting in a solution with an
analogous form to the $T^4$ case \eqref{equ_D1D5T4}. We will thus summarize
the solution for both cases in the following section.

\section{D1-D5 fuzzball solutions} \la{summary}

In this section we will summarize the D1-D5 fuzzball solutions with
internal excitations, for both the $K3$ and $T^4$ cases. In both cases
the solutions can be written as
\bea
\la{equ_D1D5K3pot}
   ds^2 &=& \frac{f_1^{1/2}}{\td{f_1} f_5^{1/2} }[-(dt-A_i
     dx^i)^2+(dy-B_i dx^i)^2] + f_1^{1/2} f_5^{1/2} dx_i dx^i
+  f_1^{1/2} f_{5}^{-1/2} ds^2_{M^4},  \nn \\
   e^{2\Phi} &=& \frac{f_1^2}{f_5 \td{f_1} }, \qquad B^{(2)}_{ty} =
   \frac{{\cal A}}{f_5 \td{f}_1},
\qquad B^{(2)}_{\bar{\mu} i} = \frac{{\cal A} {\cal B}^{\bar{\mu}}_i}{f_5 \td{f_1} }, \\
B^{(2)}_{ij} &=& \l_{ij} +
\frac{2 {\cal A} A_{[i} B_{j]}} {f_5 \td{f}_1 }, \qquad
B^{(2)}_{\rho\sigma} = f_{5}^{-1} k^{\gamma}
\w_{\rho \sigma}^{\gamma}, \qquad  C^{(0)} = - f_{1}^{-1} \cal A, \nono \\
   C^{(2)}_{ty} &=& 1-\td{f}_1^{-1}, \qquad C^{(2)}_{\bar{\mu}i} = -
   \td{f}_1^{-1} {\cal B}^{\bar{\mu}}_i, \qquad C^{(2)}_{ij} = c_{ij} - 2 \td{f}_1^{-1}
A_{[i}B_{j]}, \nono \\
   C^{(4)}_{tyij} &=& \l_{ij} + \frac{{\cal A}}{f_5 \td{f_1}}(c_{ij} + 2 A_{[i}B_{j]}),
\qquad C^{(4)}_{\bar{\mu}ijk} = \frac{3 {\cal A}}{f_5 \td{f}_1} 
{\cal B}^{\bar{\mu}}_{[i}c_{jk]}, \nono \\
   C^{(4)}_{ty \rho \sigma}
&=& f_{5}^{-1} k^{\gamma} \w^{\gamma}_{\rho \sigma},
\qquad C^{(4)}_{ij \rho \sigma} = (\l^{\gamma}_{ij} +
f_{5}^{-1} k^{\gamma}c_{ij})\w^{\gamma}_{\rho \sigma}, \qquad
   C^{(4)}_{\rho \sigma \tau \pi} = f_{5}^{-1} \cal A \e_{\rho
     \sigma \tau \pi}, \nn
\eea
where we introduce a basis of self-dual and anti-self-dual 2-forms
$\w^{\gamma} \equiv (\w^{\a_+}, \w^{\a_-})$ with $\gamma =
1,\cdots,b^2$ on the compact manifold $M^4$. For both $T^4$ and $K3$
the self-dual forms are labeled by
$\a_{+} = 1,2,3$ whilst the anti-self-dual forms are labeled by
$\a_{-} =1,2,3$ for $T^4$ and $\a_{-} = 1,\cdots 19$ for $K3$. The
intersections and normalizations of these forms are defined in
(\ref{tt1}), (\ref{tt2}) and (\ref{K3dABdef}). The solutions are expressed
in terms of the following combinations of harmonic functions
$(H,K,A_{i},{\cal A}, {\cal A}^{\a_-})$
\bea
\la{D1D5K3aux}
f_{5} &=& H; \qquad \td{f}_1 = 1 + K - H^{-1} (\cA^2 + \cA^{\a_-} \cA^{\a_-}); \qquad
f_1 = \td{f}_1 + H^{-1} \cA^2; \nn \\
k^{\gamma} &=& (0_3, \sqrt{2} \cA^{\a_-});
\qquad dB = -\ast_4 dA; \qquad
   dc = - \ast_4 df_5; \\
d\l^{\gamma} &=& \ast_4 dk^{\gamma}; \qquad d\l = \ast_4 d {\cal A}; \qquad
{\cal B}^{\bar{\mu}}_i = (-B_i,A_i), \nn
\eea
where $\bar{\mu} = (t,y)$ and the Hodge dual $\ast_{4}$ is defined over (flat)
$R^4$, with the Hodge dual in the Ricci flat metric on
the compact manifold being denoted by $\ep_{\r \s\t \pi}$.
The constant term in $C_{ty}^{(2)}$ is chosen so that the
potential vanishes at asymptotically flat infinity.
The corresponding RR field strengths are
\bea
\la{equ_D1D5K3fs}
   F_i^{(1)} &=& - \pa_i \left ( f_{1}^{-1} {\cal A} \right ) , \qquad
   F^{(3)}_{tyi} = ({f_1 \td{f}_1 f_5^2})^{-1}
\left (f_5^2 \pa_i \td{f}_1 + f_5 {\cal A} \pa_i {\cal A} -
     {\cal A}^2 \pa_i f_5 \right ) , \nono \\
   F^{(3)}_{\bar{\mu}ij} &=& ({f_5^2 f_1 \td{f}_1})^{-1}
\left (2 {\cal B}^{\bar{\mu}}_{[i}(f_5 \pa_{j]} \td{f}_1 + f_5 {\cal A} \pa_{j]}
   {\cal A} - {\cal A}^2 \pa_{j]} f_5) + 2 \td{f}_1 f_5^2
     \pa_{[i} {\cal B}^{\bar{\mu}}_{j]} \right ) , \nono \\
 %  F^{(3)}_{yij} &=& \frac{2A_{[i}(f_5 \pa_{j]} \td{f}_1 + f_5 {\cal A} \pa_{j]}
 %  {\cal A} - {\cal A}^2 \pa_{j]} f_5) + 2 \td{f}_1 f_5^2
 %    \pa_{[i}A_{j]}}{f_5^2 \td{f}_1 f_1}, \nono \\
   F^{(3)}_{ijk} &=& -\e_{ijkl}(\pa^l f_5 -  f_{1}^{-1} {\cal A} \pa^l
   {\cal A}) - 6 f_{1}^{-1} \pa_{[i}(A_j B_{k]}) \\
&& \qquad + ({f_5^2 f_1 \td{f}_1})^{-1} \left ( 6A_{[i}
       B_j (f_5 \pa_{k]}\td{f}_1 + f_5 {\cal A} \pa_{k]} {\cal A} - {\cal A}^2
\pa_{k]} f_5) \right ), \nono \\
   F^{(3)}_{i \rho \sigma} &=& f_{1}^{-1} {\cal A}
   \pa_i( f_{5}^{-1} k^{\gamma}) \w^{\gamma}_{\rho \sigma}, \nono \\
   F^{(5)}_{i \rho \sigma \tau \pi} &=& \e_{\rho \sigma \tau \pi}
\pa_i ( f_{5}^{-1} {\cal A}), \qquad F^{(5)}_{tyijk}
   = \e_{ijkl} \td{f}_1^{-1} f_5 \pa^l( f_{5}^{-1} {\cal A}), \nono \\
   F^{(5)}_{\bar{\mu} ijkl} &=& - \e_{ijkl} f_5 \td{f}_1^{-1} {\cal B}^{\bar{\mu}}_m
   \pa^m ( f_{5}^{-1} \cal A),\nn \\
%   \qquad F^{(5)}_{yijkl} = - \e_{ijkl} \frac{f_5 A_m \pa^m
%({\cal A}/f_5)}{\td{f}_1}, \nono \\
   F^{(5)}_{tyi \rho \sigma} &=& {\td{f}_1}^{-1}
   \pa_i(k^{\gamma}/f_5) \w^{\gamma}_{\rho \sigma},
\qquad F^{(5)}_{\bar{\mu}ij \rho \sigma} =
2 \td{f}_1^{-1} {\cal B}^{\bar{\mu}}_{[i}\pa_{j]}(f_{5}^{-1} k^{\gamma}) \w^{\gamma}_{\rho \sigma},
\nono \\
%   F^{(5)}_{yij\rho \sigma} &=&
%   \frac{2A_{[i}\pa_{j]}(k^{\gamma}/f_5)}{\td{f}_1}\w^{\gamma}_{\rho \sigma},
%\nono \\
   F^{(5)}_{ijk \rho \sigma} &=& \left (6 \td{f}_1^{-1}
   A_{[i}B_j\pa_{k]}( f_{5}^{-1} k^{\gamma}) + \e_{ijkl} f_5 \pa^l (
   f_{5}^{-1} k^{\gamma} ) \right ) \w^{\gamma}_{\rho \sigma}. \nn
\eea
It has been explicitly checked
that this is a solution of the ten-dimensional field equations for any
choices of harmonic functions $(H,K,A_i,{\cal A}, {\cal A}^{\a_-})$
with $\pa_{i} A^i = 0$. Note that in
the case of $K3$ one needs the identity (\ref{id-k3}) for the harmonic
forms to check the components of the Einstein equation along $K3$.

We are interested in solutions for which
the defining harmonic functions are given by
\bea
H &=& 1 + \frac{Q_5}{L} \int_0^L \frac{dv}{|x-F(v)|^2};
\qquad A_i = -\frac{Q_5}{L}\int_0^L \frac{dv \dot{F}_i(v)}{|x-F(v)|^2},
\la{harm} \\
   {\cal A} &=& -\frac{Q_5}{L}\int_0^L \frac{dv
     \dot{\cal F} (v)}{|x-F(v)|^2}; \qquad
\cA^{\a_-} = -\frac{Q_5}{L} \int_0^L \frac{dv \dot{\cal{F}}^{\a_-}(v)}{|x-F(v)|^2}, \nono \\
   K &=& \frac{Q_5}{L} \int_0^L \frac{dv (\dot{F}(v)^2 +
\dot{\cal F}(v)^2 + \dot{\cal{F}}^{\a -}(v)^2 )}{|x-F(v)|^2}. \nn
\eea
In these expressions $Q_5$ is the 5-brane charge and
$L$ is the length of the defining curve in the D1-D5 system, given
by
\be
L = 2 \pi Q_5/R,
\ee
where $R$ is the radius of the $y$ circle. Note that $Q_5$ has
dimensions of length squared and is related to the integral charge via
\be \la{q5}
Q_5 = \a' n_5
\ee
(where $g_s$ has been set to one). Assuming that the curves
$(\dot{\cal F} (v),\dot{\cal{F}}^{\a_-}(v))$ do not have zero modes,
the D1-brane charge $Q_1$ is given by
\be \la{d1-charg}
Q_1 = \frac{Q_5}{L} \int_0^L dv (\dot{F}(v)^2 +
\dot{\cal F}(v)^2 + \dot{\cal{F}}^{\a_-}(v)^2 ),
\ee
and the corresponding integral charge is given by
\be \la{q1}
Q_1 = \frac{n_1 (\a')^3}{V},
\ee
where $(2 \pi)^4 V$ is the volume of the compact manifold.
The mapping of the parameters from the original F1-P systems to
the D1-D5 systems was discussed in \cite{Lunin:2001jy} and is unchanged here.
The fact that the solutions take exactly the same form, regardless of
whether the compact manifold is $T^4$ or $K3$, is unsurprising given
that only zero modes of the compact manifold are excited.

The solutions defined in terms of the harmonic functions (\ref{harm})
describe the complete set of two-charge fuzzballs for the D1-D5 system on
$K3$. In the case of $T^4$, these describe fuzzballs with only
bosonic excitations; the most general solution would include fermionic
excitations and thus more general harmonic functions of the type
discussed in \cite{Taylor:2005db}. Solutions involving
harmonic functions with disconnected sources would be appropriate for
describing Coulomb branch physics. Note that, whilst the solutions obtained
by dualities from supersymmetric F1-P solutions are guaranteed to be
supersymmetric, one would need to check supersymmetry explicitly for
solutions involving other choices of harmonic functions.

\bigskip

In the final solutions one of the harmonic functions ${\cal A}$ describing
internal excitations is singled out from the others.
In the original F1-P system, the solutions pick out a
direction in the internal space. For the type II system on $T^4$, the
choice of $A_{\rho}$ singles out a direction in the torus whilst in
the heterotic solution the choice of $(A_{\rho}, N^{(c)})$ singles out
a direction in the 20d internal space. Both duality chains, however,
also distinguish directions in the internal space. In the $T^4$ case
one had to choose a direction in the torus, whilst
in the $K3$ case the choice is implicitly made when one uplifts type IIB
solutions from six to ten dimensions. In particular, the uplift splits
the $21$ anti-self-dual six-dimensional 3-forms into $19 + 1 + 1$
associated with the ten-dimensional $(F^{(5)}, F^{(3)}, H^{(3)})$ respectively.

When there are no internal excitations, the final solutions must be
independent of the choice of direction made in the duality chains but
this does not remain true when the original solution breaks the
rotational symmetry in the internal space. ${\cal A}$ is the component
of the original vector along the direction distinguished in the
duality chain, whilst ${\cal A}^{\a_-}$ are the components orthogonal
to this direction. When there are no excitations along the direction
picked out by the duality, i.e. ${\cal A} = 0$, the solution
considerably simplifies, becoming
\bea
  ds^2 &=& \frac{1}{ (f_1 f_5)^{1/2} }[-(dt-A_i
     dx^i)^2+(dy-B_i dx^i)^2] + f_1^{1/2} f_5^{1/2} dx_i dx^i
+  f_1^{1/2} f_{5}^{-1/2} ds^2_{M^4},  \nn \\
   e^{2\Phi} &=& \frac{f_1}{f_5},  \qquad
B^{(2)}_{\rho\sigma} = f_{5}^{-1} k^{\gamma} \w_{\rho \sigma}^{\gamma}, \qquad
   C^{(2)}_{ty} = 1-{f}_1^{-1}, \qquad C^{(2)}_{\bar{\mu}i} = -
   {f}_1^{-1} {\cal B}^{\bar{\mu}}_i, \nn \\
C^{(2)}_{ij} &=& c_{ij} - 2 {f}_1^{-1}
A_{[i}B_{j]}, \qquad
C^{(4)}_{ty \rho \sigma}
= f_{5}^{-1} k^{\gamma} \w^{\gamma}_{\rho \sigma},
\qquad C^{(4)}_{ij \rho \sigma} = (\l^{\gamma}_{ij} +
f_{5}^{-1} k^{\gamma}c_{ij})\w^{\gamma}_{\rho \sigma}. \nn
\eea
In this solution the internal excitations induce fluxes
of the NS 3-form and RR 5-form along anti-self dual cycles in the
compact manifold (but no net 3-form or 5-form charges). By contrast
the excitations parallel to the duality direction induce a field
strength for the RR axion, NS 3-form field strength in the non-compact
directions and RR 5-form field strength
along the compact manifold (but again no net charges).

Let us also comment on the $M^4$ moduli in our solutions. The
solutions are expressed in terms of a Ricci flat metric on
$M^4$ and anti-self dual harmonic two forms. The forms satisfy
\be
\w^{\g}_{\r \s} \w^{\d \r \s} = D^{\e}_{\hspace{2mm} \d} d_{\g \e} \equiv \d_{\g \d},
\ee
where the intersection matrix $d_{\d  \g}$ and the matrix
$D^{\g}_{\; \; \d}$ relating the basis of forms and dual forms are defined
in \eqref{K3dABdef} and \eqref{K3Hodge} respectively.
The latter condition on $D^{\g}_{\; \;\e}$ 
arose from the duality chain, and followed from
the fact that in the original F1-P solutions the internal manifold had a
flat square metric. Thus, the final solutions are expressed at a
specific point in the moduli space of $M^4$ because the original F1-P
solutions have specific fixed moduli. It is straightforward to extend
the solutions to general moduli: one needs to change
\be
\td{f}_1 = 1 + K - H^{-1} ({\cal A}^2 + {\cal A}^{\a_-} {\cal
  A}^{\a_-}) \rightarrow  1 + K - H^{-1} ({\cal A}^2 + \half {k}^{\g}
   {k}^{\d} D^{\e}_{\hspace{2mm} \d} d_{\g \e}),
\ee
with $k^{\g}$ as defined in \eqref{D1D5K3aux}, 
to obtain the solution for more general $D^{\g}_{\hspace{2mm} \d}$.

\bigskip

Given a generic fuzzball solution, one would like to check whether the geometry
is indeed smooth and horizon-free. For the fuzzballs with no internal
excitations this question was discussed in \cite{Lunin:2002iz}, the
conclusion being that the solutions are non-singular unless the
defining curve $F^i(v)$ is non-generic and self-intersects. In the
appendix of \cite{Lunin:2002iz}, the smoothness of fuzzballs with
internal excitations was also discussed. However, their D1-D5 solutions were
incomplete: only the metric was given, and this was effectively given
in the form {(\ref{D1D56d}) rather than (\ref{equ_D1D5K3pot}).
Nonetheless, their conclusion remains unchanged: following the same
discussion as in \cite{Lunin:2002iz} one can show that a generic fuzzball
solution with internal excitations is non-singular provided that the
defining curve $F^i(v)$ does not self-intersect and $\dot{F}_i(v)$
only has isolated zeroes. In particular, if
there are no transverse excitations, $F^i(v) = 0$, the solution will
be singular as discussed in section \ref{int}.

One can show that there are no horizons as follows. The harmonic
function $f_5$ is clearly positive definite, by its definition. The
functions $(f_1, \tilde{f}_1)$ are also positive definite, since they
can be rewritten as a sum of positive terms as
\bea
f_5 \tilde{f}_1 &=&
      \left (1+ \frac{Q_5}{L} \int_0^L \frac{dv}{|x-F|^2}  \right )
      \left (1 + \frac{Q_5}{L}
\int_0^L \frac{dv \dot{F}^2} {|x-F|^2} \right ) \\
&& + \frac{Q_5}{L}
\int_0^L \frac{dv (\dot{\cF}(v))^2 +(\dot{\cF}^{\a-}(v))^2 } {|x-F|^2}
\nn \\
&&   + \half (\frac{Q_5}{L})^2 \int_0^L \int_0^L dv dv^{\prime} \frac{
(\dot{\cF}(v) - \dot{\cF}(v^\prime))^2 + (\dot{\cF}^{\a-}(v) -
\dot{\cF}^{\a-}(v^\prime))^2} {\left | x-F(v) \right | ^2 \left |
  x-F(v^\prime) \right |^2 }, \nn
\eea
and a corresponding expression for $f_5 f_1$. Note that in the decoupling
limit only the terms proportional to $Q_5^2$ remain, and these are
also manifestly positive definite. Given that the defining functions
have no zeroes anywhere, the geometry therefore has no horizons.

\bigskip

Now let us consider the conserved charges. From the asymptotics one
can see that the fuzzball solutions have the same mass and D1-brane,
D5-brane charges as the naive solution; the latter are given in
(\ref{q5}) and (\ref{q1}) whilst the ADM mass is
\be
M = \frac{\Omega_3 L_y}{\k_6^2} (Q_1 + Q_5),
\ee
where $L_y = 2 \pi R$, $\Omega_3 = 2\pi^2$ is the volume of a unit 3-sphere, and
$2\k_6^2 = (2 \k^2)/(V(2 \pi)^4)$ with $2\k^2 = (2\pi)^7 (\a')^4$ in
our conventions. The fuzzball solutions have in addition angular
momenta, given by
\be
J^{ij} = \frac{\Omega_3 L_y}{\k_6^2 L} \int_{0}^{L} dv (F^{i} \dot
F^{j} - F^{j} \dot{F}^i).
\ee
These are the only charges; the fields $F^{(1)}$ and $F^{(5)}$ fall off
too quickly at infinity for the corresponding
charges to be non-zero. One can compute from the harmonic
expansions of the fields dipole and more generally
multipole moments of the charge distributions. A generic solution
breaks completely the $SO(4)$ rotational invariance in $R^4$, and this
symmetry breaking is captured by these multipole moments.

However, the multipole moments computed at asymptotically flat
infinity do not have a direct interpretation in the dual field
theory. In contrast, the asymptotics of the solutions in the decoupling
limit do give field theory information: one-point functions of chiral
primaries are expressed in terms of the asymptotic expansions (and
hence multipole moments) near the $AdS_3 \times S^3$ boundary. Thus it
is more useful to compute in detail the latter, as we shall do in the
next section.

\section{Vevs for the fuzzball solutions} \la{s-vevs}

In the decoupling limit all of the fuzzball solutions are asymptotic to
$AdS_3 \times S^3 \times M^4$, where $M^4$ is $T^4$ or $K3$. Therefore
one can use AdS/CFT methods to extract holographic data from the
geometries. In particular, the asymptotics of the six-dimensional
solutions near the $AdS_3 \times S^3$ boundary encode the vevs of
chiral primary operators in the dual field theory.

The precise relationship between asymptotics and vevs is however
rather subtle. A systematic method for extracting vevs from
asymptotically $AdS \times X$ solutions (with $X$ an arbitrary compact
manifold) was only recently constructed, in
\cite{Skenderis:2006uy}, building on earlier work
\cite{S2,BFS1,BFS2,P1,P2}, see also the review \cite{S1}.
This method of Kaluza-Klein holography was
then applied to the case of asymptotically $AdS_3 \times S^3$ solutions
of $d=6, N = 4b$ supergravity coupled to $n_t$ tensor multiplets
in \cite{Skenderis:2006ah, Kanitscheider:2006zf} and in what
follows we will make use of many of the results derived there.

For fuzzball solutions on $K3$, the relevant solution of
six-dimensional $N=4b$ supergravity coupled to $21$ tensor
multiplets was given explicitly in \eqref{D1D56d}. For the case of
$T^4$, we obtained the solution in ten dimensions, but there is
a corresponding six-dimensional solution of $N=4b$ supergravity
coupled to 5 tensor multiplets. This solution is of exactly the same
form as the $K3$ solution given in \eqref{D1D56d}, but with the index
$\a_{-} = 1,2,3$. Thus in what follows we will analyze both cases
simultaneously. As mentioned earlier, the $T^4$ solution
reduces to a solution of $d=6,N=4b$ supergravity rather than a
solution of $d=6,N=8$ supergravity because forms associated with
the odd cohomology of $T^4$ (and hence six-dimensional vectors) are
not present in our solutions.

\subsection{Holographic relations for vevs}

Consider an $AdS_3 \times S^3$ solution of the six-dimensional field
equations (\ref{sugraIIBK3}), such that
\bea
ds_6^2 &=& \sqrt{Q_1 Q_5} \left ( \frac{1}{z^2} (-dt^2 + dy^2 + dz^2)
+ d\Omega_3^2 \right );  \la{background} \\
G^{5} &=& H^{5} \equiv g^{o5} =  \sqrt{Q_1 Q_5} (r dr \wedge dt \wedge
dy + d\Omega_3), \nn
\eea
with the vielbein being diagonal and all other three forms (both
self-dual and anti-self dual) vanishing. In what follows it is
convenient to absorb the curvature radius $\sqrt{Q_1 Q_5}$ into an
overall prefactor in the action, and work with the unit radius $AdS_3
\times S^3$. Now express the perturbations
of the six-dimensional supergravity fields relative to the $AdS_3
\times S^3$ background as
\bea
g_{MN} &=& g^{o}_{MN} + h_{MN}; \hsp
G^{A} = g^{oA} + g^A; \\
V^{n}_{A} &=& \d^{n}_{A} + \f^{nr} \d_{A}^r + \half \f^{nr} \f^{mr}
\d_{A}^m; \nn \\
V^{r}_A &=& \d^{r}_{A} + \f^{nr} \d_{A}^n + \half \f^{nr} \f^{ns}
\d_{A}^s. \nn
\eea
These fluctuations can then be expanded in spherical harmonics as
follows:
\bea
h_{\m \n} &=& \sum h_{\m\n}^I (x) Y^{I} (y), \la{flc1} \\
h_{\m a} &=& \sum (h_{\m}^{I_v} (x) Y_a^{I_v} (y) +
h_{(s)\m}^{I} (x) D_a Y^I (y) ), \nn \\
h_{(a b)} &=& \sum (\rho^{I_t} (x) Y_{(ab)}^{I_t} (y) +
\rho_{(v)}^{I_v} (x) D_a Y_b^{I_v} (y) +
\rho_{(s)}^{I} (x) D_{(a} D_{b)} Y^{I} (y) ), \nn \\
h^{a}_{a} &=& \sum \pi^{I} (x) Y^{I} (y), \nn \\
g^{A}_{\m\n \r} &=& \sum 3 D_{[\m} b_{\n \r]}^{(A)I} (x) Y^{I} (y), \nn
\\
g^{A}_{\m \n a} &=& \sum ( b_{\m \n}^{(A)I} (x) D_{a} Y^{I} (y) + 2
D_{[\m} Z_{\n]}^{(A) I_{v}} (x) Y_{a}^{I_v} (y)); \nn \\
g^{A}_{\m a b} &=& \sum (D_{\m} U^{(A)I}(x) \ep_{abc} D^{c} Y^I(y) +
2 Z_{\m}^{(A) I_v} D_{[b} Y^{I_v}_{a]}); \nn \\
g^{A}_{a b c} &=& \sum (- \ep_{abc} \L^{I} U^{(A)I}(x) Y^{I} (y)) ; \nn \\
%b^{A}_{\m a} &=& \sum (Z_{\m}^{A I_v } (x) Y_{a}^{I_v }(y) +
%Z_{(s)\m}^{A I} (x) D_{a} Y^I (y) ); \nn \\
%b^{A}_{ab} &=& \sum \ep_{abc} (U^{A I} (x) D^{c} Y^I (y) + U_{(v)}^{A
%  I_{v} } (x)  Y^{c I_v } (y) ); \nn \\
\phi^{mr}  &=& \sum \phi^{(mr) I} (x) Y^{I} (y), \nn
\eea
Here $(\mu, \nu)$ are AdS indices and $(a,b)$ are $S^3$ indices,
with $x$ denoting AdS coordinates and $y$ denoting sphere coordinates. The
subscript $(ab)$ denotes symmetrization of indices $a$ and $b$ with
the trace removed. Relevant properties of the spherical harmonics are
reviewed in appendix \ref{sphere}. We will often use a notation
where we replace the index $I$ by the degree of the harmonic $k$
or by a pair of indices
$(k,I)$ where $k$ is the degree of the harmonic and $I$ now parametrizes
their degeneracy, and similarly for $I_v, I_t$.

Imposing the de Donder gauge condition $D^{A} h_{aM} = 0$
on the metric fluctuations removes the fields with subscripts $(s,v)$.
In deriving the spectrum and computing correlation functions, this is
therefore a convenient choice. The de Donder gauge choice is however not
always a convenient choice for the asymptotic expansion of solutions;
indeed the natural coordinate choice in our application takes us
outside de Donder gauge. As discussed in \cite{Skenderis:2006uy}
this issue is straightforwardly dealt with by
working with gauge invariant combinations of the fluctuations.

Next let us briefly review the linearized spectrum
derived in \cite{Sez98}, focusing on fields dual to chiral
primaries. Consider first the scalars. It is
useful to introduce the following combinations
which diagonalize the linearized equations of motion:
\bea
s^{(r) k}_{I} &=& \frac{1}{4(k+1)} ({\phi}^{(5r) k}_{I} +2 (k+2)
{U}^{(r)k}_{I}), \la{diageqm} \\
%t^{(r)k}_{I} &=& \frac{1}{4} ({\phi}^{(5r)k}_{I} - 2 k {U}^{(r)k}_{I}),
%\nn \\
\s^k_{I} &=& \frac{1}{12 (k+1)} (6 (k+2) \hat{U}^{(5)k}_{I} -
\hat{\pi}_I^{k}), \nn
%\t^k_{I} &=& \frac{1}{12 (k+1)} ( \hat{\pi}^k_I + 6k \hat{U}^{(5) k}_{I}). \nn
\eea
The fields $s^{(r)k}$ and $\s^k$ correspond to scalar chiral
primaries, with the masses of the scalar fields being
\be \la{masses}
m_{s^{(r)k}}^2 = m_{\s^k}^2 = k (k-2), \hsp
\ee
The index $r$ spans $6 \cdots 5 + n_t$ with $n_t = 5, 21$ respectively
for $T^4$ and $K3$.
Note also that $k \ge 1$ for $s^{(r)k}$; $k \ge 2$ for $\s^k$. The hats
$(\hat{U}^{(5)k}_{I}, \hat{\pi}_I^{k})$ denote the following. As
discussed in \cite{Skenderis:2006uy}, the equations of motion for
the gauge invariant fields are precisely the same as those in de
Donder gauge, provided one replaces all fields with the corresponding
gauge invariant field. The hat thus denotes the appropriate gauge
invariant field, which reduces to the de Donder gauge field
when one sets to zero all fields with subscripts $(s,v)$. For our
purposes we will need these gauge invariant quantities only to
leading order in the fluctuations, with the appropriate
combinations being
\bea
\hat{\pi_2}^{I} &=& \pi_2^{I} + \Lambda^{2} \rho_{2(s)}^{I}; \la{gaug}
\\
\hat{U}^{(5)I}_2 &=& U^{(5)I}_2 - \half \rho^{I}_{2(s)}; \nn \\
\hat{h}^{0}_{\m\n} &=& h^{0}_{\m\n} - \sum_{\a,\pm} h_{\m}^{1 \pm \a}
h_{\n}^{1 \pm \a}. \nn
\eea
Next consider the vector fields. It is useful to introduce
the following combinations which diagonalize the equations of motion:
\be
h^{\pm}_{\m I_v} = \half (C_{\m I_v}^{\pm} - A^{\pm}_{\m
  I_v}), \hsp
Z_{\m I_v}^{(5) \pm} = \pm \qu (C_{\m I_v}^{\pm} + A_{\m I_v}^{\pm}).
\ee
For general $k$ the equations of motion are Proca-Chern-Simons
equations which couple $(A_{\m}^{\pm}, C^{\pm}_{\m})$ via a first
order constraint \cite{Sez98}. The three dynamical fields at each degree $k$
have masses $(k-1, k+1, k+3)$, corresponding to dual operators of
dimensions $(k,k+2,k+4)$ respectively; the operators of dimension $k$
are vector chiral primaries. The lowest dimension operators
are the R symmetry currents, which couple to the $k=1$ $A^{\pm \a}_{\m }$
bulk fields. The latter satisfy the Chern-Simons equation
\be \la{cs11}
F_{\m \n}(A^{\pm \a}) = 0,
\ee
where $F_{\m\n}(A^{\pm \a})$ is the curvature of the connection and
the index $\a = 1,2,3$ is an $SU(2)$ adjoint index. We will here only
discuss the vevs of these vector chiral primaries.

Finally there is a tower of KK gravitons with $m^2 = k (k+2)$ but
only the massless graviton, dual to the stress energy tensor, will
play a role here. Note that it is the
combination $\hat{H}_{\m \n} = \hat{h}^{0}_{\m \n} + \pi^{0} g^{o}_{\m \n}$ which
satisfies the Einstein equation; moreover one needs the appropriate
gauge covariant combination $\hat{h}^0_{\m\n}$ given in (\ref{gaug}).

\bigskip

Let us denote by $({\cal O}_{S^{(r) k}_I}, {\cal O}_{\Sigma^{k}_I})$
the chiral primary operators dual to the fields $(s^{(r) k}_I,
  \s^{k}_I)$ respectively. The vevs of the scalar operators with dimension two or less
can then be expressed in terms of the coefficients in the
  asymptotic expansion as
\bea
\left < {\cal O}_{S^{(r)1}_i} \right > &=& \frac{2 N}{\pi} \sqrt{2}
[s^{(r)1}_i]_1; \hsp
\left < {\cal O}_{S^{(r)2}_I} \right > = \frac{2 N}{\pi} \sqrt{6}
[s^{(r)2}_I]_2; \la{sc-1} \\
\left < {\cal O}_{\Sigma^{2}_I} \right > &=& \frac{N}{\pi} \left (2
  \sqrt{2} [\s^2_I]_2 - \frac{1}{3} \sqrt{2} a_{Iij}
  \sum_{r} [s^{(r) 1}_{i}]_1 [s^{(r) 1}_{j}]_1 \right ). \nn
\eea
Here $[\psi]_n$ denotes the coefficient of the $z^n$ term in the
  asymptotic expansion of the field $\psi$. The coefficient $a_{Iij}$
refers to the triple overlap between spherical harmonics, defined in
  (\ref{ap-ov0}). Note that dimension one scalar spherical harmonics
have degeneracy four, and are thus labeled by $i = 1, \cdots 4$.

Now consider the stress energy tensor and the R symmetry currents. The
three dimensional metric and the Chern-Simons gauge fields admit the
following asymptotic expansions
\bea
ds_{3}^{2} &=& \frac{dz^2}{z^2} + \frac{1}{z^2} \left(g_{(0)\bar{\mu}\bar{\nu}} + z^2
\left(g_{(2) \bar{\mu}\bar{\nu}} + {\rm{log}}(z^2) h_{(2)
  \bar{\mu}\bar{\nu}} + ({\rm{log}}(z^2))^2
\tilde{h}_{(2) \bar{\mu}\bar{\nu} }\right) + \cdots\right) dx^{\bar{\mu}} dx^{\bar{\nu}}; \nn \\
A^{\pm \a} &=& {\cal A}^{\pm \a} + z^2 A_{(2)}^{\pm \a} + \cdots
\eea
The vevs of the R symmetry currents $J^{\pm \a}_u$ are then given in terms
  of terms in the asymptotic expansion of $ A^{\pm \a}_{\m}$ as
\be
\left < J^{\pm
  \a}_{\bar{\mu}} \right >  = \frac{N}{4 \pi} \left (g_{(0) \bar{\mu} \bar{\nu}}  \pm \ep_{\bar{\mu}
  \bar{\nu}} \right ) {\cal A}^{\pm \a \bar{\nu}}.  \la{j1}
\ee
The vev of the stress energy tensor $T_{\bar{\mu}\bar{\nu}}$ is given by
\bea
\< T_{\bar{\mu}\bar{\nu}} \> &=& \frac{N}{2 \pi} \left (
g_{(2)\bar{\mu}\bar{\nu}} + \half R g_{(0) \bar{\mu}\bar{\nu}}
+ 8 \sum_r [\td{s}^{(r)1}_{i}]_1^2 g_{(0) \bar{\mu}\bar{\nu}} + \qu ({\cal A}^{+ \a}_{(\bar{\mu}}
{\cal A}^{+ \a}_{\bar{\nu})} + {\cal A}^{- \a}_{(\bar{\mu}}
{\cal A}^{- \a}_{\bar{\nu})}) \right) \la{t1}
\eea
where parentheses denote the symmetrized traceless combination of indices.

\bigskip

This summarizes the expressions for the vevs of chiral primaries with
dimension two or less which were derived in
\cite{Kanitscheider:2006zf}. Note that these operators correspond to
supergravity fields which are at the bottom of each Kaluza-Klein
tower. The supergravity solution of course
also captures the vevs of operators dual to the other fields in each
tower.  Expressions for these vevs were not derived in
\cite{Kanitscheider:2006zf}, the obstruction being the non-linear
terms: in general the vev of a dimension $p$ operator will include
contributions from terms involving up to $p$ supergravity
fields. Computing these in turn requires the field equations (along
with gauge invariant combinations, KK reduction maps etc) up to
$p$th order in the fluctuations.

Now (apart from the stress energy tensor) none of the operators whose
vevs are given above is an $SO(4)$ (R symmetry) singlet. For later purposes it will
be useful to review which other operators are $SO(4)$ singlets.
The computation of the linearized spectrum in \cite{Sez98} picks out
the following as $SO(4)$ singlets:
\be \la{singlet}
\t^{0} \equiv \frac{1}{12} \pi^0; \hsp
t^{(r)0} \equiv \qu \phi^{5(r)0},
\ee
along with $\phi^{0i(r)}$ with $i = 1,\cdots,4$.
Recall $\psi^0$ denotes the projection of the field $\psi$ onto the degree zero
harmonic. The fields $(\t^0, t^{(r)0})$ are dual to operators of
dimension four, whilst the fields $\phi^{0i(r)}$ are dual to
dimension two (marginal) operators. The former lie in the same tower as
$(\s^2, s^{(r)2})$ respectively, whilst the latter are in the same
tower as $s^{(r)1}$. In total there are $(n_t + 1)$ $SO(4)$ singlet
irrelevant operators and $4 n_t$ $SO(4)$ singlet marginal operators,
where $n_t = 5,21$ for $T^4$ and $K3$ respectively.

Consider the $SO(4)$ singlet marginal operators dual to the supergravity fields
$\phi^{i(r)}$. These operators have been discussed previously in the
context of marginal deformations of the CFT, see the review
\cite{David:2002wn} and references therein. Suppose one introduces a
free field realization for the $T^4$ theory, with bosonic and fermionic
fields $(x^{i}_I(z),\psi^i_I(z))$ where $I = 1,\cdots,N $. Then some
of the marginal operators can be explicitly realized in the untwisted sector
as bosonic bilinears
\be
\pa x_I^i(z) \bar{\pa} x_I^j (\bar{z});
\ee
there are sixteen such operators, in correspondence with
sixteen of the supergravity fields. The remaining four marginal
operators are realized in the twisted sector, and are associated with
deformation from the orbifold point.

\subsection{Application to the fuzzball solutions}

The six-dimensional metric of \eqref{D1D56d} in the decoupling limit
manifestly asymptotes to
\be
   ds^2 = \frac{r^2}{\sqrt{Q_1Q_5}}(-dt^2+ dy^2) + \sqrt{Q_1Q_5}\left(\frac{dr^2}{r^2} + d\W_3^2\right).
\ee
where
\be \la{d1}
Q_1 = \frac{Q_5}{L} \int_0^L dv (\dot{F}(v)^2 +
\dot{\cal F}(v)^2 + \dot{\cal{F}}^{\a -}(v)^2 ).
\ee
Note that the vielbein \eqref{D1D5vielbein} is asymptotically constant
\be
  V^o = \W_4^T \left(\begin{array}{cccc}
                       I_2 & 0 & 0 & 0\\
                       0 & \sqrt{Q_1/Q_5} & 0 & 0 \\
                       0 & 0 & \sqrt{Q_5/Q_1} & 0 \\
                       0 & 0 & 0 & I_{22}
                     \end{array}\right) \W_4,
\ee
but it does not asymptote to the identity matrix.
Thus one needs the constant $SO(5,21)$ transformation
\be
\la{vielbshift}
   V \rightarrow V (V^o)^{-1}, \qquad G_3 \rightarrow V^o G_3.
\ee
to bring the background into the form assumed in (\ref{background}).

The fields are expanded about the background values, by
expanding the harmonic functions defining the solution in spherical harmonics as
\bea
   H &=& \frac{Q_5}{r^2} \sum_{k,I} \frac{f_{kI}^5 Y_k^I(\theta_3)}{r^k}, \qquad
   K = \frac{Q_1}{r^2} \sum_{k,I} \frac{f_{kI}^1 Y_k^I(\theta_3)}{r^k}, \\
   A_i &=& \frac{Q_5}{r^2} \sum_{k \geq 1,I} \frac{(A_{kI})_i Y_k^I(\theta_3)}{r^k}, \qquad
   {\cal A} = \frac{\sqrt{Q_1Q_5}}{r^2} \sum_{k \geq 1,I}
   \frac{({\cal A} _{kI}) Y_k^I(\theta_3)}{r^k}, \nono \\
    {\cal A}^{\a_-} &=& \frac{\sqrt{Q_1 Q_5}}{r^2} \sum_{k\geq1,I} \frac{{\cal A}^{\a_-}_{kI}
  Y_k^I(\theta_3)}{r^k}. \nn
\eea
The polar coordinates here are denoted by $(r,\theta_3)$ and
$Y_k^I(\theta_3)$ are (normalized) spherical harmonics of degree $k$
on $S^3$ with $I$ labeling the degeneracy. Note that the restriction
$k \geq 1$ in the last three lines is due to the vanishing zero mode,
see \cite{Kanitscheider:2006zf}. As in \cite{Kanitscheider:2006zf},
the coefficients in the expansion can be expressed as
\bea
   f_{kI}^5 &=& \frac{1}{L (k+1)}  \int_{0}^L dv (C^I_{i_1\cdots i_k}
   F^{i_1} \cdots F^{i_k}) , \la{coeff} \\
   f_{kI}^1 &=& \frac{Q_5}{L (k+1)Q_1} \int_{0}^L dv \left (\dot{F}^2 + \dot{\cal{F}}^2 +
     (\dot{\cal{F}}^{\a_-})^2 \right ) C^I_{i_1\cdots i_k} F^{i_1} \cdots
   F^{i_k}  , \nono \\
   (A_{kI})_i &=& -\frac{1}{L (k+1)} \int_{0}^L  dv
\dot{F}_i C^I_{i_1\cdots i_k} F^{i_1} \cdots F^{i_k}, \nono \\
   (\cA_{kI}) &=& -\frac{\sqrt{Q_5}}{\sqrt{Q_1} L(k+1)} \int_{0}^L
dv \dot{\cal{F}} C^I_{i_1\cdots i_k} F^{i_1} \cdots F^{i_k} , \nono \\
   \cA^{\a_-}_{kI} &=& -\frac{\sqrt{Q_5}}{\sqrt{Q_1} L(k+1)} \int_0^L dv
\dot{\cal{F}}^{\a_-} C^I_{i_1\cdots i_k} F^{i_1} \cdots F^{i_k} .\nn
\eea
Here the $C^I_{i_1\cdots i_k}$ are orthogonal symmetric traceless rank $k$
tensors on $\mathbb{R}^4$ which are in one-to-one correspondence with
the (normalized) spherical harmonics $Y_k^I(\theta_3)$ of degree $k$
on $S^3$. Fixing the center of mass of the whole system implies that
\be
   (f_{1i}^1 + f_{1i}^5) = 0.
\ee
The leading term in the asymptotic expansion of the transverse gauge
field $A_i$ can be written in terms of degree one vector harmonics as
\be
  A = \frac{Q_5}{r^2}(A_{1j})_iY_1^j dY_1^i \equiv
\frac{\sqrt{Q_1Q_5}}{r^2}(a^{\a-}Y_1^{\a-} + a^{\a+} Y_1^{\a+}),
\ee
where $(Y_1^{\a-},Y_1^{\a+})$ with $\a=1,2,3$ form a basis for
the $k=1$ vector harmonics and we have defined
\be
   a^{\a\pm} = \frac{\sqrt{Q_5}}{\sqrt{Q_1}} \sum_{i > j} e^\pm_{\a ij} (A_{1j})_i,
\ee
where the spherical harmonic triple overlap $e^\pm_{\a ij}$ is
defined in \ref{ap-ov3}. The dual field is given by
\be
  B = -\frac{\sqrt{Q_1Q_5}}{r^2}(a^{\a-}Y_1^{\a-} - a^{\a+} Y_1^{\a+}).
\ee
Now given these asymptotic expansions of the harmonic functions one can
proceed to expand all the supergravity fields, and extract the
appropriate combinations required for computing the vevs defined in (\ref{sc-1}),
(\ref{j1}) and (\ref{t1}). Since the details of the
computation are very similar to those in \cite{Kanitscheider:2006zf},
we will simply summarize the results as follows. Firstly the vevs of
the stress energy tensor and of the R symmetry currents are the same
as in \cite{Kanitscheider:2006zf}, namely
\bea
\left < T_{\bar{\mu}\bar{\nu}} \right > &=& 0; \\
\left < J^{\pm \a} \right > &=& \pm \frac{N}{2 \pi} a^{\a \pm} (dy \pm
dt). \la{j-char}
\eea
The vanishing of the stress energy tensor is as anticipated, since
these solutions should be dual to R vacua. As in
\cite{Kanitscheider:2006zf}, however, the cancellation is very
non-trivial. The vevs of the scalar operators dual to the fields $(s^{(6)k}_I,\s^k_I)$
are also unchanged from \cite{Kanitscheider:2006zf}:
\bea
\left < {\cal O}_{S^{(6)1}_i} \right > &=& \frac{N}{4 \pi}
(- 4 \sqrt{2} f^{5}_{1i}); \la{vv2} \\
\left < {\cal O}_{S^{(6)2}_I} \right > &=& \frac{N}{4 \pi} ( \sqrt{6} (f^1_{2I} -
f^5_{2I}) ); \nn \\
\left < {\cal O}_{\Sigma^2_I} \right > &=& \frac{N}{4 \pi} \sqrt{2}
( -  (f^1_{2I} +f^5_{2I}) + 8 a^{\a -} a^{\b +} f_{I\a\b} ). \nn
\eea
The internal excitations of the new fuzzball solutions are therefore
captured by the vevs of operators dual to the fields
$s^{(r)k}_I$ with $r > 6$:
\bea
\left < {\cal O}_{S^{(5 + n_t)1}_i} \right > &=& - \frac{N}{\pi}
\sqrt{2} (\cA_{1i}); \hsp
\left < {\cal O}_{S^{(6+ \a_-)1}_i} \right > =
\frac{N}{\pi} \sqrt{2} \cA^{\a_-}_{1i}; \la{vv-new} \\
\left < {\cal O}_{S^{(5+n_t)2}_I} \right > &=&  - \frac{N}{2 \pi}
\sqrt{6} (\cA_{2I}); \hsp
\left < {\cal O}_{S^{(6+ \a_-)2}_I} \right > =
\frac{N}{2 \pi} \sqrt{6} \cA^{\a_-}_{2I}. \nn
\eea
Here $n_{t} = 5,21$ for $T^4$ and $K3$ respectively, with $\a_{-} =
1,\cdots,b^{2-}$ with $b^{2-} = 3,19$ respectively.
Thus each curve $({\cal F}(v), {\cal F}^{\a_-}(v))$ induces
corresponding vevs of operators associated with the middle cohomology
of $M^4$. Note the sign difference for the vevs of operators which are
related to the distinguished harmonic function ${\cal F}(v)$.

\section{Properties of fuzzball solutions} \la{s-field}

In this section we will discuss various properties of the fuzzball
solutions, including the interpretation of the vevs computed in the
previous section.

\subsection{Dual field theory}

Let us start by briefly reviewing aspects of the dual CFT and
the ground states of the R sector; a more detailed review of the
issues relevant here is contained in \cite{Kanitscheider:2006zf}.
Consider the dual CFT at the orbifold point; there is a family of
chiral primaries in the NS sector associated with the cohomology
of the internal manifold, $T^4$ or $K3$. For our discussions only the
chiral primaries associated with the even cohomology are relevant; let
these be labeled as ${\cal{O}}_{n}^{(p,q)}$ where $n$ is the twist and
$(p,q)$ labels the associated cohomology class. The degeneracy of the
operators associated with the $(1,1)$ cohomology is $h^{1,1}$. The
complete set of chiral primaries associated with the even cohomology
is then built from products of the form
\be
\prod_{l}({\cal{O}}_{n_l}^{p_l,q_l})^{m_l}, \hsp
\sum_{l} n_l m_l = N, \la{oper-1}
\ee
where symmetrization over the N copies of the CFT is implicit. The
correspondence between (scalar) supergravity fields and
chiral primaries is \footnote{As discussed in
  \cite{Kanitscheider:2006zf}, the dictionary between
  $(\s_n,s^{(6)}_n)$ and
  $({\cal{O}}_{(n-1)}^{(2,2)},{\cal{O}}_{(n+1)}^{(0,0)})$ may be more
  complicated, since their quantum numbers are indistinguishable, but this
  subtlety will not play a role here.}
\bea
\s_n & \leftrightarrow & {\cal{O}}_{(n-1)}^{(2,2)}, \hsp n \ge 2; \\
s^{(6)}_n & \leftrightarrow & {\cal{O}}_{(n+1)}^{(0,0)}, \hsp
s^{(6+\td{\a})}_n \leftrightarrow {\cal{O}}_{(n) \td{\a}}^{(1,1)}, \hsp \td{\a} =
1,\cdots h^{1,1}, \hsp n \ge 1. \nn
\eea
Spectral flow maps these chiral primaries in the NS sector to R ground states, where
\bea
h^R &=& h^{NS} - j_3^{NS} + \frac{c}{24}; \nonumber \\
j_3^R &=& j_3^{NS} - \frac{c}{12}, \label{spe_fl}
\eea
where $c$ is the central charge. Each of the operators in (\ref{oper-1})
is mapped  by spectral flow to a (ground state) operator of definite R-charge
\bea \label{Rop}
& & \prod_{l=1} ({\cal {O}}^{(p_l,q_l)}_{n_l})^{m_l} \quad
  \rightarrow \qquad
\prod_{l=1} ({\cal {O}}^{R (p_l,q_l)}_{n_l})^{m_l}, \\
j_3^R &=& \half \sum_l (p_l - 1) m_l, \qquad
\bar{j}^R_3 =\half \sum_l (q_l - 1) m_l. \nn
\eea
Note that R operators which are obtained from spectral flow of those
associated with the $(1,1)$ cohomology have zero R charge.

\subsection{Correspondence between geometries and ground states}

In \cite{Skenderis:2006ah,Kanitscheider:2006zf} we discussed
the correspondence between fuzzball geometries characterized by a
curve $F^i(v)$ and R ground states (\ref{Rop}) with $(p_l,q_l) = 1\pm
1$. The latter are related to chiral primaries in the NS sector built
from the cohomology common to both $T^4$ and $K3$, namely the $(0,0)$,
$(2,0)$, $(0,2)$ and $(2,2)$ cohomology.

The following proposal was made in
\cite{Skenderis:2006ah,Kanitscheider:2006zf} for the precise correspondence between
geometries and ground states; see also \cite{Alday:2006nd}. Given a curve $F^i(v)$
we construct the corresponding coherent state in the FP system and then
find which Fock states in this coherent state have excitation number
$N_L$ equal to $n w$, where $n$ is the momentum and $w$ is the winding.
Applying a map between FP oscillators and R operators then yields the
superposition of R ground states that is proposed to be dual to the D1-D5 geometry.

This proposal can be straightforwardly extended to the new
geometries, which are characterized by the curve $F^i(v)$ along with
$h^{1,1}$ additional functions $({\cal F}(v), {\cal F}^{\a_-}(v))$. Consider
first the $T^4$ system, for which the four additional functions are
$F^{\rho}(v)$. Then the eight functions $F^I(v) \equiv
(F^i(v),F^{\rho}(v))$ can be expanded in harmonics as
\be
F^I(v) = \sum_{n > 0} \frac{1}{\sqrt{n}} (\a^I_n e^{-i n \s^+} +
  (\a^I_n)^{\ast} e^{in \s^+}),
\ee
where $\s^+ = v/w R_9$. The corresponding coherent state in the FP
system is
\be
\left | F^I \right ) = \prod_{n,I} \left | \a^I_n \right ),
\ee
where $\left | \a^I_n \right )$ is a coherent state of the left moving
oscillator $\hat{a}^I_n$,  satisfying $\hat{a}^I_n \left | \a^I_n \right ) =
\a^I_n \left | \a^I_n \right )$. Contained in this coherent state are Fock
states, such that
\be
\prod (\hat{a}^I_{n_I})^{m_I} \left | 0 \right >, \hsp
N = \sum n_I m_I.
\ee
Now retain only the terms in the coherent state involving these Fock
states, and map the FP oscillators to CFT R operators via the
dictionary
\bea
\frac{1}{\sqrt{2}} (\hat{a}^1_n \pm i \hat{a}^2_n)  & \leftrightarrow &
{\cal O}^{R (\pm 1 + 1),(\pm 1 + 1)}_{n}; \la{fp-d} \\
\frac{1}{\sqrt{2}} (\hat{a}^3_n \pm i \hat{a}^4_n)  & \leftrightarrow &
{\cal O}^{R (\pm 1 + 1),(\mp 1 + 1)}_{n}; \nn \\
\hat{a}^{\rho}_{n} & \leftrightarrow & {\cal O}^{R (1,1)}_{(\rho-4) n}. \nn
\eea
The dictionary for the case of $K3$ is analogous. Here one has four
curves $F^{i}(v)$ describing the transverse oscillations and twenty
curves ${\cal{F}}^{\td{\a}}(v)$ describing the internal
excitations. The oscillators associated with the former are mapped to
operators associated with the universal cohomology as in (\ref{fp-d})
whilst the oscillators associated with the latter are mapped to
operators associated with the $(1,1)$ cohomology as
\be \la{dict1}
\hat{a}^{\td{\a}}_{n}  \leftrightarrow  {\cal O}^{R (1,1)}_{\td{\a} n}.
\ee
This completely defines the proposed superposition of R ground states
to which a given geometry corresponds. Note that below we will
suggest that a slight refinement of this dictionary may be necessary,
taking into account that one of the internal curves is distinguished by the
duality chain. For the distinguished curve the mapping may include a
negative sign, namely $\hat{a}_{n} \leftrightarrow - {\cal
  O}^{R (1,1)}_{n}$; this mapping would explain the relative sign
between the vevs found in (\ref{vv-new}) associated with the distinguished
curve ${\cal F}$ and the remaining curves ${\cal F}^{\a}$ respectively.

Note that there is a direct correspondence between the frequency of the
harmonic on the curve and the twist label of the CFT operator. The
latter is strictly positive, $n \ge 1$, and thus in the
dictionary (\ref{fp-d}) there are no candidate CFT operators
to correspond to winding modes of the curves $({\cal F}(v),
{\cal F}^{\a_-}(v))$. In the case of $T^4$ such candidates might be provided by
the additional chiral primaries associated with the extra $T^4$ in the
target space of the sigma model, discussed in \cite{Finn}. However
the latter is related to the degeneracy of the
right-moving ground states in the dual F1-P system, rather than to winding
modes. For $K3$ all chiral primaries
have been included (except for the additional primaries which appear
at specific points in the $K3$ moduli space). Thus one confirms that
winding modes of the curves $({ \cal F}(v), {\cal F}^{\a_-}(v))$ should not be included
in constructing geometries dual to the R ground states. As discussed
in appendix \ref{wind} these winding modes may describe
geometric duals of states in deformations of the CFT.

\subsection{Matching with the holographic vevs}

In this section we will see how
the general structure of the vevs given in (\ref{vv-new}) can be
reproduced using the proposed dictionary. The holographic vevs take the form
\be \la{sugra-vev}
\left < {\cal O}^{(1,1)}_{\td{\a} kI} \right > \approx
\frac{N \sqrt{Q_5} }{\sqrt{Q_1} L} \int^{L}_{0} dv \dot{\cal F}^{\td{\a}} C^{I}_{i_1 \cdots
  i_k} F^{i_1} \cdots F^{i_k}.
\ee
Thus the vevs of the operators ${\cal O}^{(1,1)}_{\td{\a} kI}$ are zero unless
the curve ${\cal F}^{\td{\a}}(v)$ is non-vanishing and at least one of
the $F^{i}(v)$ is non-vanishing. Moreover, the dimension one operators will not acquire
a vev unless the transverse and internal curves have excitations with
the same frequency. Analogous selection rules for frequencies of curve
harmonics apply for the vevs of higher dimension operators.

These properties of the vevs follow directly from the proposed
superpositions, along with selection rules for three point functions
of chiral primaries. The superposition dual to a given set of
curves is built from the R ground states
\be
{\cal O}^{R{\cal I}} | 0 \rangle =
\prod_{l} ({\cal O}^{R (p_l,q_q)}_{n_l})^{m_l} | 0 \rangle,
\ee
with $\sum_l n_l m_l = N$ and ${\cal I}$ labeling
the degeneracy of the ground states. So this superposition
can be denoted abstractly as $| \Psi ) = \sum_{\cal I}
a_{\cal I} {\cal O}^{R{\cal I}} | 0 \rangle$ with certain coefficients
$a_{\cal I}$. In particular, if the curve ${\cal F}^{\td{\a}}(v) = 0$ the
superposition does not contain any R ground states built from ${\cal O}^{R
  (1,1)}_{\td{\a} n}$ operators. Moreover, if there are no transverse
excitations, the superposition will contain only states with zero R
charge.

Now consider evaluating the vev of a dimension $k$ operator
${\cal O}^{(1,1)}_{\td{\a} k}$ in such a superposition. This is determined by
three point functions between this operator and the chiral primary
operators occurring in the superposition. More explicitly,
the operator vev is related to three point functions via
\be \la{3pf}
( \Psi_{NS} | {\cal O}^{(1,1)}_{\td{\a}k } | \Psi_{NS} ) = \sum_{{\cal I},
  {\cal J}} a_{{\cal I}}^{\ast} a_{{\cal J}}
\langle ({\cal O}^{{\cal I}})^{\dagger}(\infty) {\cal O}^{(1,1)}_{\td{\a}k}
(\mu) ({\cal O}^{{\cal J}}) (0) \rangle.
\ee
Here ${\cal O}^{{\cal I}}$ is the NS sector operator which flows to
${\cal O}^{R{\cal I}}$ in the R sector and $| \Psi_{NS} )$ is the flow
of the superposition back to the NS sector, namely $\sum_{\cal I}
a_{\cal I} {\cal O}^{\cal I}| 0 \rangle$. The quantity $\mu$ is a mass scale.
Note we are evaluating the relevant three point function in the NS sector,
and have hence flowed the ground states back to NS sector chiral
primaries. We would get the same answer by flowing the operator
whose vev we wish to compute, ${\cal O}^{(1,1)}_{\td{\a} k}$, into
the Ramond sector and computing the three point function there.
Recall that the R charges of these operators are related by the
spectral flow formula (\ref{spe_fl}) as $j_{3}^{NS} = j_{3}^R + \half
N$. In particular, NS sector chiral primaries built only from
operators associated with the middle cohomology all have the same R
charges, namely $\half N$.

There are two basic selection rules for the three point functions
(\ref{3pf}). Firstly, as usual one has to impose conservation of the R charges.
Secondly, a basic property of such
three point functions is that they are only
non-zero when the total number of operators ${\cal O}^{(1,1)}_{\td{\a}}$ with
a given index $\td{\a}$ in the correlation function is even
\footnote{Note that this selection rule was used for the computation of
three point functions of single particle operators in
the orbifold CFT in \cite{Jevicki:1998bm}.}. From a supergravity
perspective one can see this selection rule arising as follows. One
computes $n$-point correlation functions using $n$-point couplings in the
three dimensional supergravity action, with the latter following from the
reduction of the ten-dimensional action on $S^3 \times M^4$. Since
a $(1,1)$ form integrates to zero over $M^4$, the three dimensional
action only contains terms with an even number of fields $s^{\td{\a}}$
associated with a given $(1,1)$
cycle $\td{\a}$ on $M^4$. Therefore non-zero $n$-point functions must contain
an even number of operators ${\cal O}^{(1,1)}_{\td{\a}}$, and so do
corresponding multi-particle $3$-point functions obtained by taking
coincident limits.

Expressed in terms of cohomology,
allowed three point functions contain an even number of $(1,1)_{\td{\a}}$ cycles labeled by
$\td{\a}$. Thus in single particle correlators one can have processes such as
${\cal O}^{(0,0)} + {\cal O}^{(1,1)}_{\td{\a}} \rightarrow {\cal O}^{(1,1)}_{\td{\a}}$ and
${\cal O}^{(1,1)}_{\td{\a}} +  {\cal O}^{(1,1)}_{\td{\a}} \rightarrow
{\cal O}^{(2,2)}$, but processes
such as ${\cal O}^{(0,0)} + {\cal O}^{(1,1)}_{\td{\a}} \rightarrow
{\cal O}^{(0,0)}$ which involve an
odd number of $\td{\a}$ cycles are kinematically forbidden.
This kinematical selection rule for $(1,1)$ cycle conservation immediately
explains why the operator ${\cal O}^{(1,1)}_{\td{\a}k}$ can only acquire a
vev when the curve ${\cal F}^{\td{\a}}(v)$ is non-vanishing: only then
does the ground state superposition contain operators  ${\cal
  O}^{R(1,1)}_{\td{\a}}$ such that the selection rule can be satisfied.

One can also easily see why the operator only acquires a vev if there
are transverse excitations as well. All Ramond ground states
associated with the middle cohomology have
zero R charge, with the corresponding chiral primaries in the NS
sector having the same charge  $j_{3}^{NS} = \half
N$. Thus a superposition involving only ${\cal O}^{(1,1)}$ operators has a
definite R charge, and a charged operator cannot acquire a
vev. Including transverse excitations means that the superposition
of Ramond ground states contains charged operators, associated
with the universal cohomology, and does not have definite R
charge. Therefore a charged operator can acquire a vev.

Thus, to summarize, the proposed map between curves and superpositions
of R ground states indeed reproduces the principal features
of the holographic vevs. Using basic selection rules for three point
functions we have explained why the operators ${\cal
  O}^{(1,1)}_{\td{\a} k}$ acquire vevs only when the curve ${\cal
  F}^{\td{\a}}(v)$ is non-zero and when there are excitations in
$R^4$. We will see below that using reasonable assumptions for the three point
functions we can also reproduce the selection rules for vevs relating to
frequencies on the curves. Before discussing the general case,
however, it will be instructive to consider a particular example.

\subsection{A simple example}

Consider a fuzzball geometry
characterized by a circular curve in the transverse $R^4$ and one
additional internal curve, with only one harmonic of the same frequency:
\be \la{example}
F^1(v) = \frac{\mu A }{n} \cos (2 \pi n \frac{v}{L}); \hsp
F^2(v) = \frac{\mu A }{n} \sin (2 \pi n \frac{v}{L}); \hsp
{\cal F}(v) = \frac{\mu B }{n}  \cos (2 \pi n \frac{v}{L}),
\ee
where $\mu = \sqrt{Q_1 Q_5}/R$ and the D1-brane charge constraint
(\ref{d1}) enforces
\be
(A^2 + \half B^2) = 1.
\ee
The corresponding dual superposition of R ground states is then given by
\bea
| \Psi ) &= & \sum_{l=0}^{N/n} C_l ({\cal O}^{R (2,2)}_{n})^l ({\cal O}^{R
  (1,1)}_{1 n})^{\frac{N}{n} - l} \left | 0 \right >, \la{coh} \\
C_l &=& \sqrt{\frac{ ({\frac{N}{n}})!}
{ ({\frac{N}{n}} - l)!l !}} A^{l} (\frac{B}{\sqrt{2}})^{\frac{N}{n} -
  l}, \nn
\eea
with the operators being orthonormal in the large $N$ limit.
In the case that either $A$ or $B$ are zero
the superposition manifestly collapses to a single term. In the
general case, this superposition gives the following for the
expectation values of the R charges:
\bea
\left ( \Psi | j^R_3 | \Psi \right ) &=&
\left ( \Psi | \bar{j}^R_3 | \Psi \right ) = \half  \sum_{l=0}^{N/n} C_l^2 l ; \\
&=& \frac{N}{2n} \sum_{l=0}^{N/n-1}  \frac{(\frac{N}{n} -1)!}{l!
  (\frac{N}{n} - (l+1))!} A^{2(l+1)} (\frac{B}{\sqrt{2}})^{2
  (\frac{N}{n} - (l+1))}
= \frac{N}{2n} A^2. \nn
\eea
Evaluating (\ref{j-char}) for (\ref{example}) gives
\be
\left < J^{\pm 3} \right > = \frac{N}{2n R} A^2 (dy \pm dt),
\ee
and thus the integrated R charges defined in our conventions as
\be
\left < j_3 \right > = \frac{1}{2 \pi } \int dy \left < J^{+3} \right >;  \hsp
\left < \bar{j}_3 \right > = \frac{1}{2 \pi } \int dy \left < J^{-3}
\right >,
\ee
agree with those of the superposition of R ground states.

The kinematical properties also match between the geometry and the
proposed superposition. In particular, when $B \neq 0$ the $SO(2)$ symmetry in
the 1-2 plane is broken: the harmonic functions $(K,{\cal A})$
depend explicitly on the angle $\phi$ in this plane. The asymptotic
expansions of these functions involve charged harmonics, and therefore
charged operators acquire vevs characterizing the symmetry
breaking. More explicitly, the relevant terms in (\ref{coeff}) are
\bea
f^{1}_{kI} &\propto & \int_{0}^{L} dv (A^2 + B^2 \sin^2 (\frac{2 \pi n
  v}{L})) C^{I}_{i_1 \cdots i_k} F^{i_1} \cdots F^{i_k}; \\
{\cal A}_{kI} & \propto & \int_{0}^{L} dv B \sin (\frac{2 \pi n
  v}{L})) C^{I}_{i_1 \cdots i_k} F^{i_1} \cdots F^{i_k}. \nn
\eea
Now the symmetric tensor of rank $k$ and $SO(2)$ charge
in the 1-2 plane of $\pm m$ behaves as
\be
( (F^1)^2 + (F^2)^2)^{k-m} (F^1 \pm i F^2)^m = ( \frac{\mu
  A}{n} )^k e^{\pm 2 \pi i n m \frac{v}{L}}.
\ee
Note that $m$ is related to $(j_3, \bar{j}_3)$ via $m = j_3 +
\bar{j}_3$.
Thus, when $B \neq 0$, harmonics in the expansion of $f^1$ with charges
$|m| = 2$ are excited, and terms with $|m| = 1$ are excited in the
expansion of ${\cal A}$. Following \eqref{sugra-vev}
the latter implies that the dimension $k$ operators
${\cal O}^{(1,1)}_{1(km)}$ only acquire vevs when their $SO(2)$
charge $m$ in the 1-2 plane is $\pm 1$. In particular using
\eqref{vv-new} the vevs of the dimension one
operators are
\be \la{vev-one}
\langle {\cal O}^{(1,1)}_{1 (1 \pm 1)} \rangle = \mp i \frac{N}{2 \pi n}
\mu A B,
\ee
where the normalized degree one symmetric traceless tensors are
$\sqrt{2} (F^1 \pm i F^2)$.

These properties are implied by the
superposition (\ref{coh}). The latter is a superposition of states
with different R charge, and therefore it does break the $SO(2)$
symmetry, with the symmetry breaking being characterized by the vevs
of charged operators. Moreover following (\ref{3pf})
the vev of ${\cal O}^{(1,1)}_{1 (km)}$ is given by
\be \la{exx6}
\sum_{l,l'} C_{l}^{\ast} C_{l'} \langle ({\cal O}^{(2,2)}_{n})^l ({\cal O}^{
  (1,1)}_{1 n})^{\frac{N}{n} - l} | {\cal O}^{(1,1)}_{1 (k m)} (\mu) |
({\cal O}^{ (2,2)}_{n})^{l'} ({\cal O}^{
  (1,1)}_{1 n})^{\frac{N}{n} - l'}  \rangle.
\ee
For the dimension one operators, charge conservation reduces this to
\be \la{exx7}
\sum_{l} C_{l\pm 1}^{\ast} C_{l}  \langle ({\cal O}^{
  (2,2)}_{n})^{l\pm 1} ({\cal O}^{
  (1,1)}_{1 n})^{\frac{N}{n} \mp 1 - l} | {\cal O}^{(1,1)}_{1 ( 1 \pm 1)}
(\mu) | ({\cal O}^{ (2,2)}_{n})^{l} ({\cal O}^{(1,1)}_{1 n})^{\frac{N}{n} - l}  \rangle.
\ee
Thus there are contributions only from neighboring terms in the
superposition. 
%Now charge conservation implies that the vev is only
%non-zero when $m = \pm 1$, exactly the same selection rule as for the
%holographic vevs. 
Computing the actual values of these vevs is
beyond current technology: one would need to know
three point functions for single and multiple particle chiral primaries at
the conformal point. However, as in \cite{Kanitscheider:2006zf}, the
behavior of the vevs as functions of the curve radii $(A,B)$ can be
captured by remarkably simple approximations for the correlators,
motivated by harmonic oscillators. Suppose one treats the operators as
harmonic oscillators, with the operator ${\cal O}^{(1,1)}_{1 (1 1)}$
destroying one ${\cal O}^{(1,1)}_{1 n}$ and creating
one ${\cal O}^{ (2,2)}_{n}$. For harmonic oscillators such that
$[\hat{a},\hat{a}^{\dagger}] = 1$ the normalized state with $p$ quanta is given by
$| p \rangle = (\hat{a}^{\dagger})^p/\sqrt{p!} | 0 \rangle$ and therefore
$\hat{a}^{\dagger} | p \rangle = \sqrt{p+1} | p + 1 \rangle$. Using 
harmonic oscillator algebra for the operators gives
\be
\langle ({\cal O}^{
  (2,2)}_{n})^{l + 1} ({\cal O}^{
  (1,1)}_{1 n})^{\frac{N}{n} - 1 - l} | {\cal O}^{(1,1)}_{1 (1 1) }
(\mu) |  ({\cal O}^{(2,2)}_{n})^{l}
({\cal O}^{(1,1)}_{1 n})^{\frac{N}{n} - l}  \rangle \approx \mu \sqrt{ (\frac{N}{n} - l) (l+1)}.
\ee
Then the corresponding vev in the superposition $| \Psi )$ is
\be
\langle {\cal O}^{(1,1)}_{1 (1 1) } \rangle_{\Psi} =
\mu \sum_{l=0}^{N/n-1} c_{l+1}^{\ast} c_l \sqrt{ (\frac{N}{n} - l) (l+1)}
= \mu \frac{N}{n} AB,
\ee
which has exactly the structure of (\ref{vev-one}). Given that such
simple approximations (and factorizations) of the correlators
reproduce the structure of the vevs so well, it would be interesting
to explore whether this relates to simplifications in the structure
of the chiral ring in the large $N$ limit.

Next consider the vevs of dimension $k$ operators.  Using charge
conservation and $(1,1)$ cycle conservation in \eqref{exx6} implies
that only operators with $m$ odd can acquire a vev. To reproduce the
holographic result, that vevs are non-zero only when $m = \pm 1$,
requires the assumption that only nearest neighbor terms in the
superposition contribute to one point functions. This would follow
from a stronger selection rule for $(1,1)$ cycle conservation, that
the number of $(1,1)$ cycles in the in and out states differ by at
most one. In particular, multi-particle processes such as $
({\cal O}^{(1,1)}_{\td{a}n})^3 + {\cal O}^{(1,1)}_{\td{\a}n} \rightarrow 
({\cal O}^{(2,2)}_n)^3$ would be forbidden. The selection rules for 
holographic vevs suggest that there is indeed such cycle
conservation, and it would be interesting to explore this issue further.

Let us now return to the comment made below (\ref{dict1}), that one
may need to include a minus sign in the dictionary for the
distinguished curve. Such a minus sign would introduce factors of
$(-1)^{N/n-l}$ into the superposition \eqref{coh}, and thence an overall
sign in the vevs of the associated operators ${\cal
  O}^{(1,1)}_{1 (kI)}$. This would naturally account for the relative
sign difference between the vevs associated with the distinguished
curve and those associated with the remaining curves. It is not
conclusive that one needs such a minus sign without knowing the exact
three point functions and hence vevs. However such a sign change for
oscillators associated with the direction distinguished by the duality
would not be
surprising. Recall that under T-duality of closed strings right moving
oscillators associated with the duality direction switch sign, whilst
the left moving oscillators and oscillators associated with orthogonal directions do not.

\subsection{Selection rules for curve frequencies}

Selection rules for charge and $(1,1)$ cycles are sufficient to
reproduce the general structure of the vevs. In the particular example
discussed above, these rules also implied the selection rules for the
curve frequencies: operators acquire vevs only when the transverse and
internal curves have related frequencies.

Here we will note how, with reasonable assumptions, one can motivate
the selection rules for frequencies in the general case. Consider the
computation of the vev of a dimension one operator ${\cal
  O}^{(1,1)}_{\td{\a}1 } $ for a general
superposition $|\Psi )$ using (\ref{3pf}). Using the selection
rules for charge and $(1,1)$ cycles, the contributions to
(\ref{3pf}) involve only certain pairs of operators $({\cal O}^{{\cal
    I}},{\cal O}^{{\cal J}})$. Their $SO(2)$ charges must differ by
$(\pm 1/2, \pm 1/2)$ and they must differ by an odd number of
${\cal O}^{(1,1)}_{\td{\a}}$ operators.

Now let us make the further assumption that there are contributions to
(\ref{3pf}) only from pairs of operators $({\cal O}^{{\cal I}},{\cal
  O}^{{\cal J}})$ which differ by only one term, the relevant
operators taking the form
\be \la{form}
{\cal O}^{{\cal J}} = {\cal O}^{(p,q)}_{n} {\cal O}^{
  \td{{\cal J}}},
\ee
with ${\cal O}^{\td{{\cal J}}}$ being the same for in and out states, but
  the single operator ${\cal O}^{(p,q)}_{n}$ differing between in and
  out states. Thus we are assuming that the
relevant three point functions factorize, with the non-trivial part of the
correlator arising from a single particle process.

This is indeed the structure of the three point functions arising in our example. Only
nearest neighbor terms in the superposition contribute in the
computation of the vev of the dimension one operator in (\ref{exx7}).
Moreover the $m = \pm 1$ charge selection rule for the vevs of
higher dimension operators immediately follows from restricting to
nearest neighbor terms in the three-point functions. Note further that 
this factorization structure is present in the orbifold CFT computation of the
three point functions. The operator ${\cal O}^{(1,1)}_{\td{\a} 1}
\equiv {\cal O}^{(1,1)}_{\td{\a} 1} I^{N-1}$ is the identity operator
in $(N-1)$ copies of the CFT and thus only acts non-trivially in one
copy of the CFT.

Consider the case of the vev of the operator with $SO(2)$ charges
$(1/2, 1/2)$; it would take the form
\bea \la{3pf2}
&& \sum_{ {\cal I}, {\cal J},\td{{\cal I}}} a^{\ast}_{{\cal I}} a_{{\cal J}}
{\cal N}_{\td{{\cal I}}} \left ( \langle ({\cal O}_{{n}}^{(2,2)})^{\dagger} (\infty) {\cal
  O}^{(1,1)}_{\td{\a} 1} (\mu) ({\cal O}^{ (1,1)}_{\td{\a} n}) (0)
\rangle \right .\\
&& \qquad \qquad \left . + \langle ({\cal O}_{{n}}^{(1,1)})^{\dagger} (\infty) {\cal
  O}^{(1,1)}_{\td{\a} 1} (\mu) ({\cal O}^{ (0,0)}_{\td{\a} n}) (0)
\rangle \right ), \nn
\eea
where ${\cal N}_{\td{{\cal I}}}$ is the norm of ${\cal O}^{\td{{\cal I}}}$.
Analogous expressions would hold for the dimension one operators
with other charge assignments.
Such a factorization would immediately explain the frequency selection
rule found in the holographic vevs obtained from supergravity
\eqref{sugra-vev}. The superposition contains
operators of the form (\ref{form}) with both $(p,q) = (1,1)$ and
$(p,q) \neq (1,1)$ only when the internal curve and the transverse
curves share a frequency. Extending these arguments to vevs of
higher dimension operators would be straightforward, and would 
imply selection rules for curve frequencies.

\subsection{Fuzzballs with no transverse excitations} \la{int}

Consider the case where the fuzzball geometry has only internal
excitations, $F^i(v) = 0$. Then the corresponding dual
superposition of ground states can involve only states built from
the operators ${\cal O}^{R(1,1)}_{\a n}$. Any such state
will be a zero eigenstate of both $j^R_3$ and $\bar{j}^R_3$.
Furthermore, such ground states associated with the middle cohomology
account for a finite fraction of the
entropy of the D1-D5 system. In the case of $K3$ the total entropy behaves
as
\be \la{entropy}
S = 2 \pi \sqrt{\frac{c}{6}},
\ee
with $c = 24 N$. The ground states associated with the middle
cohomology account for a central charge $c = 20 N$. In the case of
$T^4$ the entropy behaves as (\ref{entropy})
with $c= 12 N$. The states
associated with the universal cohomology account for $c=4N$, the odd
cohomology accounts for another $c=4N$ and the middle cohomology
accounts for the final $c = 4N$.

Now let us consider the properties of the corresponding fuzzball geometry.
When there are no transverse excitations and no winding modes of
the internal curves, the $SO(4)$ symmetry in $R^4$ is
unbroken, and the defining harmonic functions (\ref{harm}) reduce to
\be \la{naive}
H = 1 + \frac{Q_5}{r^2}; \hsp K = \frac{Q_1}{r^2};
\ee
with $A_i = 0$ and where $Q_1$ is defined in (\ref{d1}).
The solutions manifestly all collapse to the
standard (singular) D1-D5 solution and so, whilst one would need an
exponential number of geometries (upon quantization)
to account for dual ground states build from operators associated with
the middle cohomology, one has
only one singular geometry. Therefore {\it the relevant fuzzball solutions
  are not visible in supergravity}: one needs to take into account
higher order corrections.

One can understand this from several perspectives. Firstly, as
discussed above, R ground states associated with the middle cohomology
have zero R charge; they do not break the $SO(4)$ symmetry. A geometry
which is asymptotically $AdS_3 \times S^3$ for which the $SO(4)$
symmetry is exact can be characterized by
the vevs of $SO(4)$ singlet operators. The only such operators in
supergravity are the stress energy tensor, and the scalar operators
listed in (\ref{singlet}). Since the vev of the stress
energy tensor must be zero for the D1-D5 ground states, the geometry
would have to be distinguished by the vevs of the singlet operators
given in (\ref{singlet}).

Our results imply that these operators do not acquire vevs, and
therefore within supergravity (without higher order corrections)
geometries dual to different R ground states associated with the
middle cohomology cannot be distinguished. The reason is the following.
The $SO(4)$ singlet operators dual to supergravity fields
are related to chiral primaries
by the action of supercharge raising operators; they are the top
components of the multiplets.
Thus these $SO(4)$ singlet operators cannot acquire
vevs in states built from the chiral primaries. $SO(4)$
singlet operators associated with stringy excitations would be needed
to characterize the different ground states.

\bigskip

A heuristic argument based on the supertube picture also
indicates that geometries dual to these ground states are not to be
found in the supergravity approximation. The
geometries with transverse excitations in $R^4$
can be viewed as a bound state of D1-D5 branes, blown up by
their angular momentum in the $R^4$. Indeed, the
characteristic size of the fuzzball geometry is directly related to
this angular momentum. The simplest example, related to a circular
supertube, is to take a geometry characterized by a circular curve;
this is obtained by setting $B=0$ in (\ref{example}). The characteristic
scale of the geometry is
\be
r_{c} \sim g_s \mu/n,
\ee
where $g_s$ is the string coupling and
$\mu$ has dimensions of length, whilst the (dimensionless) angular
momentum behaves as $j^{12} = N/n$, and thus $r_{c} \sim g_s \mu (j^{12}/N)$.
Hence the size of the D1-D5 bound state increases linearly with the
angular momentum. A general fuzzball geometry will of
course not be as symmetric but nonetheless the characteristic scale
averaged over the $R^4$ is still related to the total angular momentum.
In our previous paper \cite{Kanitscheider:2006zf}
we noted that fuzzball geometries dual to vacua for which the R
charge is very small are not well described by supergravity. Here we
have found that this implies that an exponential number of
geometries dual to a finite fraction
of the Ramond ground states, with strictly zero R charge,
cannot be described at all in the supergravity approximation.

\section{Implications for the fuzzball program} \la{final}

In this section we will consider the implications of our results for
the fuzzball program, focusing in particular on whether one can find
a set of smooth weakly curved supergravity geometries
which span the black hole microstates.

We have seen in the previous sections that the geometric duals of
superpositions of R vacua
with small or zero R charge are not well-described in supergravity.
The natural basis for R ground states (\ref{Rop}) uses states of
definite R-charges, and it is therefore straightforward to work out
the density of ground states with given R-charges, $d_{N,j_3,
  \bar{j}_3}$, with the total number of ground states being given by
$d_N = \sum_{N,j_3, \bar{j}_3} d_{N,j_3,\bar{j}_3}$. This computation is
discussed in appendix \ref{dens} with the resulting density in the
large $N$ limit being
\be
d_{N,j^{1},j^{2}} \cong \frac{1}{ 4 (N + 1- j)^{31/4}} \exp \left [ \frac{2 \pi (2 N -
    j )}{ \sqrt{N + 1 - j}} \right ] \frac{1}{ \cosh^2( \frac{ \pi j^{1}}{
\sqrt{N + 1 - j}} )
\cosh^2( \frac{ \pi j^{2}}{  \sqrt{N + 1 - j }} )},
\ee
where $j^{1} = (j_3 + \bar{j}_3)$ and $j^2 = (j_3 - \bar{j}_3)$ and
$j = | j^1 | + | j^2 |$.
The key feature is that the number of states with zero R charge differs from
the total number of R ground states given in (\ref{dens3}) only by a polynomial factor:
\be \la{zerop}
d_{N,0,0} \cong d_{N}/N.
\ee
The geometries dual to such ground states are unlikely to be
well-described in supergravity, and therefore the basis of black hole
microstates labeled by R charges is not a good basis for the
geometric duals. This argument reinforces the discussion of
\cite{Kanitscheider:2006zf}, where we showed in detail that the
geometric duals of specific states (in this basis) must be characterized by
very small vevs which cannot be reliably distinguished in supergravity; they
are comparable in magnitude to higher order corrections.

The geometries that are smooth in supergravity
correspond to specific superpositions of the R charge eigenstates,
for which some vevs are atypically large. The natural basis for the
field theory description of the microstates is thus {\it not} the
natural basis for the geometric duals. This issue is likely to persist
in other black hole systems. For example, the microstates of the
D1-D5-P system are also most naturally described as $(j_3,\bar{j}_3)$
eigenstates, with a relation analogous to (\ref{zerop}) holding, so the
number of states with zero R-charge is suppressed only polynomially
compared to the total number of black hole microstates. Just as in the
2-charge system discussed here, the geometric duals are related to supertubes
whose radii depend on the R-charges. States or superpositions of
states which have small or zero R-charges are unlikely to be well-described by
supergravity solutions. Thus a given smooth supergravity geometry
should be described by a specific superposition of the
black hole microstates. Identifying the
specific superpositions for known 3-charge geometries is an open and
important question.

The issue is whether there exist enough such geometries, well-described and
distinguishable in supergravity, to span the entire set of black
hole microstates. It seems unlikely that a basis exists
which simultaneously satisfies all three requirements.
Firstly, on general grounds microstates with small quantum numbers
will not be well-described in supergravity. Even when considering
superpositions that are well described by supergravity,
to span the entire basis, one will have to include superpositions which
can only be distinguished by these small vevs.
I.e. in choosing a basis of geometries for which some vevs are
sufficiently large for the supergravity description to be valid one
will find that some of these geometries cannot be distinguished
among themselves in supergravity.

We have already seen several examples of this problem in the 2-charge
system. Let us parameterize the curves as
\bea
F^{i}(v) &=& \mu \sum_{n} (\a^{i}_n e^{2 \pi i n v/L} + (\a^i_n)^{\ast}
e^{-2 \pi i n v /L} ); \\
{\cal F}^{\td{\b}} (v) &=& \mu \sum_{n} (\a^{\td{\b}}_n e^{2 \pi i n v/L} + (\a^{\td{\b}}_n)^{\ast}
e^{-2 \pi i n v /L} ), \nn
\eea
where $\mu = \sqrt{Q_1 Q_5}/R$ and $\td{\b}$ runs from $1$ to
$h^{1,1}(M^4)$. The D1-brane charge constraint  (\ref{d1-charg}) limits
the total amplitude of these curves as
\be \la{ll1}
\sum_{n} n^2 ( \left | \a^i_{n} \right |^2  + \left | \a^{\td{\b}}_{n}
\right |^2 ) = 1.
\ee
Thus in general increasing the amplitude in one mode, to make certain
quantum numbers large, decreases the amplitudes in the
others. Moreover, the amplitude in a given mode is bounded via
$\left | \a_{n} \right |^2 \le 1/n^2$, and is thus is intrinsically very small
for high frequency modes, which sample vacua with large twist
labels in the CFT. Note also that the vevs of R-charges are given in
terms of
\be \la{ll2}
j^{ij} = i N \sum_{n} n (\a^{i}_{n} (\a^{j}_{n})^{\ast} - \a^{j}_n (\a^i_n)^{\ast})
\ee
As we have seen, to be describable in supergravity,
geometries must have transverse $R^4$ excitations, and thus some
large R-charges, requiring $j^{ij} \gg 1$.
Combining (\ref{ll2}) and (\ref{ll1}) one sees that
this restricts the amplitudes of the internal excitations, and thus of
the sampling of the black hole microstates associated with the middle
cohomology of $M^4$.

Another way to understand the limitations of supergravity is to
go back to the F1-P system where the corresponding state is the coherent
state $|\{\a_n^i\}, \{\a_m^{\tilde{\b}}\})$. These states form a complete
basis of states, so we know that there is an F1-P geometry corresponding
to every 1/2 BPS state. However, only when all $\a_n^i, \a_m^{\tilde{\b}}$
are large are the geometries well-described and distinguishable
within supergravity. Indeed, the amplitudes $\a_n^i, \a_m^{\tilde{\b}}$
are also the root mean deviations of the distribution around the mean
(which is described by the classical curve),
so only for large $\a_n^i, \a_m^{\tilde{\b}}$
is the classical string that sources the supergravity solution
a good approximation of the quantum state. Putting it differently,
when some of the amplitudes are small
the difference in the solutions for different amplitudes is
comparable with the error in the solutions due to
the approximation of the source by a classical string, so
one cannot reliably distinguish them within this approximation.

If one could not find a basis of distinguishable supergravity geometries
spanning the microstates, one might ask whether a sufficiently
representative basis exists. That is, suppose one chooses a single
representative of the indistinguishable geometries, and assigns a measure
to this geometry. Then is the corresponding basis of weighted geometries
sufficiently representative to obtain the black hole properties?
In the 2-charge system, the now complete set of fuzzball geometries along
with the precise mapping between these geometries and R vacua allows
these questions to be addressed at a quantitative level and we hope to
return to this issue elsewhere.

\section*{Acknowledgments}

The authors are supported by NWO, KS via
the Vernieuwingsimpuls grant ``Quantum gravity and particle physics''
and IK, MMT  via the Vidi grant ``Holography,
duality and time dependence in string theory''. This work was also
supported in part by the EU contract MRTN-CT-2004-512194. KS and MMT would like
to thank both the 2006 Simons Workshop and the theoretical physics
group at the University of Crete, where some of this work was completed.
The computer algebra package GRTensor was used to verify that our
solutions satisfy the supergravity field equations.

\appendix

\section{Conventions}

The following table summarises the indices used throughout the
paper. In some cases an index is used more than once, with different
meanings, in separate sections of the paper.

\begin{center}
\begin{tabular}{|l|l|l|} \hline
Index & Range & Usage \\ \hline
$(m,n)$ & $0,\cdots,9$ & 10d sugra fields \\ \hline
$(M,N)$ & $0,\cdots,5$ & 6d sugra fields \\ \hline
$(\m,\n)$ & $0,1,2$ & 3d fields \\ \hline
$(a,b)$ & $1,2,3$ & $S^3$ indices \\ \hline
$(i,j)$ & $1,2,3,4$ & $R^4$ indices \\ \hline
$(\rho,\sigma)$ & $1,2,3,4$ & $M^4$ indices \\ \hline
$(\bar{\mu},\bar{\nu})$ & $0,1$ & 2d fields \\ \hline
$(\a,\b)$ & $1,2,3$ & $SU(2)$ vector index \\ \hline
$(\g,\d)$ & $1,\cdots,b^2$ & $H_2 (M^4)$ \\ \hline
$(\td{\a}, \td{\b})$ & $1,\cdots,h^{1,1}$ & $H_{1,1} (M^4)$ \\ \hline
$(I,J)$ & $1,\cdots, 8$ & $SO(8)$ vector \\ \hline
$((c),(d))$ & $1,\cdots, 16$ & heterotic vector fields \\ \hline
$((a),(b))$ & $1,\cdots, 24$ & $SO(4,20)$ vector \\ \hline
$(A,B)$ & $1,\cdots, 26$ & $SO(5,21)$ vector \\ \hline
$(m,n)$ & $1,\cdots, 5$ & $SO(5)$ vector \\ \hline
$(r,s)$ & $6,\cdots, (n_t + 1)$ & $SO(n_t)$ vector \\ \hline
\end{tabular}
\end{center}

\subsection{Field equations} \la{fe}

The equations of motion for IIA supergravity are:
\bea
\la{conv_IIA}
   && e^{-2\Phi}(R_{mn} +2\na_m \na_n \Phi - \frac{1}{4}
H^{(3)}_{mpq}
H_n^{(3)pq}) -\hp F^{(2)}_{m p} {F^{(2)p}_{n}} -\frac{1}{2\cdot3!} F^{(4)}_{mpqr}
F_n^{(4)pqr} \nono \\
   && \qquad +\frac{1}{4} G_{m n}(\hp (F^{(2)})^2 + \frac{1}{4!} (F^{(4)})^2) = 0, \\
   && 4\na^2 \Phi - 4(\na \Phi)^2 +R - \frac{1}{12} (H^{(3)})^2 =0, \nono \\
   && dH^{(3)} = 0, \qquad dF^{(2)} =0, \qquad \na_m F^{(2)m n} - \frac{1}{6} H^{(3)}_{pqr}
F^{(4) n pqr} = 0, \nono \\
&&  \na_m(e^{-2\Phi}H^{(3)mnp}) -\hp
F^{(2)}_{qr}F^{(4) qr np}
-\frac{1}{2\cdot (4!)^2} \e^{n p m_1 \cdots m_4 n_1 \cdots n_4} F^{(4)}_{m_1
  \cdots m_4} F^{(4)}_{n_1 \cdots n_4} = 0, \nono \\
   && dF^{(4)} = H^{(3)} \wedge F^{(2)}, \qquad \na_m F^{(4)mnpq} - \frac{1}{3!
  \cdot 4!}
\e^{n pq m_1 \cdots m_3 n_1 \cdots n_4} H^{(3)}_{m_1 \cdots m_3} F^{(4)}_{n_1
  \cdots n_4} = 0. \nn
\eea
The corresponding equations for type IIB are:
\bea
\la{conv_IIB}
   && e^{-2\Phi}(R_{m n} +2 \na_m \na_n \Phi - \frac{1}{4}
H^{(3)}_{m pq} H_n^{(3)pq}) - \hp F^{(1)}_{m} F^{(1)}_{n} -\frac{1}{4} F^{(3)}_{m pq}
F_n^{(3)pq} - \frac{1}{4\cdot4!} F^{(5)}_{m pqrs} {F_n}^{(5)pqrs} \nono \\
   && \qquad +\frac{1}{4} G_{m n}( (F^{(1)})^2 + \frac{1}{3!} (F^{(3)})^2 ) =0,
\nono \\
   && 4\na^2 \Phi - 4(\na \Phi)^2 +R - \frac{1}{12} (H^{(3)})^2 =0, \nono \\
   && dH^{(3)} = 0, \qquad \na_m(e^{-2\Phi}H^{(3)m n p}) -
F^{(1)}_{m}F^{(3) m n p } -\frac{1}{3!}
F^{(3)}_{m qr } F^{(5) m q r n p} = 0,  \\
   && dF^{(1)} =0, \qquad \na_m F^{(1) m} + \frac{1}{6} H^{(3)}_{pqr}
F^{(3) pqr} = 0,
\nono \\
   && dF^{(3)} = H^{(3)} \wedge F^{(1)}, \qquad \na_m F^{(3) m n p}
+ \frac{1}{6} H^{(3)}_{m q r } F^{(5) m q r n p} = 0, \nono \\
   && dF^{(5)} = d(\ast F^{(5)}) = H^{(3)} \wedge F^{(3)}, \nn
\eea
where the Hodge dual of a $p$-form $\omega_p$ in $d$ dimensions is
given by
\be
   (\ast \, \omega_p)_{i_1 \cdots i_{d-p}} = \frac{1}{p!}\e_{i_1
  \cdots i_{d-p} j_1 \cdots j_p} \omega_p^{j_1 \cdots j_p},
\ee
with $\e_{01 \cdots d-1} = \sqrt{-g}$. The RR field strengths are defined as
\be
   F^{(p+1)} = dC^{(p)} - H^{(3)} \wedge C^{(p-2)}.
\ee
The equations of motion for the heterotic theory are:
\bea
    &&4\na^2\Phi - 4\left(\na\Phi\right)^2 + R - \frac{1}{12}(H^{(3)})^2
          - \a' (F^{(c)})^2 = 0, \nono \\
    && \na_m\left(e^{-2\Phi}H^{(3) mnr}\right) = 0, \nono \\
    &&R^{mn} + 2\na^m \na^n \Phi
          - \frac{1}{4}H^{(3) mrs}H^{(3)n}_{rs}
          - 2\a' F^{(c) mr} F^{(c) n}_{r} = 0, \nono \\
   &&{\na_m}\left(e^{-2\Phi}F^{(c) mn}\right)
          + \half e^{-2\Phi}H^{(3) nrs}F^{(c)}_{rs} = 0. \nn
\eea
$F^{(c)}_{mn}$ with $(c) = 1, \cdots 16$ are the field strengths
of Abelian gauge fields $V^{(c)}_m$; we consider here only supergravity
backgrounds with Abelian gauge fields. This restriction means that the
gauge field part of the Chern-Simons form in $H_3$,
\be
H^{(3)} = d B^{(2)} - 2 \a' \w_3 (V) + \cdots,
\ee
does not play a role in the supergravity solutions, nor does the
Lorentz Chern-Simons term denoted by the ellipses.

\subsection{Duality rules} \la{duality}

The T-duality rules for RR fields were derived in \cite{Ber95} by
reducing type IIA and type IIB supergravities on a circle and
relating the respective RR potentials in
the 9-dimensional theory. However, for calculations
involving magnetic sources, it is more convenient to
work with T-duality rules for RR field strengths, since
potentials can only be defined locally. In the following we will
rederive the T-duality rules in terms of RR field strengths.

It is slightly easier although not necessary to use the democratic
formalism of IIA and IIB supergravity introduced in
\cite{Ber01}. In this formalism one includes
$p$-form field strengths for $p > 5$ with Hodge dualities relating higher and
lower-form field strengths being imposed in the field equations.
This formalism is natural when both magnetic and electric sources are
present; moreover there is no need for
Chern-Simons terms in the field equations. The RR part of the
(pseudo)-action is simply
\be
\label{equ_RR10d}
  {S}_{RR} = -\frac{1}{2\k_{10}^2} \int d^{10} x \sqrt{-{{g}}} \sum_q
\frac{1}{4q!} ({F}^{(q)})^2,
\ee
where $q = 2,4,6,8$ is even in the IIA case and $q=1,3,5,7,9$
is odd in the IIB case. The field strengths are defined as
${F}^{(q)} = d{C}^{(q-1)} - {H}^{(3)} \wedge {C}^{(q-3)}$ for $q \geq 3$ and
${F}_q = d{C}^{(q-1)}$ for $q < 3$.
The Hodge duality relation between higher and lower form field
strengths in our conventions is
\be
\label{equ_Hodge10d}
   \ast {F}^{(q)} = (-1)^{\lfloor \frac{q}{2} \rfloor} {F}^{(10-q)},
\ee
where $\lfloor n \rfloor$ denotes the largest integer less or equal to $n$.

Now to compactify on a circle the ten-dimensional metric can be
parameterized as
%Compactifying on a circle can be most easily done using the metric ansatz
\be
   {ds}^2 = e^{2 \psi} (dy-A_\m dx^\m)^2 + \hat{g}_{\m\n} dx^\m dx^\n,
\ee
where $y$ denotes the compact direction, and
9-dimensional quantities will be denoted as hatted. An economic way to derive the
   T-duality rules for the field strengths is the following. Choose
   the vielbein to be
\be  \label{equ_vielbein}
{e}^{\underline{y}} = e^{\psi} ( dy - A_{\mu} dx^{\mu}); \hsp
{e}^{\underline{\mu}} = \hat{e}^{\underline{\mu}},
\ee
where $\underline{\mu}$ denotes a tangent space index, and
   $\hat{e}^{\underline{\mu}}$ is the 9-dimensional vielbein.
Now reduce the field strengths (in the tangent frame) as
\be
   \hat{F}^{(q)}_{\underline{\mu}_1...\underline{\mu}_q }=
       {F}^{(q)}_{\underline{\mu}_1...\underline{\mu}_q},
\qquad
   \hat{F}^{(q-1)}_{\underline{\mu}_1...\underline{\mu}_{q-1}} =
{F}^{(q)}_{\underline{\mu}_1...\underline{\mu}_{q-1} \underline{y}}.
\ee
The corresponding 9-dimensional action for the field strengths
is given by
\be
   S_{RR} = -\frac{2 \pi R}{2\k_{10}^2} \int d^{9} x \sqrt{- \hat{g}}
\sum_{q=1}^{9} \frac{1}{4q!} e^{\psi}  \hat{F}_q ^2.
\ee
Since $\psi_{IIA} = - \psi_{IIB}$ under T-duality, one can read from
this action the transformation rules for field strengths in 10d:
\bea
\label{equ_Tdualviel}
   \tilde{F}^{(q+1)}_{\underline{\mu}_1 \cdots \underline{\mu}_q \underline{y}} &=& e^{\psi}
   F^{(q)}_{\underline{\mu}_1 \cdots \underline{\mu}_q}, \\
   \tilde{F}^{(q+1)}_{\underline{\mu}_1...\underline{\mu}_{q+1}} &=& e^{\psi}
   F^{(q+2)}_{\underline{\mu}_1...\underline{\mu}_{q+1}
   \underline{y}}. \nono
\eea
Here $q$ even defines IIB fields in terms of IIA fields and $q$ odd
defines IIA in terms of IIB. Note that the field strengths on both
   sides are in the tangent frame.
%The sign in \eqref{equ_Tdualviel} is fixed up to an
%   overall minus sign in order to be consistent with the reduced
%   version of \eqref{equ_Hodge10d}
%\footnote{In order for the equations of motion of IIA resp. IIB in 9d
%   to be equivalent it would also matter how the $F_p$ are defined in
%   terms of potentials. However, this relative definition of
%   potentials just gives the usual T-duality rules of \cite{Ber95}.}
Given the T-duality rules for NSNS fields
\bea
  e^{\td{\psi}}  &=& e^{-\psi}, \qquad \tilde{A}_{\mu} = B^{(2)}_{y \mu},
\qquad \tilde{B}^{(2)}_{y m} = A_{m}, \\
\tilde{B}^{(2)}_{m n} &=&
B^{(2)}_{m n} + 2A_{[m} B^{(2)}_{n]y}, \qquad
   \tilde{\Phi} = \Phi - \psi, \nn
\eea
with the metric $g_{m n}$ invariant,
one can easily convert \eqref{equ_Tdualviel} back into
\bea
\label{equ_Tdualst}
   F^{(q)}_{m_1...m_q} &=& F^{(q+1)}_{m_1...m_q y} - q(-1)^q
   B^{(2)}_{{y} [m_1} F^{(q-1)}_{m_2...m_q ]} + q(q-1)
   B^{(2)}_{{y} [m_1} A_{m_2} F^{(q-1)}_{m_3...m_q] y } \nono \\
   F^{(q)}_{m_1...m_{q-1} {y}} &=&
   F^{(q-1)}_{m_1...m_{q-1}} - (q-1)(-1)^q A_{[m_1}
F^{(q-1)}_{m_2...m_{q-1}] {y}}.
\eea
Strictly speaking, this gives the duality rules in the democratic
   formalism. However we can obtain the usual rules by simply dropping the ($p >
   5$)-form field strengths as long as we make sure to
self-dualise $F^{(5)}$ in each IIB solution.

The S duality rules for type IIB are
\bea
\la{Sdual}
   \tilde{\t} &=& -\frac{1}{\t}, \qquad \tilde{B}^{(2)} = C^{(2)}, 
\qquad \tilde{C}^{(2)} = - B^{(2)}, \nono \\
   \tilde{F}^{(5)} &=& F^{(5)}, \qquad \tilde{G}_{mn} = |\t| G_{mn},
\eea
where $\t = C^{(0)} + ie^{-\Phi}$.

\section{Reduction of type IIB solutions on $K3$} \la{IIBK3}

The reduction of type IIB on $K3$ is very similar to the reduction of
type IIA, which was discussed in some detail in \cite{Duf95}. In the
following we will use the reduction of the NS-NS sector fields given
in \cite{Duf95}, and derive the reduction of the type IIB RR fields.
Let us first review the reduction of the NS-NS sector. Starting from
the ten-dimensional action
\be
S_{NS} = \frac{1}{2\k_{10}^2} \int d^{10}x \sqrt{- \hat{g}}\left ( e^{-2 \hat{\Phi}}
(\hat{R} + 4 (\pa \hat{\Phi})^2 - \frac{1}{12} \hat{H}_3^2) \right ),
\ee
where ten-dimensional fields are denoted by hats,
the corresponding six-dimensional field equations can be derived from
the action \cite{Duf95}
\be
S = \frac{1}{2 \k_6^2} \int d^6x \sqrt{-g} e^{-2\Phi}
   \left(R + 4 (\pa \Phi)^2 - \frac{1}{12} H_3^2 + \frac{1}{8} \tr(\pa
   M^{-1} \pa M) \right),
\ee
where the six-dimensional fields are defined as follows. Firstly the
10-dimensional 2-form potential is reduced as
\be
\hat{B}^{(2)}(x,y) = B_2(x) + b^{\gamma} (x)\w_2^{\gamma}(y),
\ee
where $(x,y)$ are six-dimensional and $K3$ coordinates respectively
and the two forms $\w_2^{\gamma}$ with $\gamma = 1, \cdots 22$ span the cohomology
$H^2(K3,\mathbb{R})$. The 2-forms $\w_2^{\gamma}$ transform under an $O(3,19)$ symmetry,
with a metric defined by the $22$-dimensional intersection matrix
\be
\la{K3dABdef}
   d_{\g\d} = \frac{1}{(2 \pi)^4 V} \int_{K3} \w_2^{\g} \wedge \w_2^{\d},
\ee
where $(2 \pi)^4 V$ is the volume of $K3$.
A natural choice for $d_{\g\d}$ is
\be
   d_{\g\d} = \left(\begin{array}{cc}
                     I_3 & 0 \\
                     0 & - I_{19} \end{array}\right),
\ee
corresponding to a diagonal basis for the $3$ self-dual and $19$
anti-self dual two forms of $K3$.
Furthermore, there is a matrix $D^{\d}_{\hspace{2mm} \g}$ defined by the action of the Hodge
operator
\be
\la{K3Hodge}
   \ast^{K3}_4 \w_2^{\g} = \w_2^{\d} D^{\d}_{\hspace{2mm} \g},
\ee
which is dependent on the $K3$ metric and satisfies
\bea
   D^{\g}_{\hspace{2mm} \d} D^{\d}_{\hspace{2mm} \e}
= \d^{\g}_{\hspace{2mm} \e}, \qquad D^{\e}_{\hspace{2mm} \d} d_{\e\z}
   D^{\z}_{\hspace{2mm} \g} = d_{\d\z}.
\eea
The $SO(4,20)$ matrix of scalars $M^{-1}_{(a)(b)}$
was derived in \cite{Duf95} to be
\be
\la{equ_M420}
M^{-1}= \W_2^T\left(\begin{array}{ccc}
   e^{-\r} + b^{\g} b^{\d} d_{\g \e} D^{\e}_{\d} + \frac{1}{4} e^\r b^4 &
\hp e^\r b^2 & \hp e^\r b^2 b^{\g} d_{\g\d} + b^{\g} d_{\g\e} D^{\e}_{\d} \\
   \hp e^\r b^2 & e^\r & e^\r b^{\g} d_{\g\d} \\
   \hp e^\r b^2 b^{\g} d_{\g\d} + b^{\g} d_{\g\e} D^{\e}_{\d}
& e^\r b^{\g} d_{\g\d} & e^\r b^{\e}
   d_{\e\g} b^{\z} d_{\z\d} + d_{\g \e} D^ {\e}_{\d}
   \                   \end{array}\right) \W_2,
\ee
with $b^2 \equiv b^{\g} b^{\d} d_{\g\d}$. Here $\r$ is the breathing mode of
$K3$, $e^{-\r} = \frac{1}{(2\pi)^4 V}\int_{K3} \ast_4 1$.
The six-dimensional dilaton is related to the 10-dimensional dilaton
via $\Phi = \hat{\Phi} + \r/2$.

The dimensional reduction of the NS sector makes manifest only an
$SO(4,20)$ subgroup of the full $SO(5,21)$ symmetry. Including the
reduction of the RR sector should thus give the equations of motion
following from the six-dimensional string frame action,
which for IIB was given in
\eqref{IIBK3string}
\bea
   S &=& \frac{1}{2 \k_6^2} \int d^6x \sqrt{-g} \left\{ e^{-2\Phi}
   \left(R + 4 (\pa \Phi)^2 + \frac{1}{8} \tr(\pa M^{-1} \pa M)\right)
   + \hp \pa l^{(a)} M^{-1}_{(a)(b)} \pa l^{(b)}\right. \nono \\
     && \left.- \frac{1}{3} G^A_{MNP} \cM^{-1}_{AB} G^{B MNP}\right\}, \nono
\eea
and in which only an $SO(4,20)$ subgroup of the total $SO(5,21)$
symmetry is manifest; recall that ${\cal M}^{-1}_{AB}$ here is an $SO(5,21)$
matrix, with $M_{(a) (b)}^{-1}$ being $SO(4,20)$. Note that the
six-dimensional coupling is related to the ten-dimensional coupling
via $(2\pi)^4 V (2 \k_6^2) = 2 \k_{10}^2$, where $(2 \pi)^4 V$ is the
volume of $K3$.

Following the same steps as \cite{Duf95} the RR potentials can be reduced as
\bea
\la{redIIBpot}
   \hat{C}^{(0)}(x,y) &=& C_0(x), \qquad
   \hat{C}^{(2)}(x,y) = C_2(x) + c_{(0,2)}^{\g}(x) \w_2^{\g}(y),  \\
   \hat{C}^{(4)}(x,y) &=& C_4(x) + c_{(2,4)}^{\g}(x)\wedge\w_2^{\g}(y) +
c_{(0,4)}(x) (e^\rho \ast_{K3} 1)(y), \nn
\eea
where $\ast_{K3}$ denotes the Hodge dual in the $K3$ metric and
the corresponding field strengths are
\bea
\la{redIIBfs}
   \hat{F}^{(1)}(x,y) &=& F_1(x), \\
   \hat{F}^{(3)}(x,y) &=& dC_2(x) - C_0(x) H_3(x) + \left (dc_{(0,2)}^{\g}(x) -
   C_0(x) db^{\g}(x) \right ) \w_2(y)  \equiv F_3 + K_1^{\g} \wedge \w_2^{\g}, \nono \\
   \hat{H}^{(3)}(x,y) &=& dB_2(x) + db^{\g}(x)\wedge \w_2^{\g} (y) \equiv
   H_3 + db^{\g} \wedge \w_2^{\g}, \nono \\
   \hat{F}^{(5)}(x,y) &=& dC_4(x) - C_2(x)\wedge H_3(x) + \left (dc_{(2,4)}^{\g}(x)
   - C_2(x)db^{\g}(x) - c_{(0,2)}^{\g} (x) H_3(x) \right ) \wedge \w_2^{\g} (y) \nono \\
    &&+ \left (dc_{(0,4)}(x) - c^{\g}_{0,2}(x) db^{\d}(x) d_{\g\d}
   \right ) \wedge (e^\r(x) \ast_{K3} 1)(y) \nono \\
    &&\equiv F_5 + K_3^{\g} \wedge \w_2^{\g} + \td{F}_1 \wedge e^\rho \ast_{K3} 1. \nn
\eea
The reduction of the potentials thus gives two three form field
strengths $H_3$ and $F_3$, 3 self-dual and 19 anti-self dual three
form field strengths $K_3^{\g}$ and 46 scalars $b^{\g}$, $c^{\g}_{(0,2)}$,
$c_{(0,4)}$ and $C_0$. After splitting the three forms $H_3$ and
$F_3$ into their self-dual and anti-self-dual parts, we obtain 5
self-dual and 21 anti-self dual tensors in total, as described in
\cite{Townsend:1983xt}.

It is then straightforward to obtain the map relating six and
ten-dimensional fields by inserting the expressions
\eqref{redIIBpot} and \eqref{redIIBfs} into the ten-dimensional
field equations \eqref{conv_IIB}. The additional RR scalars are contained in
\be
   l^{(a)} = \W_2^T \left( \begin{array}{c}
                  C_0 \\ \tilde{c}_{(0,4)} \\ \tilde{c}_{(0,2)}^{\g} \end{array} \right),
   \ee
with $\W_{2}$ defined in the appendix \ref{bc_matrices}
and the shifted fields defined as
\bea
\la{IIBscalshifts}
   \tilde{c}_{(0,2)}^{\g} &=& c^{\g}_{(0,2)} - C_0 b^{\g}, \\
   \tilde{c}_{(0,4)} &=& c_{(0,4)} - b^{\g} c_{(0,2)}^{\d}
   d_{\g\d} + \hp b^2 C_0. \nn
\eea
The fields $\Phi$, $l^{(a)}$ and the
$SO(4,20)$ matrix $M^{-1}$ given in \eqref{equ_M420} can be recombined
into the $SO(5,21)$ matrix $\cM^{-1}= V^TV$, with the latter
conveniently expressed in terms of the vielbein
\be
\la{redIIBscalars}
   V = \W_4^T \left(\begin{array}{ccccc}
                       e^{-\Phi} & 0 & 0 & 0 & 0\\
                      -e^{\Phi}(C_0 c_{(0,4)} - \hp c^2_{(0,2)}) &
e^{\Phi} & -e^{\Phi} \tilde{c}_{(0,4)} & - e^{\Phi} C_0 & e^{\Phi}\tilde{c}_{(0,2)}^{\g} d_{\g\d} \\
                       e^{-\r/2}C_0 & 0 & e^{-\r/2} & 0 & 0 \\
                       e^{\r/2}c_{(0,4)} & 0 & \hp e^{\r/2} b^2 &
                       e^{\r/2} & e^{\r/2}b^{\g} d_{\g\d} \\
                       \tilde{V}_{\d\g} c^{\g}_{(0,2)} & 0 & 
\tilde{V}_{\d\g} b^{\g} & 0 & \tilde{V}_{\g\d}
                     \end{array}\right) \W_4.
\ee
Here the $SO(3,19)$ vielbein $ \tilde{V}_{\a\b}$ is defined by
$d_{\a\b} D^{\b}_{\hspace{2mm} \g} = \td{V}_{\a \b} \td{V}_{\b \g}$,
$c^{2}_{(0,2)} \equiv c^{\g}_{(0,2)} c^{\d}_{(0,2)} d_{\g \d}$ and
the matrix $\W_{4}$ is defined in the appendix \ref{bc_matrices}.
The six-dimensional tensor fields are related to the ten-dimensional
fields as
\bea
\la{redIIBfs6d}
   H^{1}_3 &=& \frac{e^{-\Phi}}{4}(1+\ast_6)H_3, \qquad
   H^{\a_{+} +1}_3 = -\frac{1}{\sqrt{8}}(\tilde{V} K_3)^{\a_{+}}, \\
%\qquad \g = 1 \cdots 3 \textrm{ (self-dual part)}, \nono \\
   H^{5}_3  &=& -\frac{e^{-\r/2}}{4}(1+\ast_6)F_3, \qquad
   H^{6}_3 = -\frac{e^{-\r/2}}{4}(1-\ast_6)F_3, \nono \\
   H^{\a_{-} + 3}_3 &=& -\frac{1}{\sqrt{8}}(\tilde{V} K_3)^{\a_{-}},
   \qquad
% \qquad \g = 4 \cdots 22 \textrm{ (anti-self dual part)}, \nono \\
   H^{26}_3 = \frac{e^{-\Phi}}{4}(1-\ast_6)H_3. \nn
\eea
Here $\a_{+} = 1,2,3$ and $\a_{-} = 4,\cdots 22$, labeling the
self dual and anti-self dual forms respectively.
Note that using formulas \eqref{redIIBscalars} and
\eqref{redIIBfs6d} to lift a six-dimensional solution to ten
dimensions requires a specific choice of six-dimensional vielbein.

The solutions we find have $D^{\g}_{\,\d} = d_{\g\d}$; this
implies the identity
\be \la{id-k3}
(\w_{2}^{\a_-})_{\rho \sigma} (\w_{2}^{\b_-})_{\tau}^{\hspace{2mm} \sigma}
= \half g_{\rho\tau} \d^{\a_-\b_-},
\ee
where $(\rho,\tau)$ are $K3$ coordinates and $g_{\rho \tau}$ is the
K3 metric. As discussed in \cite{Aspinwall}, a choice of
$D^{\g}_{\hspace{2mm} \e}$ fixes the complex structure completely and implies
$(\w_{2}^{\g})_{\rho \sigma} (\w_{2}^{\d})^{\rho \sigma} =
D^{\e}_{\hspace{2mm} \d} d_{\g \e}$. Varying this identity with respect to the
metric results in (\ref{id-k3}).

\subsection{S-duality in 6 dimensions}

Given the map between 10-dimensional and 6-dimensional fields,
we can now obtain the action of S-duality on 6-dimensional fields
as part of the $SO(5,21)$ symmetry:
\be
\la{6dSdual}
   G_3 \rightarrow O_S G_3, \qquad \cM^{-1} \rightarrow O_S \cM^{-1} O_S^T,
\ee
where
\be
   (O_S)_{ij} = \left(\begin{array}{ccc} 0 & 0 & -1 \\ 0 & I_3 & 0
\\ 1 & 0 & 0 \end{array}\right), \qquad (O_S)_{rs} =
\left(\begin{array}{ccc} 0 & 0 & 1 \\ 0 & I_{19} & 0 \\ -1 & 0 & 0 \end{array}\right),
\ee
\comment{\begin{align}
   &(O_SG)^{i=1} = -G^{i=2}, & &(O_SG)^{r=6} = -G^{r=7}, \nono \\
   &(O_SG)^{i=2} = G^{i=1}, & &(O_SG)^{r=7} = G^{r=6}, \nono \\
   &(O_SG)^{i\geq 3} = G^{i\geq 3}, & & (O_SG)^{r\geq 8} = G^{r\geq 8}.
\end{align}
\be
   \left(\begin{array}{c} (O_SG)^{i=1} \\(O_SG)^{i=2} \\ (O_SG)^{i\geq 3} \end{array}\right)
    = \left(\begin{array}{ccc} 0 & -1 & 0 \\ 1 & 0 & 0 \\ 0 & 0 & I_3 \end{array}\right)
   \left(\begin{array}{c} G^{i=1} \\ G^{i=2} \\ G^{i\geq 3} \end{array}\right)
\ee
}
Moreover one can perform an $SO(5) \times SO(21)$ transformation
   to bring the vielbein of the S-dual solution back to the form
used by the 10-dimensional lift. Including this transformation, $H_3$ and $V$ transform as
\be
\la{6dSdualgauge}
   H_3 \rightarrow O_G H_3, \qquad V \rightarrow O_G V O_S^T,
\ee
with
\be
   (O_G)_{ij} = \frac{1}{|\t|}\left(\begin{array}{ccc} C_0 & 0 &
   -e^{\hat{\Phi}}\\ 0 & I_3 & 0 \\ e^{\hat{\Phi}} & 0 & C_0
   \end{array}\right), \qquad (O_G)_{rs} =
   \frac{1}{|\t|}\left(\begin{array}{ccc} C_0 & 0 &
   -e^{\hat{\Phi}}\\ 0 & I_{19} & 0 \\ e^{\hat{\Phi}} & 0 & C_0
   \end{array}\right),
\ee
where $\t = C_0 + i e^{-\hat{\Phi}}$, $\hat{\Phi} = \Phi - \r/2$ is
the 10-dimensional dilaton and the fields $C_0$ and $e^{\hat{\Phi}}$
are the original ones taken before the S-duality.

\subsection{Basis change matrices} \la{bc_matrices}

In defining six-dimensional supergravities there are implicit choices of
constant $SO(p,q)$ matrices. When discussing the compactification from
the ten to six dimensions, the most convenient choices for these
matrices are certain off-diagonal forms, see for example
\cite{Dabholkar:1995nc, Sen:1995cj,
  Maharana:1992my,You97,Bergshoeff:1995sq,Behrndt:1995si}.
When one is interested in specific solutions of the six-dimensional
supergravity equations, such as $AdS_3 \times S^3$ solutions,
and deriving the spectrum in such backgrounds, it is rather more
convenient to use diagonal choices for these matrices, see for example
\cite{Sez98, Arutyunov:2000by}. In this paper we both compactify from
ten to six dimensions, and expand six-dimensional solutions about a
given background. We therefore find it most convenient to use diagonal
choices for the constant matrices. To use previous results on
compactification and T-duality, we need to apply certain similarity
transformations. For the most part these may be implicitly written in terms of
basis change matrices, so that compactification and duality formulas
remain as simple as possible. Thus let us define matrices
$\W_1$ and $\W_2$ for $SO(4,20)$, and $\W_3$ and $\W_4$ for $SO(5,21)$ via:
\bea
   &&\W_1^T \left( \begin{array}{c} v^{\rho} \\ w^{\rho} \\ x^{(c)} \end{array} \right) =
\left( \begin{array}{c} \frac{1}{\sqrt{2}} (v^{\rho}-w^{\rho}) \\
\frac{1}{\sqrt{2}}(v^{\rho}+w^{\rho}) \\ x^{(c)} \end{array} \right), \qquad
%\qquad \W_1
%\left(\begin{array}{ccc} 0 & -I_4 & 0 \\ -I_4 & 0 & 0 \\ 0 & 0
%& -I_{16} \end{array}\right) \W_1^T = \left(\begin{array}{cc} I_4 &
%     0 \\ 0 & -I_{20} \end{array} \right), \\
\W_3^T \left( \begin{array}{c} v \\ w \\ x^{(a)} \end{array} \right) = \left(
\begin{array}{c} \frac{1}{\sqrt{2}} (v - w) \\ x^{(a)} \\ \frac{1}{\sqrt{2}}(v +
w) \end{array} \right), \\
&&   \W_2^T \left( \begin{array}{c} v \\ w \\ x^{\a} \\ y^{\a_-}
\end{array} \right) = \left( \begin{array}{c} x^{\a}\\ \frac{1}{\sqrt{2}}
(v - w) \\ \frac{1}{\sqrt{2}}(v + w) \\ y^{\a_-} \end{array} \right), \qquad
%\qquad
%   \W_2 \left(\begin{array}{cccc} 0 & 1 & 0 & 0\\ 1 & 0 & 0 & 0\\ 0 &
%     0 & I_3 & 0 \\ 0 & 0 & 0 & -I_{19} \end{array}\right) \W_2^T =
%\left(\begin{array}{cc} I_4 & 0 \\ 0 & -I_{20} \end{array} \right), \nono \\
%\qquad
%   \W_3 \left(\begin{array}{cccc} 0 & 1 & 0 & 0\\ 1 & 0 & 0 & 0\\ 0 &
%     0 & I_4 & 0 \\ 0 & 0 & 0 & -I_{20} \end{array}\right) \W_3^T =
%\left(\begin{array}{cc} I_5 & 0 \\ 0 & -I_{21} \end{array} \right), \nono \\
     \W_4^T \left( \begin{array}{c} v_1 \\ w_1 \\ v_2 \\ w_2 \\ x^{\a} \\
y^{\a_-} \end{array} \right) = \left( \begin{array}{c}
\frac{1}{\sqrt{2}} (v_1 - w_1) \\ x^{\a} \\ \frac{1}{\sqrt{2}} (v_2 -
w_2) \\ \frac{1}{\sqrt{2}}(v_2 + w_2) \\ y^{\a_-} \\
\frac{1}{\sqrt{2}}(v_1 + w_1) \end{array} \right),
%\quad
%   \W_4 \left(\begin{array}{cccccc} 0 & 1 & 0 & 0 & 0 & 0\\ 1 &
%0 & 0 & 0 & 0 & 0\\ 0 & 0 & 0 & 1 & 0 & 0 \\ 0 & 0 & 1 & 0 & 0 & 0 \\
%0 & 0 & 0 & 0 & I_3 & 0 \\ 0 & 0 & 0 & 0 & 0 & -I_{19}
%   \end{array}\right) \W_4^T = \left(\begin{array}{cc} I_5 & 0 \\ 0 & -I_{21}
%end{array} \right).
\nono
\eea
where  $\rho =1,\cdots4$, $(c) = 1, \cdots 16$, $(a) = 1,\cdots 24$, $\a = 1,2,3$ and $\a_{-} =
1,\cdots 19$. These satisfy the conditions:
\bea
   \W_1 \left(\begin{array}{ccc} 0 & -I_4 & 0 \\ -I_4 & 0 & 0 \\ 0 & 0
& -I_{16} \end{array}\right) \W_1^T &=& \left(\begin{array}{cc} I_4 &
     0 \\ 0 & -I_{20} \end{array} \right), \\
   \W_2 \left(\begin{array}{ccc} \s_1 & 0 & 0\\
     0 & I_3 & 0 \\ 0 & 0 & -I_{19} \end{array}\right) \W_2^T &=&
\left(\begin{array}{cc} I_4 & 0 \\ 0 & -I_{20} \end{array} \right), \nono \\
   \W_3 \left(\begin{array}{ccc} \s_1 & 0 & 0 \\ 0 &
     I_4 & 0 \\ 0 & 0 & -I_{20} \end{array}\right) \W_3^T &=&
\left(\begin{array}{cc} I_5 & 0 \\ 0 & -I_{21} \end{array} \right), \nono \\
   \W_4 \left(\begin{array}{cccc} \s_1 & 0 & 0 & 0 \\ 0 & \s_1 & 0 & 0
     \\ 0 & 0 &  I_3 & 0 \\ 0 & 0 & o & -I_{19}  \end{array} \right )
\W_4^T &=& \left(\begin{array}{cc} I_5 & 0 \\ 0 & -I_{21} \end{array}
\right). \nn
\eea
Here $\s_1$ is the Pauli matrix $\left(\begin{array}{cc} 0 & 1 \\ 1 &
  0 \end{array} \right )$. 

\section{Properties of spherical harmonics} \la{sphere}

Scalar, vector and tensor spherical harmonics satisfy the following
equations
\bea
\Box Y^{I} &=& - \Lambda_{k} Y^{I}, \\
\Box Y_{a}^{I_v} &=& (1 - \Lambda_{k}) Y_{a}^{I_v}, \hsp D^{a}
Y_{a}^{I_v} = 0, \nn \\
\Box Y_{(ab)}^{I_t} &=& (2 - \Lambda_{k}) Y_{(ab)}^{I_t}, \hsp D^{a}
Y_{k(ab)}^{I_t} = 0, \nn
\eea
where $\Lambda_k = k (k+2)$ and the tensor harmonic is traceless. It
will often be useful to explicitly indicate the degree $k$ of the harmonic;
we will do this by an additional subscript $k$, e.g. degree $k$ spherical
harmonics will also be denoted by $Y_k^I$, etc.
$\Box$ denotes the d'Alambertian along the three sphere. The
vector spherical harmonics are the direct sum of two irreducible
representations of $SU(2)_L \times SU(2)_R$ which are characterized by
\be
\ep_{abc} D^b Y^{c I_v \pm} = \pm (k+1) Y_{a}^{I_v \pm}
\equiv \l_k Y_{a}^{I_v \pm}. \la{vec-dual}
\ee
The degeneracy of the degree $k$ representation is
\be
d_{k,\ep} = (k+1)^2 - \ep,
\ee
where $\ep = 0,1,2$ respectively for scalar, vector and tensor harmonics.
For degree one vector harmonics $I_v$ is an adjoint
index of $SU(2)$ and will be denoted by $\a$.
We use normalized spherical harmonics such that
\be
\int Y^{I_1} Y^{J_1} = \Omega_3 \d^{I_1 J_1}; \hsp
\int Y^{a I_v} Y_a^{J_v} = \Omega_3 \d^{I_v J_v}; \hsp
\int Y^{(ab) I_t} Y_{(ab)}^{J_t} = \Omega_3 \d^{I_t J_t},
\ee
where $\Omega_3 = 2 \pi^2$ is the volume of a unit 3-sphere.
We define the following triple integrals as
\bea
\int Y^{I} Y^{J} Y^K &=& {\Omega_3} a_{IJK}; \la{ap-ov0} \\
\int (Y^{\a \pm}_{1})^a Y^{j}_{1} D_a Y^{i}_1 &=& \Omega_3 e^{\pm}_{\a ij};
\la{ap-ov3}
\eea

\section{Interpretation of winding modes} \la{wind}

In the fundamental string supergravity solutions \eqref{cnm} the null curves describing
the motion of the string along a torus direction $x^{\rho}$ (whose
periodicity is $2 \pi R_{\rho}$) could have winding modes
such that $F_{\rho}(v) = w_{\rho} R_{\rho} v/R_y$, with $w_{\rho}$
integral. Consider now the correspondence with quantum string states.
Such winding modes are not consistent with both supersymmetry
and momentum and winding quantization for a string propagating in flat
space, with no $B$ field. Recall that the zero
modes of a worldsheet compact boson field can be written as
\be
X (\s^{+},\s^{-}) = x + \frac{1}{2} (\a' \frac{p}{R} + n R) \s^+ +
\frac{1}{2} (\a' \frac{p }{R} - n R)
\s^- \equiv x + \td{w} \s^{+} + {w} \s^{-},
\ee
where $R$ is the radius and $(p,n)$ are the quantized momentum and
winding respectively; note that we define $\s^{\pm} = (\t \pm \s)$.
BPS left-moving states with no right-moving excitations have
${w} = 0$ and hence $\a' p = - n R^2$. However the 
latter condition has no solutions at generic
radius and so states with winding along the torus directions
cannot be BPS. Therefore winding modes should not be included to
describe the F1-P states and corresponding dual D1-D5 ground states of
interest here.

Now consider switching on constant $B^{(2)}_{\rho v} \equiv b_{\rho}$ on the
worldsheet. The constant B field shifts the momentum charges, and thus
there are BPS left-moving states with winding
around the torus directions. To be more precise, following the
discussion of \cite{Kanitscheider:2006zf}, one can describe a
string with left-moving excitations using a null lightcone gauge.
The relevant terms in the worldsheet fields are then
\bea
V &=& w^{v} \s^-; \hsp
U = w^{u} \s^{-} + \td{w}^u \s^+ + \sum_{n} \frac{1}{\sqrt{n}}
a_n^- e^{-i n \s^-}; \\
X^{I} &=& \d_{I \rho} w^{I} \s^- + \sum_{n} \frac{1}{\sqrt{n}}
a_n^I e^{-i n \s^-}, \nn
\eea
where winding modes are included only along torus directions, labeled
by $\rho$. The $L_0$ constraint implies
\be
w^v w^u = (w^{\rho})^2 + 2 \sum_{n >0} | n | a^{I}_{n} a^{I}_{-n}
\equiv (w^v)^2 | \pa_V {X}^{I} |^2_{0},
\ee
where $|A|_0$ denotes the projection onto the zero mode. The momentum
and winding charges are given by
\be
P^{m} = \frac{1}{4 \pi} \int d \s (\pa_{\t} X^m + B^{(2)}_{mn} \pa_{\s}
X^n); \qquad
W^m = \frac{1}{2 \pi} \int d \s \pa_{\s} X^m,
\ee
respectively, where $\a' = 2$. Requiring no winding in the time
direction and no momentum along the $x^{\rho}$ directions imposes
$\td{w}^u = w^u + w^v$ and $w^{\rho} = b_{\rho} w^v$. The conserved
momentum and winding charges are then
\be \la{wsch}
P^M = \half w^v \left ( ( 1 +  | \pa_V {X}^{I} |^2_{0} + b_{\rho}^2),
(| \pa_V {X}^{I} |^2_{0} - b_{\rho}^2), 0 \right ); \qquad
W^{M} = w^v (0, 1, 0, b_{\rho}).
\ee
Note that the integral quantized momentum charge $p_y$ along the $y$
direction is therefore 
\be \la{mom-fp}
p_{y} = R_{y} (w^{u} - (w^v)^{-1} (w^{\rho})^2).
\ee
Now consider the solitonic string supergravity solution (\ref{cnm})  
with defining curves
$F^{I}(v)$ where $F^{\rho}(v) = b_{\rho} v + \bar{F}^{\rho}(v)$,
with $\bar{F}^{\rho}(v)$ having no zero mode. The ADM charges of this
solitonic string were computed in \cite{Dabholkar:1995nc}, and are
given by 
\be
P_{\rm{ADM}}^{M} = k Q \left ( ( 1 +  | \pa_v {F}^{I} |^2_{0}),
| \pa_v {F}^{I} |^2_{0}, 0, b_{\rho} \right ),
\ee
where the effective Newton constant is 
$k = \Omega_3 L_y/2 \k_6^2$. When $b_{\rho} = 0$ these charges match
the worldsheet charges \eqref{wsch} provided that $w^v = 2 k Q$ as in
\cite{Dabholkar:1995nc} but when $b_{\rho} \neq 0$
they do not quite agree with the
worldsheet charges. The reason is that in the supergravity solution
$B^{(2)}_{\rho v}$  approaches zero at infinity, but to match 
with the constant $B^{(2)}_{\rho v}$ background
on the worldsheet, $B^{(2)}_{\rho v}$ should approach $b_{\rho}$ at
infinity. This can be achieved via a constant gauge transformation $A_{\rho} \rightarrow
A_{\rho} - b_{\rho}$, combined with a coordinate shift $u \rightarrow u + 2
b_{\rho} x^{\rho}$. The ADM charges of this shifted background 
indeed exactly match the worldsheet charges (\ref{wsch}). 
The harmonic functions $A_{\rho}$ then take the form
\be
A_{\rho} = - b_{\rho} H - \frac{Q}{L_y} \int_{0}^{L_y} dv \frac{\pa_{v}
  \bar{F}^{\rho}}{ | x - F|^2}, 
\ee
where in the latter expression $ | x - F|^2$ denotes $\sum_{i} (x^i -
F^i(v))^2$; the harmonic function has been smeared over the $T^4$ and
the $y$ circle. Note that when $F^{i}(v) = 0$ the supergravity solution
collapses to 
\bea
ds^2 &=& H^{-1} dv (- du + K dv) + dx^I dx_I; \qquad K = (1 + \frac{Q
  | \pa_v {F}^{\rho} |^2_{0}}{r^2}), \\
e^{-2 \Phi} &=& H \equiv (1 + \frac{Q}{r^2}); \hsp
B^{(2)}_{uv} = \half (H^{-1} -1); \hsp
B^{(2)}_{v \rho} = - b_{\rho}. \nn
\eea
This is the naive $SO(4)$ invariant F1-P solution, with an additional 
constant $B$ field. Finally let us note that one can similarly switch
on winding modes for the curves $q^{(c)}(v)$ 
characterizing the charge waves in the
heterotic solution \eqref{het1} by including constant 
$A^{(c)}_{v}$ on the worldsheet.

\bigskip

Now let us consider solutions in the D1-D5 system, and the
interpretation of including winding modes of the internal curves. In
particular, it is interesting to note that the general 
$SO(4)$ invariant solutions include harmonic functions
\be \la{po}
{\cal A} = a_o + \frac{a}{r^2}; \hsp
{\cal A^{\a_-}} = a^{\a_{-}}_o + \frac{a^{\a_-}}{r^2},
\ee
in addition to the harmonic functions $(H,K)$ given in (\ref{naive}).
The non-constant terms in these harmonic functions are related to the
winding modes of
the internal curves, with the quantities $a^{\td{\a}} = (a,a^{\a_-})$ being given by
\be \la{zero}
a = - \frac{Q_5}{L} \int_{0}^L dv \dot{{\cal F}}(v); \hsp
a^{\a_-} = - \frac{Q_5}{L} \int_{0}^L dv \dot{{\cal F}}^{\a_-}(v).
\ee
Following the duality chain, these constants are given by $a^{\td{\a}}
= -Q_{5} b^{\td{\a}}$ where for the $T^4$ case $b^{\td{\a}} \equiv B^{(2)}_{\r v} =
b_{\rho}$ and for the $K3$ case $b^{\td{\a}} \equiv (B^{(2)}_{\r v} =
  b_{\rho}, A^{(c)}_{v} = b^{(c)})$.
The constant terms $(a_o,a^{\a_{-}}_o)$ are related to the boundary
conditions at asymptotically flat infinity, as we will discuss below. 

When these functions $({\cal A}, {\cal A^{\a_-}})$ are non-zero,
the geometry generically differs from the naive D1-D5 geometry. 
The functions $(f_1, \td{f}_1)$ appearing in the metric behave as
\bea
\td{f}_1 &=& 1 + \frac{Q_1}{r^2} - (1 + \frac{Q_5}{r^2})^{-1} \left (
(a_o + \frac{a}{r^2})^2 + (a^{\a_{-}}_o + \frac{a^{\a_-}}{r^2})^2
\right ) \nn \\
{f}_1 &=& 1 + \frac{Q_1}{r^2} - (1 + \frac{Q_5}{r^2})^{-1}  
\left ( (a^{\a_{-}}_o + \frac{a^{\a_-}}{r^2})^2
\right ). \la{func}
\eea
In the decoupling limit these functions become
\be
\td{f}_1 \rightarrow r^{-2} (Q_1 -
Q_5^{-1} (a^2 + a^{\a_-} a^{\a_-}) ) \equiv \frac{\td{q}_1}{r^2}; \qquad
{f}_1 \rightarrow r^{-2} (Q_1 -
Q_5^{-1} (a^{\a_-} a^{\a_-}) ) \equiv \frac{q_1}{r^2},
\ee
and thus $(a_o,a^{\a_{-}}_o)$ drop out. Note that $\td{q}_1$
corresponds to the conserved momentum charge in the F1-P system
\eqref{mom-fp}.  
Substituting the decoupling region functions into (\ref{equ_D1D5K3pot}), one finds
that the near horizon region of the solution is $AdS_3 \times S^3
\times M^4$, supported by both $F^{(3)}$ and $H^{(3)}$ flux:
\bea
ds^2 &=& \frac{r^2 \sqrt{q_1}}{\td{q_1} \sqrt{Q_5}} (-dt^2 + dy^2) +
\sqrt{q_1 Q_5} (\frac{dr^2}{r^2} + d\Omega_3^2) +
\frac{\sqrt{q_1}}{\sqrt{Q_5}} ds^2_{M^4}; \\
e^{2 \Phi} &=& \frac{q_1^2}{Q_5 \td{q}_1}, \qquad
F^{(3)}_{tyr} = - \frac{2r}{\td{q}_1}, \qquad
F^{(3)}_{\Omega_3} = 2 q_{1}^{-1} \td{q}_1 Q_5; \nn \\
H^{(3)}_{tyr} &=& 2 a Q_{5}^{-1} \td{q}_1^{-1} r, \qquad
H^{(3)}_{\Omega_3} = - 2 a. \nn
\eea
%Note that this solution is supersymmetric only when $a/Q_{5} < 0$;
%this is a necessary condition for the projections in the Killing
%spinor equation related to $F^{(3)}$ and $H^{(3)}$ to be identical.
%This same condition was required in the original F1-P system also. 
The field strengths $F^{(1)}$ and $F^{(5)}$ vanish, but there are
non-vanishing potentials:
\bea
B_{\r \s}^{(2)} &=& \sqrt{2} Q_{5}^{-1} a^{\a_-} \w^{\a_{-}}_{\r \s},
\qquad
C^{(0)} = - q_{1}^{-1} a, \qquad
C^{(4)}_{\r \sigma \tau \pi} = Q_{5}^{-1} a \ep_{\r \s \t \pi}; \\
C^{(4)}_{t y \a \b} &=& a (1 + \td{q}_{1}^{-1} r^2) \e_{\a\b}, \qquad
C^{(4)}_{\a \b \r \s} = 2 \sqrt{2} \e_{\a\b} a^{\a_-} \w^{\a_{-}}_{\r
  \s}, \qquad
C^{(4)}_{t y \r \s}= \sqrt{2} Q_{5}^{-1}  a^{\a_-} \w^{\a_{-}}_{\r
  \s}, \nn 
\eea
where $\e$ is a 2-form such that $d\e$ is the volume form 
of the unit 3-sphere. The conserved
charges therefore include Chern-Simons terms; using the equations of
motion \eqref{conv_IIB} one finds that they are given by 
\bea
D5 &:& Q_5 = \half \int_{S^3} (F^{(3)} + H^{(3)} C^{(0)});  \nn \\
D1 &:& \td{q}_1 = \half \int_{S^3 \times M^4} (\ast F^{(3)} + H^{(3)} \wedge C^{(4)} ); \\
D3 &:& a^{\a_{-}} = \frac{1}{2 \sqrt{2}} \int_{S^3 \times \w^{\a_{-}}} B^{(2)}
\wedge (F^{(3)} + H^{(3)} C^{(0)}); \nn \\
NS5 &:& a = - \half \int_{S^3} H^{(3)}, \nn
\eea
where we drop terms which do not contribute to the charges. 
The curvature radius of the $AdS_3 \times S^3$ is $l = (q_1 Q_5)^{1/4}$,
and the three-dimensional Newton constant is 
\be
\frac{1}{2 G_{3}} = \frac{8 \pi V_4 \Omega_3}{\k_{10}^2}
\frac{\td{q_1}}{q_1} (q_1 Q_5)^{3/4},
\ee
with the volume of $M^4$ being $(2 \pi)^4 V$ and $2 \k_{10}^2 = (2
\pi)^7 (\a')^4$. Then using \cite{Brown:1986nw,HK}
the central charge of the dual CFT is 
\be
c = \frac{3l}{2 G_{3}} = 6 \frac{V}{(\a)'^4} \td{q_1} Q_5 \equiv 6 \td{n}_1 n_5
\ee
where the integral charges $(\td{n}_1, n_5)$ are given by
\be
Q_{5} = \a' n_5; \hsp
\td{q_1} = \frac{(\a')^3 \td{n}_1}{V}.
\ee
Now consider the relation between this system and the F1-P system
discussed previously. The conserved charges here are $(Q_5, \td{q}_1,
a, a^{\a_{-}})$, which correspond to the winding, momentum along the
$y$ circle and winding along the internal manifold in the original system. 
The fact that $(a,a^{\a_{-}})$ measure NS5-brane and D3-brane charges in
the final system is consistent with the duality chains from the F1-P
systems: applying the standard duality rules along 
the chains given in (\ref{chain1}),(\ref{chain2}) and (\ref{chain3}), one
indeed finds that the original winding charges become NS5-brane and D3-brane
charges.  

Finally let us comment on the constant terms in the harmonic
functions, $(a_{o},a_{o}^{\a_-})$. These clearly determine the behavior of the
solution at asymptotically flat infinity: the $B$ field and RR
potentials at infinity depend on them. Now consider how these constant
terms can be described in the CFT. In the context of the pure D1-D5
system it was noted in \cite{Kanitscheider:2006zf} that
(infinitesimal) constant terms in the harmonic
functions $(f_1,f_5)$ can be reinstated by making (infinitesimal)
irrelevant deformations of the CFT by $SO(4)$ singlet operators.
See also \cite{Skenderis:2006di} for a related discussion in the context of the $AdS_5/CFT_4$
correspondence. It seems probable that a similar
interpretation would hold here: the $(n_t - 1)$ parameters
$(a_o,a_o^{\a_-})$  (where $n_t = 5,21$ for $T^4$ and $K3$ respectively) would be
related to the parameters of deformations of the
CFT by irrelevant $SO(4)$ singlet operators.  In total taking into
account these $(n_t - 1)$ zero modes, plus the two constant terms in the
$(f_1,f_5)$ harmonic functions, one gets $(n_t + 1)$ parameters. This
agrees exactly with the count of the number of irrelevant $SO(4)$
singlet operators\footnote{Such deformations may also be related to
the attractor flow of moduli; this idea is currently being developed
by Kyriakos Papadodimas and collaborators.}. How to describe these deformations
in the field theory beyond the infinitesimal level is not known,
however.

\section{Density of ground states with fixed R charges} \la{dens}

In this appendix we will derive an asymptotic formula for the number
of R ground states with given R charges. Our derivation follows
closely that of \cite{Russo:1994ev} for the density of fundamental
string states with a given mass and angular momentum. In fact, we will
consider the case of $K3$, so the relevant counting is precisely that
of the density of left moving heterotic string states with a given
excitation level $N$ and (commuting) angular momenta $(j^{12},j^{34})$ in
the transverse $R^4$. For this purpose we can consider the following
Hamiltonian
\be
H = \sum_{n=1}^{\infty} \left (\sum_{(a)=1}^{24} \a^{(a)}_{-n} \a^{(a)}_n
\right ) + \lambda_1 j^{1} +
\lambda_2 j^{2},
\ee
where $(\lambda_1,\lambda_2)$ are Lagrange multipliers and
\be
j^{1} = j^{12} =  - i \sum_{n=1}^{\infty} n^{-1} ( \a^{1}_{-n} \a^{2}_n -
\a^{2}_{-n} \a^{1}_n); \hsp
j^{2} = j^{34} = - i \sum_{n=1}^{\infty} n^{-1} ( \a^{3}_{-n} \a^{4}_n -
\a^{4}_{-n} \a^{3}_n).
\ee
Here the oscillators satisfy the standard commutation relations,
namely $\left [ \a_n^{(a)}, \a_m^{(b)} \right ] = n \d_{n+m}
\d^{(a)(b)}$. In \cite{Russo:1994ev} the partition function was
computed in the case $\l_2 =0$, and thus the partition function
of interest here can be computed by generalizing their results. The
first step is to diagonalize the Hamiltonian by introducing
combinations
\be
a_{n}^{12} = \frac{1}{\sqrt{2n}} (\a^1_n + i \a^2_n); \hsp
b_{n}^{12} = \frac{1}{\sqrt{2n}} (\a^1_n - i \a^2_n)
\ee
and analogously $(a_n^{34},b_n^{34})$. Then the Hamiltonian takes the
form
\bea
H &=& \sum_{n=1}^{\infty} \left (\sum_{(a)=5}^{24} \a^{(a)}_{-n} \a^{(a)}_n
+ (n - \lambda_1) (a_{n}^{12})^{\dagger} a_{n}^{12}
+ (n + \lambda_1) (b_{n}^{12})^{\dagger} b_{n}^{12} \right . \\
&& \qquad \left . + (n - \lambda_2) (a_{n}^{34})^{\dagger} a_{n}^{34}
+ (n + \lambda_2) (b_{n}^{34})^{\dagger} b_{n}^{34}
\right ) \nn
\eea
The partition function $Z = \rm{Tr} (e^{-\b H})$ is then
\be
Z = \prod_{n=1}^{\infty} \left [ (1- w^n)^{-20} (1 - c_1 w^n)^{-1} (1-
  c_1^{-1} w^n)^{-1} (1 - c_2 w^n)^{-1} (1 - c_2^{-1} w^n)^{-1} \right ]
\ee
with $w = e^{-\b}$ and $c_1 = e^{\b \l_1}$, $c_2 = e^{\b \l_2}$. To
estimate the asymptotic density of states, one as usual expresses the
partition function in terms of modular functions and then uses the
modular transformation properties. Here one needs the Jacobi theta
function
\be
\q_1(z | \t) = 2 f(q^2) q^{1/4} \sin (\pi z) \prod_{n=1}^{\infty} (1
- 2 q^{2n} \cos (2 \pi z) + q^{4n} ),
\ee
with
\be
f(q^2) = \prod_{n=1}^{\infty} (1 - q^{2n}), \qquad q = e^{i \pi \t},
\ee
and the modular transformation property
\be
\q_1(- \frac{z}{\t} | - \frac{1}{\t}) = e^{ i \pi/4} \sqrt{\t} e^{i
  \pi z^2/\t} \q_1(z | \t)
\ee
Rewriting the partition function in terms of the modular functions,
applying this modular transformation and then taking the high
temperature limit results in
\be
Z(\b,\l_1,\l_2) = C \b^{12} e^{4 \pi^2/\b} \frac{\l_1 \l_2}{\sin(\pi
  \l_1) \sin(\pi \l_2) },
\ee
with $C$ a constant. From this expression one can extract the density
of states with level $N$ and angular momenta $(j^{1},j^{2})$ by
expanding
\be
Z (w,k_1,k_2) = \sum_{N,j} d_{N,j^{1},j^{2}} w^{N} e^{i k_1 j^{1} + i k_2
  j^{2}},
\ee
where $k_{1} = - i \b \l_1$ and $k_2 = - i \b \l_2$, and projecting
out the $d_{N,j^1,j^2}$. Integrating over $(k_1,k_2)$ can be done exactly,
since
\be
\int_{-\infty}^{\infty} dk e^{i k y} \frac{k}{\sinh(\pi k/\b)} = \half
\b^2 \frac{1}{\cosh^2 (\b y/2)},
\ee
resulting in the following contour integral over a circle around $w=0$
for $d_{N,j^{1},j^{2}}$:
\be
d_{N,j^{1},j^{2}} = C' \oint \frac{dw}{w^{N+1}} \b^{14} e^{4 \pi^2/\b} \frac{1}
{\cosh^2 (\b j^{1}/2) \cosh^2 (\b j^{2}/2)}.
\ee
Assuming $N$ is large the integral can be approximated by a saddle
point evaluation, with the saddle point defined by the solution of
\be
\frac{4 \pi^2}{\b^2} = N + 1 - j^{1} \tanh (\half j^{1} \b) - j^{2}
\tanh (\half j^{2} \b).
\ee
For small angular momenta, which is the case of primary interest here, the
solution is $\b \cong 2 \pi/\sqrt{N+1}$. For $(\left | j^{1} \right |,
\left | j^{2} \right |) = {\cal{O}}(N)$ the stationary point is at
\be
\b \cong \frac{2 \pi}{\sqrt{N + 1 - \left | j^{1} \right | - \left |
    j^{2} \right |}}.
\ee
Note that $\left | j^{1} \right | + \left | j^{2} \right | \le N$.
This latter stationary point is equally applicable to small angular
momenta, and thus one can write the asymptotic density of states as
\be
d_{N,j^{1},j^{2}} \cong \frac{1}{ 4 (N + 1- j)^{31/4}} \exp \left [ \frac{2 \pi (2 N -
    j )}{ \sqrt{N + 1 - j}} \right ] \frac{1}{ \cosh^2( \frac{ \pi j^{1}}{
\sqrt{N + 1 - j}} )
\cosh^2( \frac{ \pi j^{2}}{  \sqrt{N + 1 - j }} )}, \la{dens2}
\ee
where $j = \left | j^{1} \right | + \left | j^{2} \right |$. The
constant of proportionality is fixed by the state with $j^{1} = N$,
$j^2 =0$ being unique. Note that the commuting generators
$(j_3,\bar{j}_3)$ of $(SU(2)_L,SU(2)_R)$ respectively are related to the rotations in the
1-2 and 3-4 planes via $j_3 = \half (j^1 + j^2)$ and $\bar{j}_3 = \half (j^1 - j^2)$.
The total number of states at level $N$ is
\be
d_{N} \cong \frac{1}{N^{27/4}} \exp (4 \pi \sqrt{N}), \la{dens3}
\ee
and thus the density of states with zero angular momenta differs from
the total number of states only by a factor of $1/N$; the exponential
growth with $N$ is the same.

\end{document}